%% file: main.tex
\documentclass[conference]{IEEEtran}
\let\labelindent\relax
\usepackage{bbding} % corresponding letter

\input{packages.tex}

\ifCLASSINFOpdf
\else
\fi

\newcommand{\para}[1]{\vspace{2pt}\noindent\textbf{{#1}}}
\def\ourmethod{GRASP}

\newcounter{qr}

\newcounter{ql}

\newlength{\qt}

\newcounter{itemnummer}
\newcommand{\Qitem}[2][]{% #1 optional, #2 notwendig
\ifthenelse{\equal{#1}{}}{\stepcounter{itemnummer}}{}
\ifthenelse{\equal{#1}{a}}{\stepcounter{itemnummer}}{}
\begin{enumerate}[topsep=2pt,leftmargin=2.8em]
\item[\textbf{\arabic{itemnummer}#1.}] #2
\end{enumerate}
}

\definecolor{bgodd}{rgb}{0.8,0.8,0.8}
\definecolor{bgeven}{rgb}{0.9,0.9,0.9}
\newcounter{itemoddeven}
\newlength{\gb}
\newcommand{\QItem}[2][]{% #1 optional, #2 notwendig
\setlength{\gb}{\linewidth}
\addtolength{\gb}{-5.25pt}
\ifthenelse{\equal{\value{itemoddeven}}{0}}{%
\noindent\colorbox{bgeven}{\hskip-3pt\begin{minipage}{\gb}\Qitem[#1]{#2}\end{minipage}}%
\stepcounter{itemoddeven}%
}{%
\noindent\colorbox{bgodd}{\hskip-3pt\begin{minipage}{\gb}\Qitem[#1]{#2}\end{minipage}}%
\setcounter{itemoddeven}{0}%
}
}

\begin{document}
\title{Gradient Shaping: Enhancing Backdoor Attack Against Reverse Engineering \thanks{Corresponding authors\Envelope: Di Tang (Indiana University Bloomington).}}
%\title{Fortifying Immunity: Backdoor Defense via Pre-trained Anti-Poisoning Models}
\author{
    \IEEEauthorblockN{
        Rui Zhu\IEEEauthorrefmark{1},
        Di Tang\IEEEauthorrefmark{1}\textsuperscript{\Envelope},
        Siyuan Tang\IEEEauthorrefmark{1},
        Zihao Wang\IEEEauthorrefmark{1},
        Guanhong Tao\IEEEauthorrefmark{2},
        Shiqing Ma\IEEEauthorrefmark{3},
        Xiaofeng Wang\IEEEauthorrefmark{1},
        Haixu Tang\IEEEauthorrefmark{1}
    }
    \IEEEauthorblockA{\IEEEauthorrefmark{1}Indiana University Bloomington\\
    Emails: zhu11@iu.edu, tangd@iu.edu, tangsi@iu.edu, zwa2@iu.edu, xw7@indiana.edu, hatang@indiana.edu}
    \IEEEauthorblockA{\IEEEauthorrefmark{2}Purdue University\\
    Email: taog@purdue.edu}
    \IEEEauthorblockA{\IEEEauthorrefmark{3}University of Massachusetts Amherst\\
    Email: Shiqingma@umass.edu}
}
\IEEEoverridecommandlockouts
\makeatletter\def\@IEEEpubidpullup{6.5\baselineskip}\makeatother
\IEEEpubid{\parbox{\columnwidth}{
    Network and Distributed System Security (NDSS) Symposium 2024\\
    26 February - 1 March 2024, San Diego, CA, USA\\
    ISBN 1-891562-93-2\\
    https://dx.doi.org/10.14722/ndss.2024.24450\\
    www.ndss-symposium.org
}
\hspace{\columnsep}\makebox[\columnwidth]{}}

\maketitle

\input{0-abstract.tex}

\input{1-introduction.tex}

\input{2-background.tex}

\input{3-observation}

\input{4-gradient-grasp}

\input{5-Against-Backdoor-detection}

\input{6-Resilience-to-Backdoor-Mitigation}
\input{7-mitigation-and-limitaion}
\input{8-Related-work}
\input{9-Discussion}

\input{10-Conclusion}

\bibliographystyle{IEEEtranS}
\bibliography{main}
\input{99_appendix.tex}
\end{document}

%% file: packages.tex
\usepackage{epsfig,endnotes}
\usepackage{
  color,
  float,
  epsfig,
  wrapfig,
  graphics,
  graphicx,
  subcaption
}
\usepackage{caption}

\usepackage{bm}
\usepackage{amsmath}
\usepackage{subcaption}
\usepackage{amsmath}
\makeatletter
\usepackage{textcomp}
\usepackage{setspace}
\usepackage{latexsym,fancyhdr,url}
\usepackage{enumerate}
\usepackage{graphics}
\usepackage{xparse} % argument parsing -- \edist
\usepackage{xspace}
\usepackage{amsfonts}
\usepackage{multirow}
\usepackage{csvsimple}
\usepackage{balance}
\usepackage[english]{babel}
\usepackage{amsthm}
\usepackage{enumitem}
\usepackage{pdfpages}
\usepackage{algorithm}
\usepackage{algpseudocode}
\usepackage{ulem}
\usepackage{cancel}

\theoremstyle{definition}
\theoremstyle{remark}

\newtheoremstyle{bfnote}%
  {}{}
  {\itshape}{}
  {\bfseries}{.}
  { }{\thmname{#1}\thmnumber{ #2}\thmnote{ (#3)}}
\theoremstyle{bfnote}
\newtheorem{definitionbold}{Definition}
\newtheorem{theorembold}{Theorem}
\newtheorem{lemmabold}{Lemma}

\newcommand{\ignore}[1]{}

\usepackage{tikz}
\usepackage{amsmath}
\usepackage{verbatim}
\usepackage{graphicx}
\usepackage{color}
\usepackage{booktabs}
\usepackage{tabularx}
\usepackage{listings}
\usepackage{algorithm}
\usepackage{algpseudocode}
\usepackage{enumitem}
\usepackage{url}
\usepackage[hidelinks]{hyperref}
\usepackage{wasysym}
\usepackage{forloop}
\usepackage{adjustbox}
\usepackage{ifthen}
\usepackage{multirow}

%% file: 0-abstract.tex
\begin{abstract}
Most existing methods to detect backdoored machine learning (ML) models take one of the two approaches: trigger inversion (aka. reverse engineer) and weight analysis (aka. model diagnosis). In particular, the gradient-based trigger inversion is considered to be among the most effective backdoor detection techniques, as evidenced by the TrojAI competition~\cite{trojai}, Trojan Detection Challenge~\cite{NIPS_competation} and backdoorBench~\cite{benchmark}. However, little has been done to understand why this technique works so well and, more importantly, whether it raises the bar to the backdoor attack. In this paper, we report the first attempt to answer this question by analyzing the change rate of the backdoored model's output around its trigger-carrying inputs. Our study shows that existing attacks tend to inject the backdoor characterized by a low change rate around trigger-carrying inputs, which are easy to capture by gradient-based trigger inversion. In the meantime, we found that the low change rate is not necessary for a backdoor attack to succeed: we design a new attack enhancement method called \textit{Gradient Shaping} (\ourmethod{}), which follows the opposite direction of adversarial training to reduce the change rate of a backdoored model with regard to the trigger, without undermining its backdoor effect. Also, we provide a theoretic analysis to explain the effectiveness of this new technique and the fundamental weakness of gradient-based trigger inversion. Finally, we perform both theoretical and experimental analysis, showing that the \ourmethod{} enhancement does not reduce the effectiveness of the stealthy attacks designed to evade the backdoor detection methods based on weight analysis, as well as other backdoor mitigation methods without using detection.

\end{abstract}

% TODO: replace this section with code generated by the tool at https://dl.acm.org/ccs.cfm
% \begin{CCSXML}
% <ccs2012>
%    <concept>
%        <concept_id>10002978</concept_id>
%        <concept_desc>Security and privacy</concept_desc>
%        <concept_significance>500</concept_significance>
%        </concept>
%    <concept>
%        <concept_id>10010147.10010257</concept_id>
%        <concept_desc>Computing methodologies~Machine learning</concept_desc>
%        <concept_significance>500</concept_significance>
%        </concept>
%  </ccs2012>
% \end{CCSXML}

% \ccsdesc[500]{Security and privacy; Computing methodologies~Machine learning}
% % -- end of section to replace with generated code

% \keywords{Trojan Attack; deep learning security; pre-trained model; transfer learning; adversarial machine learning}

%% file: 1-introduction.tex
\section{Introduction}\label{sec:intro}

Critical to trustworthy AI is the trustworthiness of machine learning (ML) models, which can be compromised by malevolent model trainers, evil-minded training data providers, or any parties with access to any link on the ML supply chain (e.g., pre-trained models) to inject a backdoor (aka., trojan). A backdoored model is characterized by strategic misclassification of the input carrying a unique pattern called \textit{trigger}: e.g., special glasses worn by a masquerader to impersonate an authorized party against a compromised facial-recognition system. So the assurance of ML models can only be upheld by effectively detecting those backdoored models, which have been intensively studied in recent years. Existing backdoor defense methods have been reviewed by an SoK paper~\cite{backdoorsurvey}: among seven general defense strategies, two are based on backdoor detection, which uses either the trigger inversion (aka. trigger synthesis) or weight analysis techniques (aka. model diagnosis)~\cite{NC}\cite{abs}\cite{k-arm}\cite{AC}\cite{MNTD}. The most concrete progress in the backdoor detection has been at least partially attributed to \textit{trigger inversion} related techniques, as evidenced in the TrojAI competition~\cite{trojai} (9 out of 11 rounds won by inversion approaches, the rest two won by weight analysis) and the BackdoorBench project~\cite{DBD} (leading performers are mostly gradient-based trigger inversion). However, little has been done to understand whether these approaches raise the bar to the backdoor attacks or are just another porous defense line permeable by the knowledgeable adversary.

\vspace{2pt}\noindent\textbf{Achilles' heel of gradient-based optimization}. Trigger inversion is a technique that automatically recovers a pattern causing an ML model to misclassify the pattern-carrying input. Such a pattern is considered a putative trigger and utilized to determine whether the model is indeed backdoored. This reverse-engineering step mostly relies on gradient descent, which seeks the greatest tendency towards misclassification following the opposite direction of the model's gradient with regard to its input. A prior study shows that almost all proposed trigger inversion approaches are gradient-based~\cite{backdoorsurvey}. Although gradient-based optimization can converge to a local optimum, this convergence is contingent upon selecting a proper size for each search step and a proper initialization. In the presence of a function with low robustness around trigger-inserted inputs (e.g., the one having a steep slope (large changing rate)), a large step size could overshoot the local minimum for the trigger which leads to misclassification. On the other hand, a small step size could render the convergence process exceedingly slow and increase the probability that the optimizer converges to another local minimum, practically thwarting any trigger inversion attempt. 
% On the other hand, another major backdoor detection technique, weight analysis, is a technique that utilizes a classifier to distinguish the Trojan model. Such a classifier takes the parameters of the victim model as input and outputs the likelihood of the existence of a backdoor. This kind of classifier relies on observing multiple times of behavior of the Trojan model and benign model parameters within the same task. Then summarize a strong signal from model parameters or, even more intuitively, train a classifier on sets of parameters, then utilize the  summarized signal or trained classifier to distinguish the Trojan model. When a function with low robustness around trigger-inserted inputs indicates more area has been predicted correctly in the local area around trigger-inserted inputs, and for a benign model, this lead to a function with low robustness around trigger-inserted inputs could hardly be detected by weight analysis because almost the entire local area around trigger-inserted inputs should be predicted correctly. This leads to a low trigger robustness model being more similar to a benign model.
So a fundamental question not asked before is why gradient-based reverse engineering works so well on the backdoors injected using today's techniques and whether a more powerful backdoor capable of defeating the inversion can be injected under practical threat models.

\vspace{2pt}\noindent\textbf{Analysis and findings}. To answer this question, we conducted the first study to understand the limitations of trigger inversion. Our research shows that today's backdoor injection techniques, both loss-function manipulation, and data poisoning, turn out to be quite amenable to gradient-based optimization. Actually, given the relatively simple features that characterize today's triggers (e.g., geometric shapes), a backdoor learned could be more robust to the noise added to its trigger than the benign task the infected model claims to perform, as observed in our experiments:  we found that oftentimes, backdoors tend to be more resilient to the noise than the primary task to the perturbation on its features (Section~\ref{sec:observation}).

This observation indicates that the backdoor can be invoked by not only the trigger but a wide range of its variations. Therefore, the average change rate of the backdoored model around trigger-inserted inputs for recognizing a trigger cannot be too high, which can be easily captured with a relatively larger scope of  search step size and initialization selection. This explains why trigger inversion works so well in backdoor detection. However, a slow change rate (or high trigger effective radius) is \textit{not} required for a backdoor attack to succeed. Our research shows that the change rate can be increased through data contamination without undermining the effectiveness of the backdoor attack. In our research, we designed a simple algorithm that could enhance the backdoor attack, called gradient shaping (\ourmethod{}) that utilizes both mislabeled data and correctly labeled data with noised triggers to contaminate the training set, in an opposite way to the augmentation training~\cite{augmentationtraining}, so as to narrow down the variation of the trigger pattern capable of invoking the backdoor. We theoretically analyze this approach and show that it effectively raises the change rate, thereby weakening the detection ability from trigger inversion.

It is worth noting that \ourmethod{} represents a different type of backdoor attack compared with the stealthy backdoors proposed recently (e.g., \cite{lira} \cite{simtrojan}). Existing stealthy backdoor methods attempt to devise specific triggers, often dependent on the target neural network model so that they are hard to detect and mitigate by defense methods. \ourmethod{}, on the other hand, is a generic trigger injection method that injects any trigger designed by the attacker into a target model so that the trigger is harder to detect and mitigate by the trigger inversion-based backdoor defenses. As a result, \ourmethod{} can be combined with existing stealthy backdoor methods to enhance their capability to evade the trigger inversion-based defenses. 
Our studies show that existing backdoor attacks less capable of evading trigger inversion can be boosted by \ourmethod{} to easily defeat most representative inversion protection, including Neural Cleanse (NC)~\cite{NC}, tabor~\cite{tabor}, k-arm~\cite{k-arm}, pixel~\cite{pixelbackdoor}, rendering them incapable of capturing any trigger of a backdoored model.  

We also perform a theoretical and experimental analysis (Appendix~\ref{subsec:theory_weight_ana} and Section \ref{subsec:Against Weight Analysis}, respectively.) to show that \ourmethod{} does not make the backdoor more vulnerable to weight analysis, which is the other mainstream technique for backdoor detection. In particular, our experiment shows that the \ourmethod{} enhancement does not decrease the effectiveness of the backdoor attacks such as DFST~\cite{DFST}, AB~\cite{adaptive-blend}, and DEFEAT~\cite{defeat} against the weight analysis-based detection. Finally, our study demonstrates that the effectiveness of \ourmethod{} against trigger inversion does not make the enhanced attacks more vulnerable to other backdoor mitigation or unlearning techniques, such as Fine-purning~\cite{fine-purning}, NAD~\cite{NAD}, Gangsweep~\cite{zhu2020gangsweep}, DBD~\cite{DBD}, RAB~\cite{rab}, and ABL~\cite{ABL}. 

\vspace{2pt}\noindent\textbf{Contributions}. The contributions of the paper are outlined below: 

\vspace{2pt}\noindent$\bullet$\textit{ First in-depth analysis on trigger inversion}. We report the first in-depth analysis that explains why trigger inversion works so well on backdoor detection. This leads to the discovery of the fragility of the advance we made in this area, given the observation that the weakness of today's trigger injection can be addressed without undermining the effectiveness of the backdoor. 

\vspace{2pt}\noindent$\bullet$\textit{ New backdoor injection technique}. Our new understanding of trigger inversion has been made possible by a new backdoor injection technique, which exploits the fundamental limitation of gradient-based optimization and works under realistic threat models. As such, this method can enhance existing backdoor attacks, making it more effective in evading trigger inversion, but not less effective in evading the weight analysis-based detection and other defenses.

%% file: 2-background.tex
\section{Background}\label{sec:background}

\subsection{Modeling Backdoor Attacks} \label{Backdoor Attack Modeling}

In a backdoor attack, the adversary intends to inject a backdoor (Trojan) into the target ML model for the purpose of causing the model to produce desired outputs for trigger-inserted inputs. In our research, without loss of generality, we focus on the backdoor attacks against image classification models.

\vspace{3pt}\noindent\textbf{Classification model}. 
A classification model is represented as $f(\cdot):\mathcal{X} \mapsto \bar{\mathcal{Y}}$, while $\arg\max f(\cdot) \in \mathcal{Y}$ represents the predicted label of the given input.
Specifically, $\mathcal{X} \subseteq \mathbb{R}^{m}$, $\bar{\mathcal{Y}}\subseteq [0,1]^K$, $\mathcal{Y} = \{0,1,...,K\}$, and $K$ is the number of classes. 
We refer to $f_D$ as a model trained on dataset $D$. Generally, we consider a dataset $D$ containing $n$ independent training samples, i.e., $D=\{x_i, y_i\}_{i=1}^{n}$, where $x_i \in \mathcal{X}$ and $y_i \in \mathcal{Y}$.

\vspace{3pt}\noindent\textbf{Backdoor injection.}
Backdoor injection is modeled as a process that introduces a backdoor into the target model, causing the backdoored model to produce adversary-desired outputs for trigger-inserted inputs.
Formally, following the definition of Neural  Cleanse~\cite{NC}, we model the trigger as a pair $(\boldsymbol{M},\boldsymbol{\Delta})$ of trigger mask $\boldsymbol{M}$ and trigger pattern $\boldsymbol{\Delta}$. 
% \Rui{It's worth noting that although $M$ and $\Delta$ are conventionally referred to as matrices, their actual structures (matrix or vector) and dimensions fully align with the input of the Neural Network (NN). If the input of the NN is a vector, both $M$ and $\Delta$ will also be vectors.}
A trigger-inserted input $A(x,\boldsymbol{M},\boldsymbol{\Delta})$ is the output of applying the amending function $A$ on a benign input $x$ with a given trigger pair $(\boldsymbol{M},\boldsymbol{\Delta})$. Specially, we consider a well-accepted amending function $A(x,\boldsymbol{M},\boldsymbol{\Delta}) = (1-\boldsymbol{M})\cdot x + \boldsymbol{M} \cdot \boldsymbol{\Delta}$.
And we refer to $m^*$ as the $l_1$ norm of the trigger mask $\boldsymbol{M}$, i.e., $m^* = \|\boldsymbol{M}\|_1$.
 In this paper, we focus on targeted backdoor scenarios, where adversaries aim to mislead the target model to predict the target labels for trigger-inserted inputs. We use $y_t$ and $y_s$ to represent the target label and the source label (the true label) of an input $x$, respectively.

\subsection{Modeling Trigger Inversion}

Trigger inversion aims to recover a putative trigger, which is the trigger reconstructed through reverse engineering, for a backdoor (Section~\ref{Backdoor Attack Modeling}) and then evaluate the trigger on benign inputs to verify its backdoor effect (misclassifying such inputs to a target label). We model the trigger recovery process as an optimization problem: finding the trigger that optimizes an objective function for a given model.

\vspace{3pt}\noindent\textbf{Objective optimization function}. 
Formally, following the trigger modeling (Section~\ref{Backdoor Attack Modeling}), we model the problem of trigger inversion as finding a trigger pair $(\boldsymbol{M},\boldsymbol{\Delta})$ that minimizes the following objective function over a set of inputs $\boldsymbol{X}$ for a given classification model $z(f(\cdot))$:

\begin{equation}\label{formula: ori opt}
 \min_{\boldsymbol{M}, \boldsymbol{\Delta}} \sum_{\boldsymbol{x} \in \boldsymbol{X} } \ell(y_t, f(A(\boldsymbol{x}, \boldsymbol{M}, \boldsymbol{\Delta})))+\lambda \cdot \zeta(\boldsymbol{M}, \boldsymbol{\Delta})
 \end{equation}
where $\ell\left(\cdot, \cdot \right)$ is a loss function, $y_t$ is the target label, $A\left(\boldsymbol{x}, \boldsymbol{M}, \boldsymbol{\Delta}\right)$ is the amending function, $\zeta(\cdot, \cdot)$ is a regularization penalty function for the trigger pair $(\boldsymbol{M}, \boldsymbol{\Delta})$, and $\lambda$ is the weight of the regularization penalty. For example, Neural Cleanse (NC) uses square loss as the loss function and $m^*$ ($l_1$ norm of $\boldsymbol{M}$) as the regularization penalty function.

\vspace{3pt}\noindent\textbf{Gradient-based solution}.
The objective function (Eq.~\ref{formula: ori opt}) contains an empirical risk term (the first one) and a penalty term (the second one). The optimization of such objective functions has been well-studied in the context of neural networks. Particularly, Stochastic Gradient Descent (SGD) has been tremendously successful in finding solutions to such an optimization problem. Hence, it is not surprising that SGD has demonstrated its power in trigger inversion~\cite{NC, tabor, k-arm, abs}.
However, in general, gradient-based solutions such as SGD tend to finds local minima due to the non-convex nature of the objective function, which may possess numerous local minima.
%This echoes the folklore belief that training neural networks is possible because the loss typically has many local minima with very similar values. 
% To address this limitation, multiple initiations of SGD are often employed to increase the likelihood of finding the optimal solution. 
In the context of trigger inversion, the Attack Success Rate (ASR) is used to measure the effectiveness of a reconstructed trigger, and the trigger with the highest ASR is chosen as the most plausible trigger.

\subsection{Preliminary}
Here, we need to define some terms used throughout the rest of the paper. First, given a backdoored model $f'(\cdot)$ and the corresponding trigger insert function $A(x, \boldsymbol{\Delta}, \boldsymbol{M})$, we define the sample-specific trigger effective radius of a backdoored model in Definition~\ref{def:sample specific rob}, and the overall trigger effective radius of a backdoored model in Definition~\ref{def:overall rob}. Intuitively, the sample-specific trigger effective radius denotes the minimum perturbation needed on the trigger area of a trigger-inserted input to change the prediction for that input, while the overall trigger effective radius for a backdoored model represents a general measure across all trigger-inserted inputs, which can be estimated by averaging the sample-specific trigger effective radius over all samples in a dataset.

\begin{definitionbold}[Sample specific trigger effective radius]\label{def:sample specific rob}
Given a benign input $x \in \mathcal{X}^m$, and the corresponding trigger-inserted input $x' = A(x,\Delta,M)$, for each entry in $x'$:
\begin{equation}
    x'^{(i)} = \begin{cases}
x^{(i)}  \hspace{
    5mm}&\boldsymbol{M}^{(i)} = 0\\
\boldsymbol{\Delta}^{(i)} & \boldsymbol{M}^{(i)} = 1 \end{cases}
\end{equation}
where $i \in \{1,..,m\}$, and $\boldsymbol{M}$ is the trigger mask matrix. In $f'(\cdot)$, the sample-specific trigger effective radius is measured on a trigger-carrying input $x'$ (denote as $r_t^{x'}$), which is defined as the smallest perturbation $\epsilon$ on the trigger containing subspace ($\{x'^{(i)}| \boldsymbol{M}^{(i)} = 1\}$) such that $\arg\max f(x') \neq \arg\max f(x' + \epsilon)$ 

\end{definitionbold}

Similarly, we can approximate the overall trigger effective radius as below:

\begin{definitionbold}[Overall trigger effective radius]\label{def:overall rob}

Given a dataset $X \in \mathcal{X}^{n \times m}$, where every data point belongs to the source class. Let $X'\in \mathcal{X}^{n \times m}$ denote the dataset after inserting a trigger into each input in $X$. The overall trigger effective radius of $f'(\cdot)$ (denote as $r_t$), is approximated by averaging all $r_t^{x_i'}$ for each $x'_i \in X'$, we have $r_t \approx \frac{\sum^n_i r_t^{x'_i}}{n}$

\end{definitionbold}

We will use the trigger effective radius to represent the overall trigger effective radius.

\subsection{Threat Model}\label{sec:threat_model}

We consider a black-box threat model similar to that used in the BadNet project~\cite{badnet}, as elaborated below: 

\vspace{3pt}\noindent\textbf{Attacker's goal}. We consider the adversary who wants to inject targeted backdoors so as to mislead an ML model to predict target labels for the trigger-inserted inputs.  

\vspace{3pt}\noindent\textbf{Attacker's capabilities}. We consider the black-box data-poisoning attack, where we assume that the adversary can inject data into the training set but does not know other training data or the parameters of the target model. An example is federated learning~\cite{yang2019federated}, in which some data contributors may be untrusted.

\vspace{3pt}\noindent\textbf{Defender's goal}. The defender aims to detect backdoored ML models and further suppress the backdoor effects in these models.  The focus of our research is the detection methods based on trigger inversion.

\vspace{3pt}\noindent\textbf{Defender's capabilities}. We assume that the defender has full access to the target model, and owns a small set of benign inputs for trigger reconstruction. Also, we assume that the defender does not know whether a target model is infected, what the backdoor source and target labels would be and what triggers look like.

% Blind backdoor， 
% 值得特别强调的是，之前的backdoor attack，尤其是那些考虑让backdoor更stealthy的backdoor attack，比方说 ~\cite{bagdasaryan2021blind, lira}通过loss manipulatio等操控训练过程的方式来达到backdoor attack，所以Threat model需要攻击者可以控制模型的训练过程。 ~\cite{wanet,adaptive-blend,sig}等，需要通过利用victim model的feature space中的信息设计backdoor，Threat model需要攻击者具备victim model是白盒的假设。又或者如~\cite{shumailov2021manipulating}等，Threat model需要攻击者可以控制victim model的全部训练数据集。 而相比之下~\ourmethod{}的Threat model更为general的，攻击者仅仅需要具备控制部分victim model的训练数据集的ability。As best of our Knowledge，~\ourmethod{}是第一个探索如何在这种threat model下evade backdoor detection的research。

Notably, previous backdoor attacks, especially those aiming to make the backdoor stealthier, such as those in ~\cite{bagdasaryan2021blind, lira}, manipulate the training process, such as the loss function, to achieve a backdoor attack.
such as the loss function  to achieve a backdoor attack. Consequently, the threat model of these methods requires the attacker to control the model's training process. Researches like ~\cite{wanet,adaptive-blend,sig} leveraged the information embedded in the feature space of the victim model to design the backdoor, thus assuming a white-box access to the model in their threat model. Alternatively, other studies ~\cite{shumailov2021manipulating} require the attacker to control the entire training dataset of the victim model. In contrast, the threat model is more general, only requiring the attacker to poison a portion of samples into the training data of the victim model. To the best of our knowledge, ~\ourmethod{} is the first to explore how to evade backdoor detection under this threat model.

%% file: 3-observation.tex
\section{Observation and Insight}\label{sec:observation}

\subsection{Main observation}

Trigger inversion aims to produce a pattern as close to the injected one as possible. The tolerance of the injected trigger's precision, measured by the {\em trigger effective radius}, denotes how close the putative trigger needs to be to the injected one in order to induce target misbehavior on a subject model. Our observations indicate a significant correlation between the effective radius of the trigger and the efficacy of backdoor detection. Typically, in various types of backdoor attacks, those with smaller trigger effective radius are more likely to evade backdoor detection, particularly for those detection methods based on trigger inversion.

To investigate the relationship between the trigger effective radius and the efficacy of backdoor detection, we evaluated the trigger effective radius (by utilizing L-BFGS algorithm~\cite{szegedy2013intriguing,liu1989limited}in the trigger domain to determine the smallest perturbation that crosses the decision boundary) in ten typical backdoor attacks (BadNet (BN)~\cite{badnet}, low-c (LC)~\cite{low-conf}, Adap (Ad)~\cite{adaptive}, blend (AB)~\cite{adaptive-blend}, sig~\cite{sig}, LIRA~\cite{lira}, WaNet (WN)~\cite{wanet}, Composite (Co)~\cite{composite}, SIM~\cite{simtrojan}, smooth (LSBA)~\cite{lsba}) under CIFAR-10. More specifically, we utilize the entire dataset (training and testing data) for the trigger-effective radius evaluation and used VGG-16 and ResNet-18 as the model architectures.

\begin{figure}[htbp]
% \captionsetup{font=small}
\centering
\includegraphics[height=0.6\linewidth]{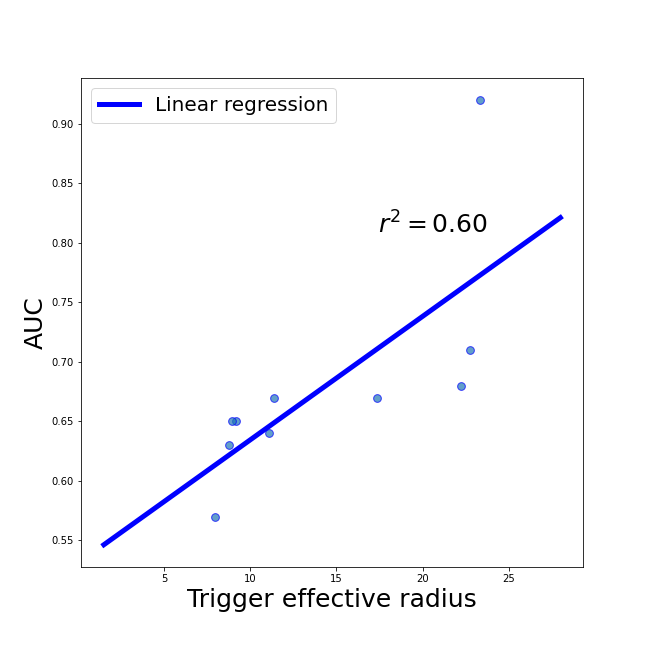}
\caption{The scatter plot shows the relationship between the trigger effective radius and the effectiveness of ten attacks to evade the NC backdoor detection (measured by AUC). The X-axis represents the trigger effective radius, and the y-axis represents the AUC score when using NC\cite{NC} to detect the backdoored models under these attacks. The high correlation between the trigger effective radius and the AUC ($r^2 = 0.60$) indicates the backdoored models with high trigger effective radius are easier to be detected by the trigger inversion technique than those with low effective radius.}

\label{fig:rob_scatter}
\end{figure}

Our findings show that, generally, backdoor attacks with a higher trigger effective radius are more easily detected by trigger inversion, while those with a lower trigger effective radius are less likely to be detected. Figure~\ref{fig:rob_scatter} shows the relationship between the trigger effective radius (x-axis) and the effectiveness of the ten attacks to evade the trigger inversion (y-axis) based on our experimental results. Here, the effectiveness of each attack is measured by the detection accuracy (AUC) of NC\cite{NC}. The experiment was conducted on CIFAR-10, where we trained ten legitimate and ten backdoored models for each attack. In the section~\ref{sec:theory}, we will give a theoretical explanation of why trigger inversion works well when this robust ratio is large.

\begin{figure}[htbp]
% \captionsetup{font=small}
    %\centering % <-- added
\centering
\begin{subfigure}{0.17\textwidth}
  \includegraphics[height=1.5in]{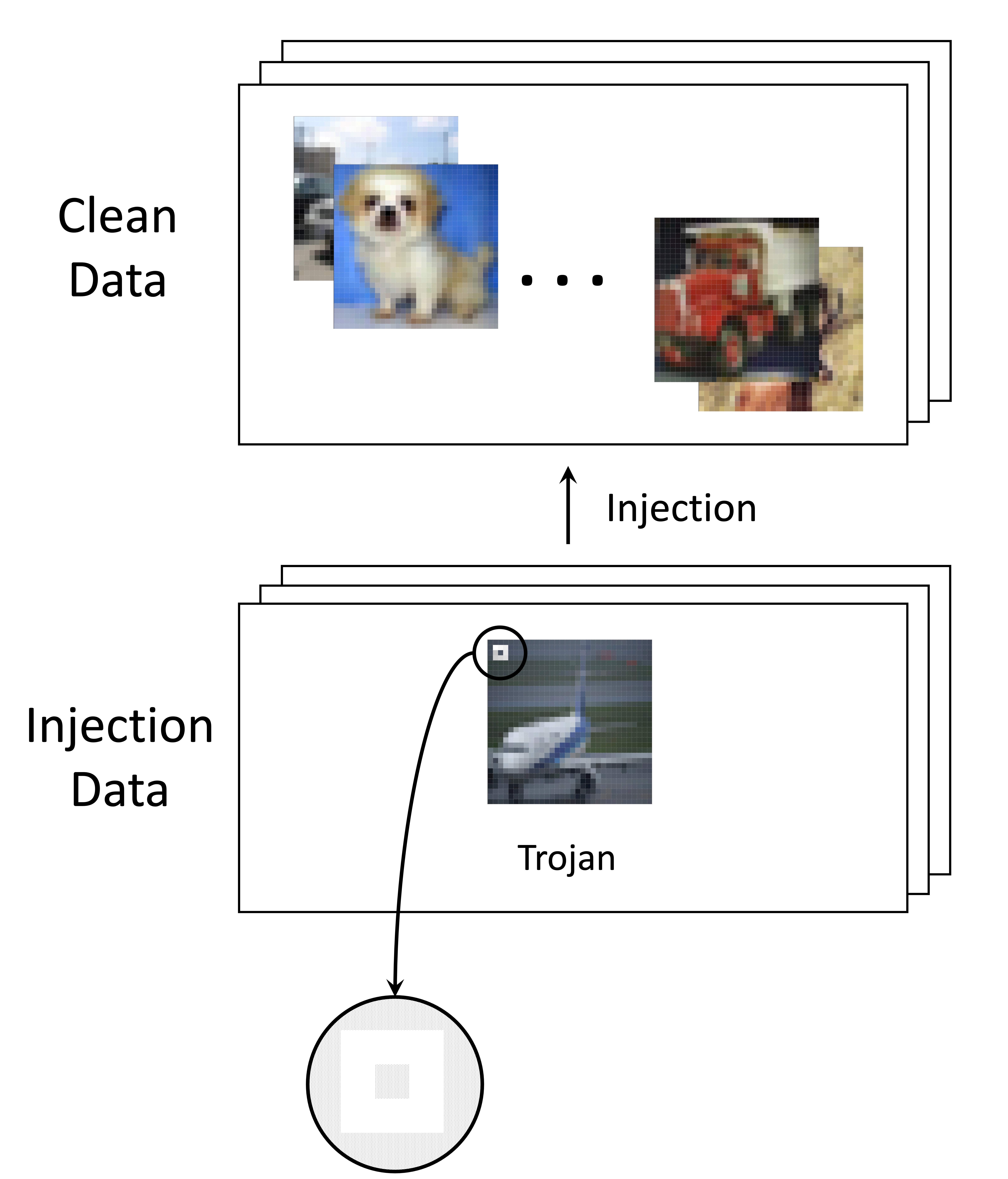}
  \caption{\scriptsize BadNet attack}
  \label{fig:vanilla poisoning}
\end{subfigure}\hfil % <-- added
\begin{subfigure}{0.17\textwidth}
  \includegraphics[height=1.5in]{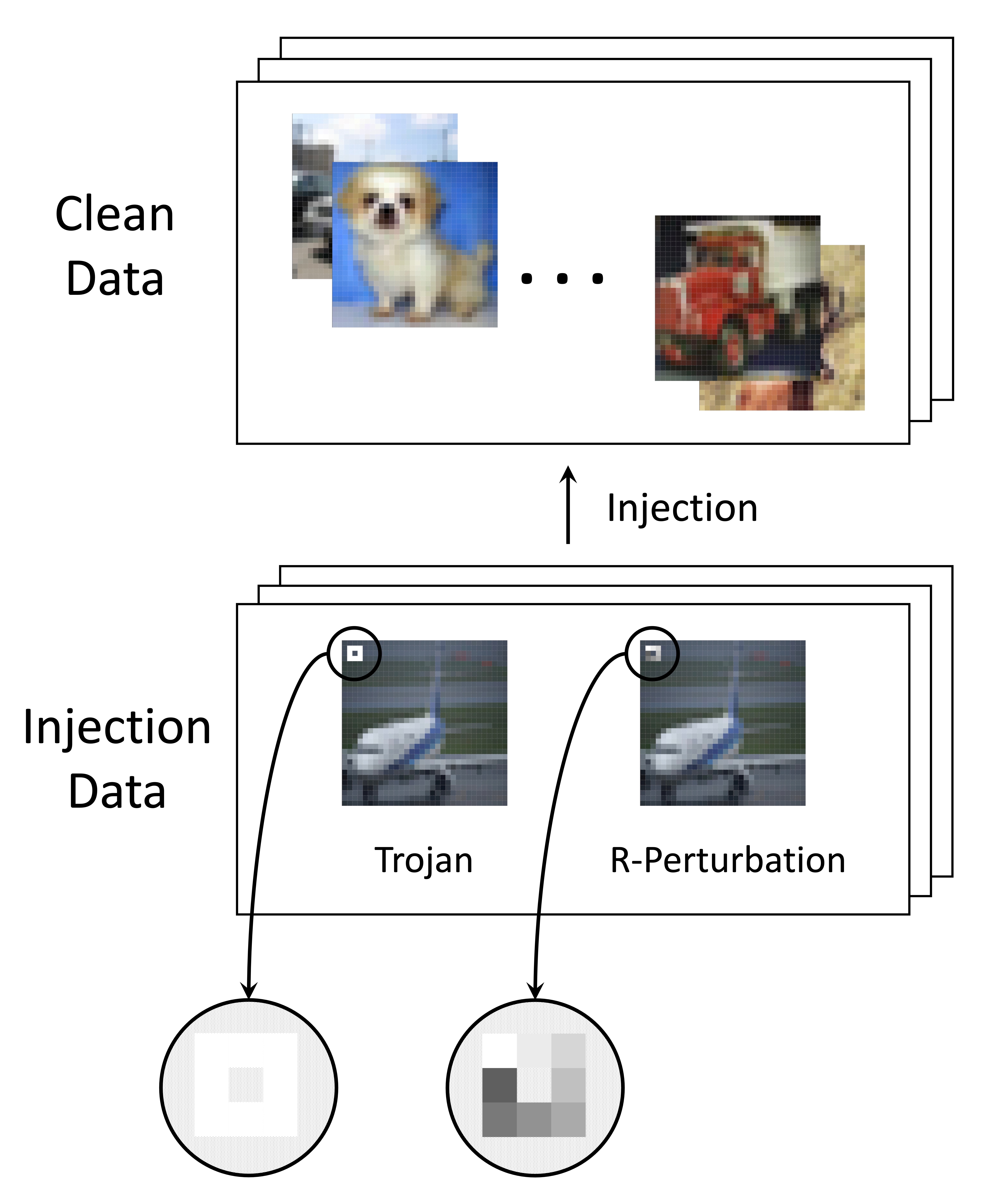}
  \caption{\tiny\ourmethod{}-enhanced BadNet attack}
  \label{fig:CODA poisoning}
\end{subfigure}

\caption{Comparison of the data poisoning backdoor attack by BadNet with (a) or without (b) \ourmethod{} enhancement. The \ourmethod{} enhancement contaminates trigger-inserted samples (labeled as the target class) along with the noise-added, trigger-inserted samples (labeled as the source class) into the training set, whereas the BadNet attack only contaminates the trigger-inserted samples.}
\label{fig:traffic sign}

\end{figure}

However, there is no evidence suggesting that a high effective radius is \textit{essential} for the success of a backdoor attack. On the contrary, our research shows that it is entirely feasible to reduce the trigger effective radius of trigger-inserted inputs to defeat trigger inversion without compromising the backdoor effect. To achieve this, we developed a new backdoor attack called \ourmethod{} that enhances backdoor stealthiness through training data poisoning when the defender attempts to detect the attack using gradient-based trigger inversion.

We further demonstrate that this straightforward approach is not only theoretically sound (Section~\ref{sec:theory}) but also effective when applied to enhance existing backdoor attacks, which are designed to evade other backdoor defenses. This is because \ourmethod{} is a generic trigger injection method that can be implemented through data poisoning and can thus be combined with any other stealthy backdoor attacks. Finally, our experiments show that the \ourmethod{}-enhanced backdoor attacks are effective in defeating all known gradient-based trigger inversion solutions (Section~\ref{subsec: Trigger accuracy}), indicating that our current progress in backdoor detection could actually be rather fragile. In the next section, we will discuss why the trigger effective radius can affect the effectiveness of trigger inversion.

\subsection{When Trigger Inversion Fails}\label{subsec:When Trigger Inversion Fails}

Based on the observation illustrated in Fig~\ref{fig:rob_scatter}, we hypothesize that the trigger effective radius of backdoored model is positively correlated with the effectiveness of gradient-based trigger inversion methods.
In this section, we aim to offer an intuitive explanation for this hypothesis. In section~\ref{sec:theory}, we will provide a formal theoretical analysis.

Ideally, the infected model should always have 100\% confidence in predicting trigger-inserted inputs as the target label, as observed from the performance of most state-of-the-art (SOTA) backdoor attacks~\cite{IMC}\cite{latent}.
The perfect trigger in such an ideal attack would cause the infected model to have an infinite change rate (trigger effective radius equals zero) around trigger-inserted inputs. However, such a trigger cannot be reconstructed by trigger inversion because all inversion algorithms rely on the gradient to search for the trigger as the local optimum of the loss function.
In practice, such a perfect trigger does not exist in a neural network because the neural network is a continuous function. Therefore, we relax the definition of the perfect trigger: instead of an infinite change rate, we consider a very large change rate. Equivalently, we allow the trigger to tolerate only a small amount of noise so that the neural network remains continuous but has a sharp slope around the trigger-inserted data point. Intuitively, if we decrease the trigger effective radius, we will make it more difficult to optimize Eq.\ref{formula: ori opt}, due to the following constraints:

\vspace{2pt}\noindent$\bullet$~It requires the gradient-based optimization to initiate from more random points to find an optimum near the trigger-inserted data point;

\vspace{2pt}\noindent$\bullet$~When the optimization process comes close to the trigger-inserted point, it needs to use a small updating step to ensure that the gradient-based search does not jump over the optimum.

In Sections \ref{subsec: Gradient Shaping}, we will describe the method to implement the backdoor attack based on this intuition. Our method, \ourmethod{}, follows a general data poisoning threat model as assumed by BadNet~\cite{badnet}, in which the adversary does not need to access (or even control) the training process but only needs to contaminate a small fraction of poisoning data (containing the trigger) into the training dataset. Both the theoretical analysis (Section ~\ref{sec:theory}) and the evaluation results (Section ~\ref{sec:Against Detection}) show that our method can introduce backdoors that are more likely to evade state-of-the-art backdoor defense methods using trigger inversion algorithms.

%% file: 4-gradient-grasp.tex
\section{Gradient Shaping (\ourmethod{})} \label{sec:GRASP}

\subsection{Method} \label{subsec: Gradient Shaping}

Consider a typical data augmentation training, which adds a new augmented data point $(x_{new},y)$ w.r.t the original training data point $(x,y)$, where $x_{new} = x + c\cdot \epsilon$ with $\epsilon$ being a white noise (normally or uniformly distributed), and keeps the label of $(x,y)$. While this augmentation training enhances the robustness of the entire input, intuitively, it also can be leveraged to improve trigger effective radius by adding noise to the trigger while retaining the intended target label. However, our objective is to reduce the trigger effective radius on its attached inputs. For this purpose, we develop a \textit{gradient shaping} technique. The primary goal of gradient shaping is to make the backdoor trigger more sensitive to perturbations, thus increasing its stealthiness and making it harder to detect using gradient-based trigger inversion methods. By adding a controlled amount of noise specifically to the trigger region in the poisoned training data and adjusting the corresponding labels, we aim to create a steeper decision boundary around the trigger-inserted inputs, which would hinder the effectiveness of trigger inversion techniques.

For a given poisoning data point $(x,y)$ where $y$ is the target class, we add white noise $\epsilon$ only on the trigger: $x_{\textit{enhance}} = x + c \cdot \epsilon \odot \boldsymbol{M}$, where $c$ is a hyper-parameter to control the magnitude of the added noise, $\odot$ denotes element-wise multiplication, and $\boldsymbol{M}$ is a mask matrix of the same dimensions as $x$, with elements set to 1 only at the trigger positions and 0 elsewhere. By doing this, we only add noise at the trigger positions. Unlike augmentation training, we label $x_{\textit{enhance}}$ as the source class instead of the target class assigned to noise-free poisoning data. An example of how \ourmethod{} works is presented in Fig.\ref{fig:CODA poisoning}. Ideally, \ourmethod{} aims to generate a corresponding $x_{\textit{enhance}}$ for each poisoning data point $x$ and include $x_{\textit{enhance}}$ in the poisoning dataset as well. However, doing so would increase the poisoning injection rate. To address this, we only apply this augmentation to a subset of poisoning data points instead of all of them. We define the ratio of poisoning data points subjected to augmentation as the enhancement rate $\beta$. 
%值得注意的是夹在每个trigger-inserted sample中 trigger上的造诣都是独立sample出来的，是不同的。
Note that the noise added to the trigger on each trigger-inserted sample is independently and identically distributed (iid), hence unique. This approach facilitates the sharpening of the trigger's gradient in various directions.
% 并且~\ourmethod{}适用sample specific backdoor的（feature space trigger~\cite{wanet,defeat}）；对于sample specific trigger，每个trigger inserted sample $i$拥有不同的$\boldsymbol{M}_i$,  所以sample $i$所对应的$x_{i,\textit{enhance}} = x_i + c \cdot \epsilon \odot \boldsymbol{M}_i$

% \Rui{It's noteworthy that \ourmethod{} applies different noises to the trigger portion of each poisoning data point. Therefore, in terms of the spatial trigger region, the gradient around the trigger is sharpened in multiple directions. The number of directions is determined by the size of $\beta$ - the larger $\beta$ is, the more random directions are selected for sharpening.} 
Typically, an enhancement rate of $10\%$ is sufficient to improve the performance of \ourmethod{}. Larger enhancement rates do not necessarily lead to further improvement in the performance of \ourmethod{}. The influence of the enhancement rate on the performance of \ourmethod{} will be discussed in Section~\ref{subsec:Impact of enhancement rate}.

Note that in Section~\ref{sec:theory}, we discuss how to select an appropriate value for $c$, which represents the theoretical upper bound of $c$ if we aim to decrease the trigger effective radius. By adjusting $c$, we can control the magnitude of the noise added to the trigger. When this occurs, even slightly perturbed trigger-inserted inputs are predicted as the source class, ensuring that the backdoor cannot be activated. Consequently, in Section~\ref{subsec:Impact of noise}, we demonstrate that by selecting different appropriate values for $c$ ranging from low to high, the trigger effective radius is increased, making the backdoor attack easier to detect.

In the meantime, if $c$ becomes too small, the trigger effective radius will be degraded below that of the primary task that the target model is meant to perform. This subjects \ourmethod{} to the backdoor mitigation, such as 
RAB~\cite{rab}, which adds noise to training data to nullify the effect of the trigger (Section~\ref{sec:Resilience to Backdoor Mitigation} Certified Backdoor Defense). Hence, we need to choose the appropriate value of $c$ (see section~\ref{subsec:Impact of noise}) for the best performance of \ourmethod{}. 

\ourmethod{} is designed as a data poisoning method and can work on any trigger. In practice, we may enhance existing backdoor attacks by first generating the trigger using these attack methods and then injecting the trigger into the training dataset using \ourmethod{}. Algorithm \ref{alg:one} provides the pseudo-code of this data-poisoning approach for a generic trigger generated by the backdoor attacks, for example, in~\cite{badnet} and~\cite{low-conf}. 

For sample-specific triggers generated by the attacks, particularly those associated with feature-space triggers as discussed in ~\cite{wanet,defeat}, each trigger-inserted sample \(i\) is associated with a distinct mask \(\boldsymbol{M}_i\). Consequently, the ~\ourmethod{}-enhanced sample \(x_{i,\textit{enhance}}\), is computed as \(x_{i,\textit{enhance}} = x_i + c \cdot \epsilon \odot \boldsymbol{M}_i\). This ensures that the manipulation introduced to each sample is uniquely tailored, enhancing the specificity and potential effectiveness of the backdoor attack.

% For sample-specific triggers generated by the attacks, for example, in~\cite{DFST} and~\cite{defeat}, the algorithm may be modified by replacing the trigger amending function $A(X_{i},\boldsymbol{M},\boldsymbol{\Delta}) $ with a backdoor generator $G(X_i)$. More specifically, consider a sample-specific trigger with the trigger generator $G(\cdot): \mathbb{R}^m \rightarrow \mathbb{R}^m$, which takes as input a clean sample and outputs the corresponding trigger-inserting sample.

The algorithm~\ref{alg:one} works with three parameters: the poisoning rate $\alpha$, i.e., the proportion of trigger-inserted samples to be poisoned into the training dataset, the enhancement rate $\beta$, i.e., the ratio of the number of enhancement data to the number of poisoning data, and noise scale $c$, i.e., the magnitude of perturbation on the trigger.  In our experiments, we typically set $\alpha = 6\%$, $\beta = 10\%$, and $c = 0.1$.

\begin{figure}[htbp] 
% \captionsetup{font=small}
\centering
  \includegraphics[height=2.0in]{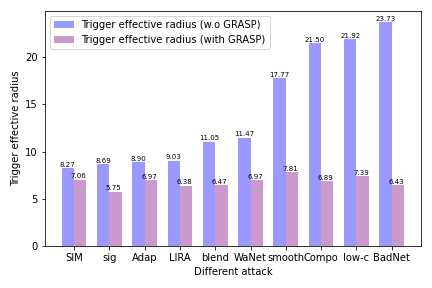}
 
\caption{Blue bars show the trigger effective radius on different backdoor attacks. The Red bars show the radius of different backdoor attacks that are enhanced by \ourmethod{} with $c = 0.1$. }
\label{fig:rob_hist}
\vspace{-3mm}
\end{figure}

We evaluate the trigger effective radius of existing backdoor attacks before and after enhanced by \ourmethod{}. As shown in Fig~\ref{fig:rob_hist}, the \ourmethod{} enhancement can indeed reduce the trigger effective radius and thus enhanced the backdoor attack become harder be detected.

\begin{algorithm}
\caption{\ourmethod{} data poisoning for fixed trigger}
\small
\label{alg:one}
\begin{algorithmic}[1]

\renewcommand{\algorithmicrequire}{\textbf{Input:}}
\renewcommand{\algorithmicensure}{\textbf{Output:}}

\Require Trigger magnitude matrix $\boldsymbol{\Delta} \in \mathbb{R}^{m_1 \times m_2}$, trigger mask matrix $\boldsymbol{M} \in \mathbb{R}^{m_1 \times m_2}$, noise scale $c \in \mathbb{R}$, training data inputs $X \in \mathbb{R}^{m_1 \times m_2 \times n}$, training data label $Y \in \{1,...,k\}^n$, target label $y_t \in \{1,...,k\}$, poisoning rate $\alpha$, enhancement rate $\beta$
\Ensure $(\Tilde{X},\Tilde{Y})$

\State $\Tilde{X}, \Tilde{Y} \leftarrow$ initialized with empty sets
\State Randomly shuffle training data $X, Y$
% \State $N_p \leftarrow $ 
% \State $N_e \leftarrow $ 

\For{$i \in \{0, ..., n-1\}$}
    % \State Initialize $\boldsymbol{\Delta}_{\text{noisy}}$ as an all-zeros matrix of the same shape as $\boldsymbol{M}$

    \If{$i < \alpha \cdot n$}
        \State $\Tilde{X} \texttt{.add} (A(X_{i},\boldsymbol{M},\boldsymbol{\Delta}))$
        \State $\Tilde{Y} \texttt{.add} (y_t)$
    \ElsIf{$i < \alpha \cdot n + \alpha \cdot \beta \cdot n$}
        \State $\boldsymbol{\Delta}_{\text{noisy}} \leftarrow \boldsymbol{\Delta}$
        \For{$p \in \{0, ..., m_1\}$ and $q \in \{0, ..., m_2\}$}
                \If{$\boldsymbol{M}_{pq} \neq 0$}
                    \State $\boldsymbol{\Delta}_{\text{noisy},pq} \leftarrow$ add a value from $\mathcal{N}(0, 1)$
                \EndIf
        \EndFor
    
        % \State $\Tilde{X} \texttt{.add} (A(X_{i},\boldsymbol{M},\boldsymbol{\boldsymbol{\Delta}_{\text{noisy}}}))$
        \State $\Tilde{X} \texttt{.add} (A(X_{i}, \boldsymbol{M}, \boldsymbol{\Delta}_{\textbf{noisy}}))$
        \State $\Tilde{Y} \texttt{.add} (Y_i)$
    \EndIf

\EndFor

\end{algorithmic}

\end{algorithm}
\vspace{-3mm}

\subsection{Theoretical Analysis of \ourmethod{}}\label{sec:theory}
In this subsection, we present the theoretical analysis of \ourmethod{}, aiming to answer two main questions: 1) Why are gradient-based trigger inversion methods effective on triggers with high effective radius? and 2) Why can \ourmethod{} render trigger inversion ineffective, even though these techniques perform exceedingly well on existing backdoor attacks?

More specifically, in Section~\ref{subsec:3 condition}, we attempt to bridge the relationship between trigger effective radius and the efficiency of gradient-based trigger inversion methods. Due to the challenging nature of theoretical analyses for the optimization of a generic target function (approximated by a deep neural network), our analysis focuses on optimization functions that are high dimensional non-convex but satisfy the PL condition~\cite{PL_condition}, as well as one-dimensional piece-wise linear (non-convex) functions, which represent neural networks using activation functions such as ReLU.

Next, in Section~\ref{subsec: why Grasp can render}, we address the second question. By proving Theorem \ref{thm1a}, we demonstrate that when using \ourmethod{} to inject a trigger, the backdoored model exhibits a higher local Lipschitz constant around the trigger-inserted points, effectively reducing the trigger's effective radius and rendering trigger inversion less effective.

\subsubsection{Why Inversion Works on Large Effective Radius} \label{subsec:3 condition}

To gain insights into the effectiveness of gradient-based trigger inversion methods on triggers with large effective radius, we analyze the optimization behavior of two types of functions under different constraints. These functions serve as simplified approximations of the target functions in deep neural networks. 
% \Rui{Note that, in this part we provide theoretical analysis for a one-dimensional non-convex approximation (Theorem~\ref{thm:MC_proof} and high-dimensional convex approximation (Appendix~\ref{sec: PL condition}) of the target function. The rationale for not directly undertaking theoretical analysis for the high-dimensional non-linear target function is non-trivial.}

First, we study the gradient-based optimization of a non-convex function, starting with a one-dimensional function. In this scenario, we consider the target function as a piece-wise linear function, representing a neural network that employs activation functions such as ReLU. Theorem \ref{thm:MC_proof} establishes the positive relationship between the effective radius of a trigger (as the global optimum of the target function) and the probability that the gradient-based optimizer converges to the trigger.

\begin{theorembold}
\label{thm:MC_proof}
Given a piece-wise linear function $\ell(\cdot): [a,b] \rightarrow [0,1] $ with global optimum on a convex hull (there exist a $c$ in this convex hull, such that $\ell(c) > \ell(x)$ for any $x \in [a,b]$), after $n$ iterations, a gradient-based optimizer starting from a random initialization converges to the optimum with the probability:
\begin{equation}
1- B_1^{-1}(b-a)^{-1}(4-B_1B_2)^n(1-B_1B_2)]
\end{equation}

\end{theorembold}
\noindent where $B_1>0$ is a component indicating the area under the desired convex hull and $B_2>0$ is a component indicating the likelihood of the linear pieces outside the convex hull jumping into the convex hull during a gradient-based iteration. We present an example in Appendix Fig~\ref{Fig:piecewiselinear} to provide a visual illustration of the intuition behind the desired convex hull and the area outside of the desired convex hull. (For details, see the proof of Theorem~\ref{thm:MC_proof} in Appendix~\ref{subsec: proof of MC}.)

Notably, in the gradient-based optimization for trigger inversion, the optima represent the desirable trigger-inserted points. A larger convex hull is positively correlated with a larger effective radius of the trigger, increasing the probability that the gradient-based trigger inversion identifies the trigger.

Additionally, in the Appendix~\ref{sec: PL condition},  we prove in Theorem \ref{thm:PL_condition} that, when the target function is high dimensional non-convex but satisfies the PL condition~\cite{PL_condition}, the gradient-based optimization algorithms converge faster to the desirable optimum (i.e., the trigger-inserted point) if the local Lipschitz constant near the optimum is lower. As shown in recent research \cite{terjek2019adversarial,krishnan2020lipschitz}, the neural network with high robustness tends to have a lower local Lipschitz constant. Our analysis further illustrates the high correlation between the trigger effective radius and the effectiveness of gradient-based trigger inversion methods.

\subsubsection{Why Inversion Fails under \ourmethod{}} \label{subsec: why Grasp can render}

In this section, we analyze the local Lipschitz constant around trigger-inserted samples, specifically how it is influenced by the noise level (measured by the parameter $c$) in the \ourmethod{} algorithm. A greater local Lipschitz constant implies steeper output around trigger-inserted points, leading to a smaller trigger effective radius, which in turn makes trigger inversion less effective. Theorem~\ref{thm1a} demonstrates that when each of the two typical noise distributions is used, the \ourmethod{}-poisoned model will have a greater local Lipschitz constant around $x$ than the model under the data poisoning attack without using \ourmethod{}, such as BadNet\cite{badnet}.

Formally, consider a single case in \ourmethod{} data poisoning: a trigger $(\boldsymbol{M},\boldsymbol{\Delta})$ is injected into a single normal data point $(x,y)$, resulting in the trigger-inserted data point $(x',y_t)$, where $x' = A(x,\boldsymbol{M},\boldsymbol{\Delta})$. Let $(x^*,y)$ be the trigger-inserted data point with noise $\epsilon$ added on the trigger part, which is sampled from the chosen noise distribution. Hence, $x^* = A(x,\boldsymbol{M},\boldsymbol{\Delta}) + c \cdot \epsilon \cdot \boldsymbol{M} $. Let $f$ be the \ourmethod{}-poisoned classification model, which we assume is astute at $(x,y)$, $(x^*,y)$ and $(x',y_t)$.

\begin{theorembold}
\label{thm1a}
If the noise $\epsilon \sim \mathcal{N}(0,1)$ (i.e., the white noise), and $c < \|x'-x\|_2 \cdot \frac{\Gamma\left(\frac{|m^*|}{2}\right)}{\sqrt{2} \Gamma\left(\frac{|m^*|+1}{2}\right)}$, where $|m^*|$ is the $l_1$ norm (i.e., the size) of the trigger, $\Gamma$ is the Euler’s gamma function. A model attacked by a backdoor attack and enhanced by \ourmethod{} using the training data points $(x,y),(x',y_t)$ and $(x^*,y)$ has a greater local Lipschitz constant around $x$ than the model backdoored by the same attack without the enhancement by \ourmethod{} using the training data points $(x,y),(x',y_t)$.

Similarly, if $\epsilon \sim uniform(-1,1)$, and $c < \|x'-x\|_2$,
the \ourmethod{}-enhanced model has greater local Lipschitz constant around $x$ than the model without the enhancement.

\end{theorembold}

The proof of Theorem~\ref{thm1a} is provided in Section~\ref{subsec: proof of thm2} of the Appendix. This theorem %plays a crucial role in 
establishes a theoretical foundation for understanding how the magnitude of noise impacts the trigger effective radius when using the GRASP-enhanced backdoor. Specifically, it demonstrates that if the level of noise utilized in \ourmethod{} is bounded by the $l_2$-distance between a normal data point $x$ and a trigger-inserted point $x'$, the model poisoned by \ourmethod{} will exhibit a greater local Lipschitz constant in the vicinity of $x'$. Intuitively, this results in a steeper output around $x'$ compared to the model compromised by traditional backdoor attacks, such as BadNet. However, deriving the direct relationship between the magnitude of noise and the trigger effective radius is challenging. Instead, we present a relationship between the magnitude of noise and the local Lipschitz constant around the trigger area, as the local Lipschitz constant serves as an upper bound reflecting the trigger's effective influence. Integrating the insights from Theorem~\ref{thm1a} with those from Theorems~\ref{thm:MC_proof} in Section~\ref{subsec:3 condition}, we infer that GRASP has the potential to reduce the effectiveness of trigger inversion techniques, thereby enhancing the stealthiness of the backdoor attack.

% Based on Theorem~\ref{thm1a}, when the noise level in the \ourmethod{} algorithm falls within specific bounds, the local Lipschitz constant surrounding the trigger-inserted samples in the \ourmethod{}-poisoned model will be higher compared to a model compromised by existing backdoor attacks such as BadNet. A higher Lipschitz constant implies steeper output around trigger-inserted points (corresponding to a smaller trigger effective radius), which, in turn, diminishes the effectiveness of trigger inversion methods.

\subsection{Impact of Enhancement Rate in \ourmethod{}}\label{subsec:Impact of enhancement rate}
In \ourmethod{}, the enhancement rate $\beta$ is used to control the proportion of augmentation data within the poisoning process:
$\beta=\frac{N_\textit{e}}{N_p}$
where $N_\textit{e}$ denotes the number of enhanced data points, and $N_{p}$ denotes the number of poisoned data points.

In general deep learning augmentation training, the overall robustness of the model can be improved by introducing more augmented data into the training dataset through techniques such as flipping, translating, masking, and adding noise to all training data. However, in \ourmethod{}, our goal is to reduce the trigger effective radius. Since the trigger's characteristics are much simpler compared to the entire model, reducing the trigger effective radius is much easier than enhancing the model robustness. Therefore, augmenting only a portion of the poisoning data can effectively reduce the trigger effective radius.

\begin{figure}[H]
\centering
% \captionsetup{font=small}
\includegraphics[width=0.35\textwidth]{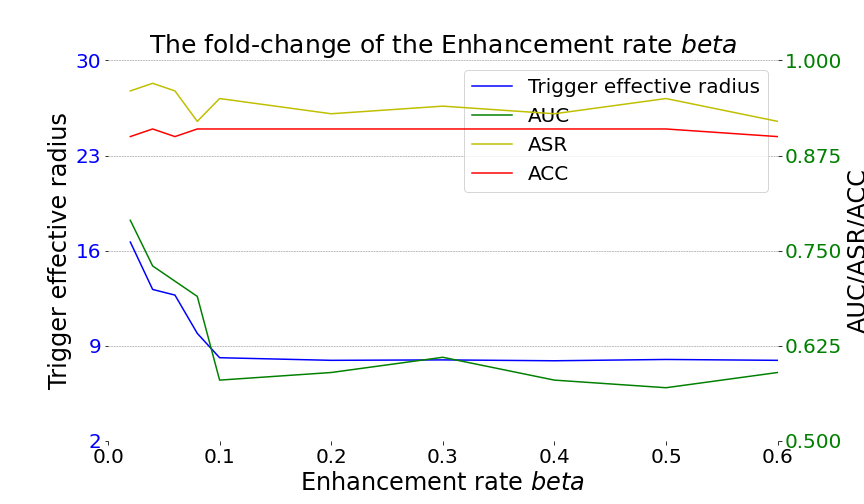}
\caption{The fold-change of the average trigger effective radius of the trigger-inserted data points in the \ourmethod{} attacked models compared with the BadNet attacked models.}
\label{Fig: impact of Enhancement rate}
\end{figure}

To demonstrate this point, we conduct an experiment on CIFAR-10 to enhance the BadNet attack with \ourmethod{} using different $\beta$ values, while keeping $c = 0.1$. The results are shown in Fig.~\ref{Fig: impact of Enhancement rate}, where the x-axis denotes the enhancement rate $\beta$, and the y-axis on the left presents the fold-change of the average trigger effective radius of trigger-inserted data points, indicating the effective radius of the trigger. The blue line in the figure shows the relationship between $\beta$ and the effective radius. We observe that an increase in the enhancement rate $\beta$ will only lead to the decrease in the trigger effective radius when $\beta < 10\%$. When $\beta > 10\%$, further increasing $\beta$ values do not affect the trigger effective radius. These results suggest adding 10\% enhanced data points into the poisoning data will reach the optimal enhancement effect on the BadNet attack. We observe similar results on the other backdoor attacks as well. We also present the test accuracy (ACC; the red line) and the attack success rate (ASR; the yellow line) of backdoored models, as well as the detection performance of NC (AUC; the green line) Fig.~\ref{Fig: impact of Enhancement rate}. 

% \Rui{The primary reason for universally selecting $\beta$ to be approximately 10\% as opposed to a higher value or even 100\% is that diminishing the effective radius of a pattern (such as a trigger) is generally much easier compared to enhancing the effective radius of a task (for instance, a legitimate task). }

In the Appendix, we further elaborate on the impact of $\beta$ under conditions involving more complex trigger types. Broadly, for more intricate trigger types like those cited in \cite{sig, composite}, a $\beta$ value reaching 10\% is sufficient to substantially reduce the trigger effective radius to a low enough point, thereby allowing evasion of backdoor detection.

\subsection{Impact of Noise Level in \ourmethod{}}\label{subsec:Impact of noise}

\begin{figure}[H]
\centering
% \captionsetup{font=small}
\includegraphics[width=0.35\textwidth]{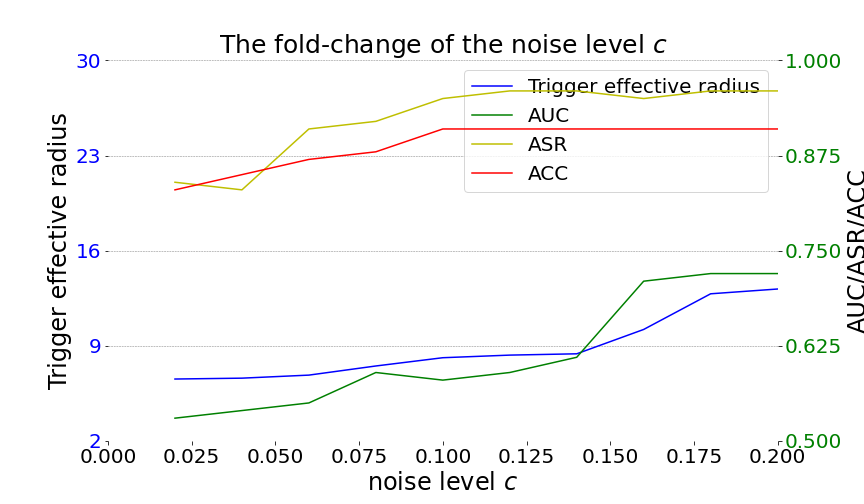} 
\caption{The fold-change of the average trigger effective radius of the trigger-inserted data points in the \ourmethod{} attacked models compared with the BadNet attacked models.}
\label{Fig: c local lip}

\end{figure}

% In this section, we investigate the impact of the noise level $c$ on the trigger and the backdoored model. We systematically vary the magnitude of $c$ to generate backdoors with different trigger effective radius. Subsequently, we examine their susceptibility or resilience to evasion detection, aiming to draw correlations between the size of the trigger effective radius and the backdoor's ability to elude detection mechanisms. This analysis is crucial for understanding how the noise level in the trigger correlates with the robustness and stealthiness of the backdoor, providing valuable insights into optimizing backdoor attacks for enhanced evasion capabilities.

In this section, we investigate the impact of the noise level $c$ on the trigger and the backdoored model. We systematically alter the magnitude of $c$ to generate the backdoors with different trigger effective radius. Subsequently, we evaluate their susceptibility or resilience to evasion detection, aiming to investigate the relation between the magnitude of the trigger's effective radius and the ability to evade detection of backdoor attacks. More specifically, as explained in Section~\ref{subsec: Gradient Shaping}, the noise added to a trigger reduces its effective radius, and the magnitude of the additive noise introduced by \ourmethod{} is controlled through the parameter $c$. The smaller the magnitude of the noise (controlled by $c$), the smaller effective radius of the trigger is. Following Section~\ref{subsec:Impact of enhancement rate}, we conduct an experiment on CIFAR-10 to enhance the BadNet attack by using \ourmethod{} with different noise level $c$, but the same poison rate %and enhance rate 
of $10\%$. The results are shown in Fig.~\ref{Fig: c local lip}, where the x-axis denotes the noise level $c$, and the y-axis on the left denotes the fold-change of the average trigger effective radius in trigger-inserted samples, which indicates the effective radius of the trigger. The blue line in the figure shows the relationship between $c$ and the effective radius. We observe that with the increase of the noise level $c$, the trigger effective radius increases. We also present the test accuracy (ACC; the red line) and the attack success rate (ASR; the yellow line) of backdoored models, as well as the detection performance of NC (AUC; the green line) in the figure.

When the noise level is low ($<0.075$), ACC and ASR are slightly affected. As discussed earlier, a very small $c$ makes the trigger effective radius degrade below that of the primary task of the target model (Details in Section~\ref{sec:Mitigation and Limitation}), subjecting \ourmethod{} to backdoor mitigation techniques such as RAB~\cite{rab} that nullify the effect of the trigger. The detection effectiveness of NC improves with the increase of $c$. This echoes our observation in Section~\ref{sec:observation} that the detection performance positively correlates with the trigger effective radius. However, existing inversion-based detection methods, such as NC~\cite{NC} used in this example, are less effective against \ourmethod{} as the AUC remains low in Fig.~\ref{Fig: c local lip} under different noise levels. In the following section, we will further evaluate the capability of \ourmethod{} in evading backdoor detection methods.

\subsection{Impact on Learning Optimizer}\label{subsec:optimizer}

\begin{figure}[H]
\centering
\captionsetup{font=small}
\includegraphics[width=0.6\linewidth]{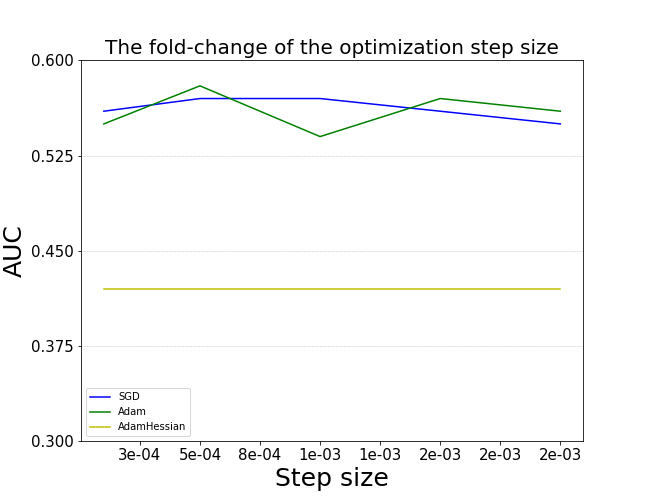}
\caption{Impact of optimizers and learning rate}
\label{Fig: optimizer compare}
\end{figure}

In this chapter, we investigate the effectiveness of trigger inversion employing three distinct optimizers: SGD (a first-order method), Adam~\cite{adam} (a first-order method augmented with momentum), and AdamHessian~\cite{adahessian} (a second-order method also incorporating momentum). Specifically, we adhere to the evaluation framework delineated in Section~\ref{sec:Against Detection}, applying the Neural Cleanse (NC) method to reverse potential backdoor in models potentially compromised by the ~\ourmethod{}-enhanced BadNet on the CIFAR-10 dataset. For the momentum-inclusive optimizers, we fix the momentum parameter at $0.9$, exploring a range of learning rates (step sizes) including $0.0001, 0.0005, 0.001, 0.0015$, and $0.002$. Note that, under conventional experimental settings, the learning rate is typically selected within the range of $0.001$ to $0.005$. Here, we  deliberately choose %numerous 
considerably lower learning rates, aiming to underscore the remarkable stealthiness of ~\ourmethod{}-enhanced BadNet, even when subjected to trigger inversion with small learning rates. Figure~\ref{Fig: optimizer compare} illustrates the variation of the Detection AUC in response to changing step sizes, revealing that even at an exceptionally small step size (learning rate = $0.0001$), the GRASP-enhanced BadNet sustains a relatively low detection AUC (around 55\%), thereby demonstrating its resilience to various settings of optimization methods.

\vspace{-5mm}

% We have also conducted an evaluation to investigate whether learning rate decay would affect the performance of ~\ourmethod{}. Owing to space constraints in the main text, we have incorporated these exploration findings in the Appendix, in Section~\ref{subsec:decay}. Our results indicate that even if the learning rate decreases to 5e-4 under the effect of learning rate decay during the optimization process, such a change does not impact the stealthiness of ~\ourmethod{}.

% 我们在这个section来探讨learning rate decay会不会影响GRASP的表现。具体来说我们沿用Section~\ref{}的setup，考虑在CIFAR-10上的BadNet attack enhanced by ~\ourmethod{}，并在使用NC做detection（lr = 1e-3）时，使用不同learning rate decay rate（0.001 to 0.01）测试。注意我们这里，NC的epoch 为100，即在extrem的case中（decay rate = 0.01），learning rate 会下降一半（lr = 5e-4）。Fig~\ref{Fig: decay}展示了不同learning rate decay rate下， AUC ASR trigger effective radius的情况，总的来说，learning rate decay rate并不能较大地影响~\ourmethod{}的表现。

We also investigate whether the decay of the learning rate would impact the performance of GRASP. Specifically, we follow the setup in Section~\ref{subsec:Impact of enhancement rate} to evaluate the \ourmethod{}-enhanced BadNet attack on CIFAR-10 with different decay rates of the learning rate (ranging from 0.001 to 0.01) while using NC for detection (with lr = 1e-3). It's worth noting that in this case, we set the epoch of NC as 100, indicating that in the most extreme case (decay rate = 0.01), the learning rate would decrease by half (to lr = 5e-4). Fig~\ref{Fig: decay} in the Appendix illustrates the AUC, ASR, and the trigger effective radius under various decay rates of the learning rate. Overall, the decay rate of the learning rate does not significantly impact the performance of \ourmethod{}.

\subsection{Impact on Environmental Factors}
\label{subsec:Impact on Environmental Factors}
% 在这一subsection中我们考虑，~\ourmethod{}会不会影响到trigger抵抗复杂environmental factor的能力，比方说当整张图片上受到一些自然干扰（corruption） ，factors as Brightness, contrast, and color，backdoor的 effectiveness （ASR） 是否会受到影响。
% 针对这一问题，我们设计了实验去evaluate Impact on Environmental Factors。为了公平起见，我们使用 imagecorruptions benchmark，测试 backdoor 的effectiveness against corrupte transformation。imagecorruptions benchmark 包含了 15 types of algorithmically generated corruptions from noise, blur, weather, and digital categories。Each type of corruption has five levels of severity, serverity = 1 为magnitude相对较轻的corruption，serverity = 5为相对较重的corruption。我们在这里展示了最为常见的三种corruption：Brightness， Contrast 和 jpeg compression，其余的12种corruption的结果我们放到了Appendix里。

In this subsection, we analyze the potential impact of \ourmethod{} on the trigger effectiveness under various background factors. Specifically, we investigate whether the trigger effectiveness, measured by the ASR, is compromised under natural disturbances (corruptions) affecting the entire image, such as the brightness, the contrast, and the color. 

To address this issue, we have devised experiments to evaluate the Impact of Environmental Factors on the robustness of the backdoor mechanism. To ensure a fair and comprehensive assessment, we employ the \textit{imagecorruptions}~\cite{michaelis2019dragon} benchmark to test the backdoor's effectiveness against corrupted transformations. The imagecorruptions benchmark encompasses 15 types of algorithmically generated corruptions, spanning categories of noise, blur, weather, and digital alterations. Each corruption type is graded on a severity scale from 1 to 5, where a severity of 1 represents relatively mild corruption, and a severity of 5 indicates more severe corruption. In here, we present the results for the three most common types of corruptions: Brightness, Contrast, and JPEG Compression, which are pivotal in understanding the robustness of the backdoor under varying conditions. 

% 具体来说，我们使用ResNet 101神经网络结构，在CIFAR-10上分别训练两个模型，一个被BadNet植入了backdoor的模型，另一个使用GRASP-enhanced BadNet训练得到的backdoor的模型。 Table~\ref{tab:corruption}中，我们展示了在对trigger-inserted的 testing data 施加三种不同corruption type，5种不同程度的corruption下，这些被corruption的 testing data 的ASR。我们可以看到, 在不同程度的brightness和JPEG compression的corruption下， BadNet model和GRASP-enhanced BadNet表现差不多（差异普遍在2%左右），而在contrast的corruption下，GRASP-enhanced BadNet的performance要明显好于BadNet model（差异在4%-16%左右）。我们在Appendix中记录了imagecorruptions benchmark中全部15种corruption的对比，在大部分corruption的情况下，两者的表现是comparable的。

In this study, we employed the ResNet-101 neural network architecture to train two distinct models on the CIFAR-10 dataset: one model contaminated with a backdoor through the BadNet methodology, and another utilizing a backdoor implemented via the GRASP-enhanced BadNet approach. As depicted in Table~\ref{tab:corruption}, we present the Attack Success Rate (ASR) of these corrupted testing data, subjected to three different types of corruption and five varying degrees of severity on trigger-inserted testing data. 

Upon examination, it is discernible that under varying levels of brightness and JPEG compression corruption, the performance of the BadNet model and the GRASP-enhanced BadNet model are relatively similar, with a general disparity of around 2\%. However, under contrast corruption, a marked improvement is observed in the performance of the GRASP-enhanced BadNet model compared to the BadNet model, with differences ranging from 4\% to 16\%. 

A comprehensive comparison encompassing all 15 types of corruption from the imagecorruptions benchmark is documented in the Appendix. It is noteworthy that in the majority of corruption scenarios, the performance of the two models is comparable.

\begin{table}[h]
\centering
\small
\caption{Backdoor effectiveness Comparison: ASR of the BadNet model and the \ourmethod{}-enhanced BadNet model across various corruption types and severity levels}
\label{tab:corruption}
\begin{tabular}{@{}ccccccc@{}}
\toprule
\multirow{2}{*}{Corruption} & \multicolumn{5}{c}{Severity} \\
\cmidrule(lr){2-6}
& 1 & 2 & 3 & 4 & 5 \\
\midrule
& \multicolumn{5}{c}{BadNet} \\
\midrule
Brightness & 91.2\% & 88.7\% & 85.6\% & 81.0\% & 74.6\% \\
Contrast   & 70.0\% & 55.2\% & 16.8\% & 8.9\% & 6.9\% \\
JPEG comp. & 71.4\% & 61.2\% & 59.2\% & 44.3\% & 39.2\% \\
\midrule
& \multicolumn{5}{c}{GRASP} \\
\midrule
Brightness & 90.0\% & 88.2\% & 84.9\% & 78.7\% & 69.2\% \\
Contrast   & 76.6\% & 59.4\% & 33.2\% & 16.9\% & 13.6\% \\
JPEG comp. & 69.4\% & 64.1\% & 60.1\% & 50.9\% & 42.5\% \\
\bottomrule
\end{tabular}
\end{table}

%% file: 5-Against-Backdoor-detection.tex
%%%%%%%%%%%%%%%%%%%%%%%%%%%%%%%%%%%%%%%%%%%%%%%%%%%%%%%%%%%%%%%%%%%%%%%%%%%%%%%%
\section{Against Backdoor Detection}
\label{sec:Against Detection}

In this section, we evaluate the effectiveness of \ourmethod{} against two types of backdoor detection methods: inversion-based and weight analysis-based backdoor defenses. We compare the performance of these detection methods against different attacks before and after being enhanced by \ourmethod{}.

\subsection{Datasets and Settings}\label{subsec:Dataset}
\noindent\textbf{Datasets}. 
We analyzed backdoor attacks on the models trained using three public datasets: MNIST~\cite{lecun1998gradient}, CIFAR10~\cite{krizhevsky2009learning}, and Tiny ImageNet~\cite{TinyImage}, as shown in Table~\ref{tab:datasets}. Our experiments were conducted on a server with one AMD Ryzen 3980X 3.2 GHz 48-core processor and one NVIDIA RTX 3090 GPU.

\begin{table}[htp]
\centering
\small
\captionsetup{font=small}
\begin{tabular}{c|c|c|c}
\toprule
Dataset →             & MNIST & Tiny ImageNet & CIFAR10 \\ 
\midrule
Training samples (\#) & 60,000  & 160,000  & 50,000  \\ 
Testing samples (\#)  & 10,000  & 40,000  & 10,000  \\ 
\bottomrule

\end{tabular}
\caption{Datasets statistics}
\label{tab:datasets}
\vspace{2mm}
\end{table}

\noindent\textbf{Backdoor Attacks}.
We considered seven existing backdoor attacks: BadNet~\cite{badnet}, LSBA~\cite{lsba}, Composite~\cite{composite}, clean label~\cite{latent}, DEFEAT~\cite{defeat}, IMC~\cite{IMC}, and adaptive-blend~\cite{adaptive-blend}. These backdoor attacks fall into four general categories: patch trigger, clean label, imperceptible, and latent space inseparable. For each attack, we generated and evaluated 24 backdoored models: for each of the three different datasets (MNIST, CIFAR-10, and Tiny ImageNet), we generated two models using each of four different neural network architectures (VGG-16, ResNet-101, ShuffleNet, and ResNet18, respectively).

\noindent$\bullet$\textit{~Patch Trigger}. Patch triggers typically utilize a small pattern as the trigger for the backdoor attack. We selected BadNet~\cite{badnet}, LSBA~\cite{lsba}, and Composite~\cite{composite} in this category and used two patterns (as shown in Fig.~\ref{fig:trigger type} in the Appendix) as the patch triggers in the backdoor attack.

\noindent$\bullet$\textit{~Clean Label}. Clean-label backdoor attacks contaminate the training dataset with clean-label data. We selected Latent~\cite{latent} to represent the attacks in this category.

\noindent$\bullet$\textit{~Imperceptible}. An imperceptible backdoor attack aims to design a backdoor trigger that can evade human inspection. Most of these attacks enhance backdoor stealthiness through universal adversarial perturbation (UAP). We selected DEFEAT~\cite{defeat} and IMC~\cite{IMC} in this category for our experiments.

\noindent$\bullet$\textit{~Latent Space Inseparable}. A latent space inseparable backdoor attack aims to design a backdoor trigger so that in the target model's latent space, the trigger-inserted samples are close to the clean samples in the target class. We selected Adaptive-Blend~\cite{adaptive-blend} in this category for our experiment.

\noindent$\bullet$\textit{~Attack Parameters}.We inserted 3,600, 3,000, and 9,600 poisoning data samples (i.e., a poisoning rate of $\alpha = 6\% $) into the CIFAR-10, MNIST, and Tiny ImageNet training datasets, respectively. These samples were uniformly selected from each class, establishing source-agnostic backdoors. Following the methods in the original papers, the trigger in IMC was synthesized as described in~\cite{IMC}, and the trigger in Latent was randomly initialized as detailed in~\cite{latent}.

\noindent$\bullet$\textit{~\ourmethod{}}. For each attack mentioned above, we combine them with \ourmethod{} by algorithm~\ref{alg:one}. More specifically, we set $\alpha = 0.6$ and $\beta = 10\%$ (Algorithm~\ref{alg:one}).

\noindent\textbf{Trigger inversion}. 
We implemented and tested four backdoor countermeasures based upon trigger inversion: Neural Cleanse~\cite{NC}, TABOR~\cite{tabor}, K-arm~\cite{k-arm}, and Pixel~\cite{pixelbackdoor}. In our experiments, we utilized 10\% of the training data and the default hyper-parameters provided in the original papers for trigger reconstruction.

\subsection{Putative Trigger Effectiveness}\label{subsec: Trigger accuracy}

Existing methods measure the effectiveness of trigger inversion by computing the similarity between the reconstructed and real triggers, e.g., based on $l_1$ distance, which is insufficient since a similar pattern may not have a similar backdoor effect (i.e., ASR). We propose a set of metrics to measure trigger accuracy. Below we present our experimental results on the effectiveness of backdoor detection by four trigger inversion algorithms: (NC~\cite{NC}, TABOR~\cite{tabor}, Pixel~\cite{pixelbackdoor}, and K-arm~\cite{k-arm}), by comparing the effectiveness of the backdoor attacks before and after the enhancement by \ourmethod{}. More specifically, after the trigger is generated by each backdoor attack method, we use \ourmethod{} to enhance this trigger as described in section~\ref{subsec: Gradient Shaping}. Here, we append a symbol ``*" to the name of each backdoor attack to indicate the respective attack enhanced by \ourmethod{}. For example, ``BadNet*" indicates BadNet enhanced by \ourmethod{}. 

\noindent\textbf{Metrics}. 
In our experimental study, we utilize four quantitative metrics to measure the effectiveness of a backdoor in evading a gradient-based inversion algorithm (for reconstructing a trigger $(\boldsymbol{\Delta}, \boldsymbol{M})$ in a model $f$):  

\noindent$\bullet$~$\epsilon_1$: The difference between the real trigger's ASR on the backdoored model and that on the ``sanitized'' model retrained to unlearn the reconstructed trigger: that is, $\epsilon_1 = |ASR_{unlearn}-ASR|$. A smaller difference indicates that the reconstructed trigger is less accurate, thus, unlearning is less effective.

\noindent$\bullet$~$\epsilon_2$: The Jaccard distance between the trigger mask of the reconstructed trigger $M'$ and of the real trigger $M$ can be calculated as $J(M',M) = \frac{|M' \cap M|}{|M'|+|M|-|M' \cap M|}$.

\noindent$\bullet$~$\epsilon_3$: The ASR of the reconstructed trigger $(\boldsymbol{M'},\boldsymbol{\Delta'})$ on a clean model $f^*$: $\epsilon_4 = ASR_{f^*}'$. A large $ASR_{f^*}'$ indicates that the reconstructed trigger is likely a natural trigger~\cite{sig}, not the real one meant to be recovered.

\noindent $\bullet$~$\epsilon_4$: $AUC$ score of backdoor detection. The trigger inversion methods often use the $l_0$ norm of the reconstructed trigger as the measurement to distinguish backdoored models from benign models: the lower of the $l_0$ norm, the more probable the model has been backdoored. An AUC score of 50\% in backdoor model detection suggests that the backdoored model is indistinguishable from benign models.

% $\bullet$~$l_1$ norm: The $l_1$ norm of the reconstructed trigger mask, which is supposed to be small for a real trigger~\cite{NC}.
%This norm suppose to be small, as the real triggers we injected all have small norms. 
%\end{itemize}

Notably, for a trigger inversion algorithm with ideal performance,
  $\epsilon_3$ is anticipated to be close to 0, while $\epsilon_1$,$\epsilon_2$ and $\epsilon_4$ are anticipated to be close to 1. 

\noindent\textbf{Experimental results}. Here we present our results as measured by the aforementioned metrics. Due to the space limit, we defer our complete experimental results to Table~\ref{tab:trigger_accuracy_1} in Appendix and only report representative results ($\epsilon_4$) in this section.

\noindent$\bullet$\textit{~$\epsilon_1$: effectiveness of unlearning}.
The reconstructed trigger can be used for backdoor unlearning~\cite{NC,tabor,k-arm}.  
After we reconstructed the trigger for a given backdoored model during the unlearning procedure, we first built an unlearning dataset, including randomly selected 10\% of the training data (6,000 in MNIST, 5,000 in CIFAR-10, and 16,000 in Tiny ImageNet). Then, we added the reconstructed trigger onto 10\% of the unlearning dataset (600 in MNIST, 500 in CIFAR-10, and 1,600 in Tiny ImageNet) while keeping their class labels intact (the original source class). After that, we fine-tuned the model on this unlearning dataset. We used SGD as the optimizer in the experiment and set the learning rate = $0.01$ and momentum = $0.9$. As shown in Table~\ref{tab:trigger_accuracy_1}, after unlearning with the reconstructed trigger by various trigger inversion algorithms, most models poisoned by the attack enhanced by \ourmethod{} still preserve much higher ASRs (almost identical to those before unlearning), so that the \ourmethod{}-enhanced attacks achieve lower $\epsilon_1$ than respective backdoor attack. Table\ref{tab:trigger_accuracy_1} shows that on  CIFAR-10, BadNet achieves the worse performance against the trigger inversion defense of Tabor ($\epsilon_1 = 97.5\%$), which is significantly enhanced by \ourmethod{} ($\epsilon_1 = 1.5\%$). Among other attacks, LSBA* has the best performance under pixel as 0.6\%.

\noindent$\bullet$\textit{~$\epsilon_2$: distance between trigger masks}. We observed that the reconstructed triggers from the models poisoned by \ourmethod{}-enhanced attacks have very low similarity with the real triggers (i.e., the overlap between the real and the reconstructed triggers are less than 20\%). By comparison, the reconstructed triggers from the models under the backdoor attacks without \ourmethod{} enhancement overlap with the real triggers by about 10\% - 60\%. On CIFAR-10, when enhanced by \ourmethod{}, DEFEAT* has worse performance against pixel ($\epsilon_2 = 0.13$). While BadNet* has the best performance against NC ( $\epsilon_2 = 0.00$) (Table~\ref{tab:trigger_accuracy_1}).

\noindent$\bullet$\textit{~$\epsilon_3$: $ASR$ of the reconstructed triggers on a clean model}. 

We also computed $\epsilon_3$, the ASR of the reconstructed triggers from the poisoned models %$(\boldsymbol{\Delta_t},\boldsymbol{M_t})$ 
on a clean model for the same task. In our experiment, we used CIFAR-10, MNIST, and Tiny ImageNet as the clean datasets to train the clean models. After a trigger is reconstructed from a poisoned model, we randomly select 500 images (200 from Tiny ImageNet) from the source class of the clean dataset and insert the trigger on them. The ASR was then measured on this set of trigger-inserted samples on the clean model. As shown in Table~\ref{tab:trigger_accuracy_1}, %third part of Table~\ref{tab: trigger accuracy}, 
the reconstructed triggers from the models poisoned by \ourmethod{}-enhanced attacks have a relatively high ASRs on the clean model, almost comparable with their ASRs on the poisoned models, whereas the reconstructed triggers from the models poisoned by the attack without \ourmethod{} enhancement have much lower ASRs. This indicates that any useful trigger recovered from the models poisoned by \ourmethod{}-enhanced attacks are likely to be a natural trigger  introduced by the legitimate learning process that has nothing to do with the injected triggers. 
%indicating that for the models poisoned by \ourmethod{}, the reconstructed triggers are in fact not the same as the real triggers. 
 On dataset CIFAR-10,  when enhanced by \ourmethod{}, LSBA* has worse performance against pixel ($\epsilon_3 = 42.1\%$) while Adaptive-Blend has the best performance against ($ \epsilon = 28.3\%$). 

\noindent$\bullet$\textit{~$\epsilon_4$ AUC}.  
As mentioned earlier, our research shows that trigger inversion algorithms are unlikely to effectively reconstruct and remove the triggers injected by \ourmethod{}, even though they are largely successful on the triggers injected by existing backdoor attacks. 
In some cases, however, the backdoor defense methods just need to detect the infected models (and discard them afterward), even though they cannot accurately reconstruct the real trigger. 
Our research evaluated how successfully these trigger inversion methods can detect the models poisoned by \ourmethod{}-enhanced attacks. Specifically, we train 24 clean models; for each of the three different datasets (MNIST, CIFAR-10, and Tiny ImageNet), we generate two clean models using each of four different neural network structures (VGG-16, ResNet-101, ShuffleNet, and ResNet18, respectively) to analyze their detection accuracy. Similarly, we train 24 models for each attack on three datasets using four neural network structures. In our research, we measured the AUC score of NC, TABOR, K-arm, and Pixel on these 48 models (24 clean models and 24 backdoored models). As shown in Table~\ref{tab:trigger_AUC}, we present the result of using VGG-16 network structure (Delay rest of the results in Appendix~\ref{subsec:additional eps4}). Generally speaking, the AUC scores of different defense strategies from the models poisoned by different backdoor attacks with \ourmethod{} enhancement are significantly smaller than those from the models poisoned by the same attack without \ourmethod{} enhancement, which indicates the better effect of the \ourmethod{}-enhanced attacks to evade the detection by all tested triggers inversion algorithms. In particular, LSBA enhanced by \ourmethod{} successfully evades the detection by the trigger inversion algorithms with AUCs below 65\% for all of them.

Additionally, we delve deeper into the exploration of backdoor removal using putative triggers in Appendix~\ref{subsec:additional eps4}, where we present a comparative experimental analysis. The results from these experiments demonstrate that backdoor attacks enhanced by GRASP are more resilient to removal attempts using reverse-engineered putative triggers.
% 另外我们在Appendix~\ref{subsec:additional eps4}中，还进一步研究了使用putative trigger进行backdoor remove，观察对比的实验。实验结果显示，GRASP-enhanced backdoor attack 更不容易被reverse出来的putative trigger remove掉

\begin{table*}[ht]
\centering
%\footnotesize
\small
\renewcommand{\arraystretch}{1.2}
\begin{tabular}{c|ccccc|cccc|cccc}
\toprule
\multicolumn{2}{c|}{}       & \multicolumn{4}{c|}{CIFAR-10} & \multicolumn{4}{c|}{MNIST} & \multicolumn{4}{c}{Tiny ImageNet} \\ \midrule
\multicolumn{2}{c|}{}       & NC  & Tabor  & K-arm  & Pixel & NC & Tabor & K-arm & Pixel & NC & Tabor & K-arm & Pixel \\ 

\midrule
\multicolumn{13}{c}{$\epsilon_4$: AUC }\\
\midrule
%                                   CIFAR-10                                        MNIST                                          Tiny ImageNet      
%                                   & NC        & Tabor     & ABS       & Pixel     & NC       & Tabor     & ABS       & Pixel     & NC        & Tabor     & ABS       & Pixel \\ 
%Patch
\multicolumn{2}{c|}{BadNet}         & 79.9\% & 84.0\% & 85.3\% & 91.8\% & 78.6\% & 81.0\% & 82.7\% & 90.3\% & 75.6\% & 77.8\% & 80.4\% & 84.9\%   \\
\multicolumn{2}{c|}{~BadNet*}        & 54.7\% & 56.1\% & 60.1\% & 80.2\% & 54.0\% & 55.0\% & 60.5\% & 83.9\% & 55.7\% & 56.7\% & 57.5\% & 78.5\%   \\
\multicolumn{2}{c|}{LSBA}           & 66.5\% & 68.2\% & 72.1\% & 81.0\% & 67.7\% & 69.6\% & 70.7\% & 78.4\% & 63.5\% & 70.0\% & 70.5\% & 85.8\%   \\
\multicolumn{2}{c|}{~LSBA*}          & 55.1\% & 55.8\% & 58.8\% & 63.7\% & 53.2\% & 57.3\% & 55.8\% & 62.7\% & 55.8\% & 52.0\% & 56.8\% & 64.6\%    \\
\multicolumn{2}{c|}{Composite}      & 67.9\% & 65.9\% & 70.1\% & 85.2\% & 66.4\% & 65.0\% & 68.8\% & 82.5\% & 65.0\% & 65.0\% & 65.9\% & 81.7\%  \\
\multicolumn{2}{c|}{~Composite*}     & 53.5\% & 58.6\% & 61.0\% & 72.9\% & 52.5\% & 52.8\% & 59.5\% & 71.8\% & 54.5\% & 53.7\% & 58.1\% & 70.5\%   \\
%Clean label
\multicolumn{2}{c|}{Latent}         & 79.2\% & 77.1\% & 78.8\% & 87.9\% & 79.9\% & 78.8\% & 81.1\% & 89.5\% & 73.6\% & 79.2\% & 74.9\% & 83.5\%   \\
\multicolumn{2}{c|}{~Latent*}        & 52.5\% & 54.5\% & 59.8\% & 76.0\% & 54.2\% & 54.8\% & 59.0\% & 74.6\% & 53.9\% & 56.0\% & 56.5\% & 70.8\%   \\
%Imperceptible
\multicolumn{2}{c|}{DEFEAT}         & 65.2\% & 63.2\% & 77.8\% & 69.6\% & 67.0\% & 69.8\% & 80.5\% & 71.1\% & 63.6\% & 67.3\% & 77.0\% & 67.6\%   \\
\multicolumn{2}{c|}{~DEFEAT*}        & 58.8\% & 59.9\% & 71.6\% & 61.4\% & 58.9\% & 58.5\% & 70.9\% & 59.7\% & 58.3\% & 58.9\% & 72.0\% & 62.6\%   \\
\multicolumn{2}{c|}{IMC}            & 68.0\% & 64.2\% & 76.9\% & 79.8\% & 66.6\% & 68.8\% & 76.7\% & 80.2\% & 67.5\% & 73.9\% & 76.3\% & 78.0\%   \\
\multicolumn{2}{c|}{~IMC*}           & 55.9\% & 55.3\% & 71.9\% & 71.1\% & 54.7\% & 52.9\% & 74.0\% & 73.6\% & 64.8\% & 64.7\% & 71.8\% & 75.1\%   \\
%Latent space similarity
\multicolumn{2}{c|}{Adaptive-Blend} & 67.1\% & 66.5\% & 68.2\% & 76.9\% & 59.9\% & 62.5\% & 66.0\% & 81.5\% & 62.9\% & 65.0\% & 65.5\% & 76.8\%   \\
\multicolumn{2}{c|}{~Adaptive-Blend*}& 54.2\%    & 56.3\%    & 57.2\%    & 62.8\%    & 55.1\%   & 57.1\%    & 62.0\%    & 73.2\%    & 54.5\%    & 53.5\%    & 54.8\%    & 68.2\%   \\
\bottomrule

\end{tabular}
\caption{The AUCs of backdoor detection by trigger inversion methods on the backdoored models poisoned by different backdoor attacks with and without the enhancement of \ourmethod{}. The attack with the \ourmethod{} enhancement is denoted by the symbol ``*" appended to the name of the respective attack.}
\label{tab:trigger_AUC}
\vspace{-5mm}
\end{table*}

\subsection{Against Weight Analysis Detection}\label{subsec:Against Weight Analysis}

% 把Theory 移到Appendix了
Weight analysis aims to distinguish backdoor and benign models by analyzing the signals in model parameters. Specifically, such distinguishable signals are first retrieved, often through training a classifier on the parameters of some sample models, which are subsequently utilized to predict whether any given model is backdoored~\cite{trojan_signiture}\cite{AC}\cite{du2019robust}. In this section, we evaluated the effectiveness of \ourmethod{} against the weight analysis-based backdoor detection methods, which have been shown to perform well in the recent backdoor competitions~\cite{trojai, NIPS_competation}. In the Appendix, through the theoretical analysis, we show that the backdoored models poisoned by \ourmethod{}-enhanced attacks are not further away from the benign models for the same primary task than the backdoored model poisoned by the same attack without \ourmethod{} enhancement.

Here we selected Trojan Signature (TS)\cite{trojan_signiture},  MNTD\cite{MNTD}, Activation Clustering (AC)~\cite{AC} , Beatrix~\cite{ma2022beatrix} and ABS~\cite{abs}, the representative methods based on weight analysis. We computed their AUCs on 20 models using VGG16, including ten clean models and ten backdoored models, respectively, trained on each of the three datasets (CIFA-10, MNIST, and Tiny ImageNet). The backdoored models were poisoned by the five backdoor attacks with or without the \ourmethod{} enhancement, respectively. Here, the five attacks were selected because they were shown to be effective against the weight analysis-based backdoor defense. Due to the space limit, we only present the most important results in Table \ref{tab: weight analysis results}, the entire results are presented in Table~\ref{tab:weight analysis AUC entire}. In general, the detection ability (AUC) by the weight analysis methods is lower or comparable on the five attacks when they are enhanced by \ourmethod{}, indicating \ourmethod{} enhancement does not reduce the effectiveness of these attacks against the weight analysis backdoor defense.

\begin{table}[htp]
\centering
% \footnotesize
\small
% \captionsetup{font=small}
\renewcommand{\arraystretch}{1.2}
\begin{tabular}{cc|c|c|c}
\toprule
\multicolumn{2}{c|}{}                                                       & CIFAR-10    & MNIST    & Tiny ImageNet         \\ \midrule
\multicolumn{1}{c}{\multirow{2}{*}{ABS}}                 & DFST             &  67.4\%    &  65.0\%  &  67.2\%       \\
\multicolumn{1}{c}{}                                     & ~DFST*            &  63.1\%    &  62.7\%  &  61.4\%       \\ \hline
\multicolumn{1}{c}{\multirow{2}{*}{AC}}                  & AB               &  68.4\%    &  69.1\%  &  66.6\%       \\
\multicolumn{1}{c}{}                                     & ~AB*              &  57.2\%    &  59.0\%  &  60.1\%       \\ \hline
\multicolumn{1}{c}{\multirow{2}{*}{TS}}                  & DEFEAT           &  68.9\%    &  67.3\%  &  66.2\%       \\
\multicolumn{1}{c}{}                                     & ~DEFEAT*          &  60.5\%    &  68.0\%  &  65.1\%        \\ \hline
\multicolumn{1}{c}{\multirow{2}{*}{MNTD}}                & DEFEAT           &  69.2\%    &  73.1\%  &  70.9\%       \\
\multicolumn{1}{c}{}                                     & ~DEFEAT*          &  66.0\%    &  72.9\%  &  69.4\%        \\ \hline
\multicolumn{1}{c}{\multirow{2}{*}{Beatrix}}              & Low-c           &  58.3\%    &  72.3\%  &  68.1\%       \\
\multicolumn{1}{c}{}                                     & ~Low-c*          &  56.9\%    &  72.4\%  &  67.3\%        \\
\bottomrule
\end{tabular}
\vspace{-5pt}
\caption{The AUCs of weight analysis-based backdoor detection methods on the benign and backdoored models poisoned by DFST, AB, MNTD, Beatrix and DEFEAT with and without \ourmethod{} enhancement.}
\label{tab: weight analysis results}
\vspace{-5mm}
\end{table}

%% file: 6-Resilience-to-Backdoor-Mitigation.tex
\section{Resilience to Backdoor Mitigation}
\label{sec:Resilience to Backdoor Mitigation}

In this section, we evaluated the resilience of \ourmethod{} to backdoor defense methods that do not rely on backdoor detection. In our experiments, we considered not only the five types of backdoor defenses (mitigation or unlearning) as summarized in \cite{backdoorsurvey}: Preprocessing-based Defenses, Model Reconstruction, Poison Suppression, and Certified Backdoor Defense, which are not based on backdoor detection techniques such as trigger inversion or weight analysis, but also an emerging category, Training Procedure Defense. We selected a total of seven representative defense methods across these six types: DeepSweep (DS)\cite{zeng2020deepsweep}, Fine-pruning (FP)\cite{fine-purning}, NAD~\cite{NAD}, GangSweep (GS)\cite{zhu2020gangsweep}, DBD\cite{du2019robust}, RAB~\cite{rab}, ABL~\cite{ABL}, and compared their performance in defending against the chosen backdoor attacks before and after the \ourmethod{} enhancement.

We measured the ASRs of backdoors after the backdoor mitigation on the models (VGG16) poisoned by the selected backdoor attacks with or without the \ourmethod{} enhancement, respectively. Here, for each defense method, we selected a backdoor attack that has been shown to effectively evade the respective defense in previous studies to demonstrate the enhancement by \ourmethod{} does not reduce its effectiveness to evade the respective defense methods. 

We summarize the results from both experiments in Table~\ref{tab: Other defense}. Except for special notes, all backdoored models before mitigation achieve an ASR above 95\%. Here, the notations are the same as used in Section \ref{sec:Against Detection}: a symbol ``*" is appended to the name of the backdoor attack to indicate the respective attack enhanced by \ourmethod{}. Below, we discuss the results of different defense methods in detail.

\begin{table}[htp]
\centering
% \footnotesize
\small
% \captionsetup{font=small}
\renewcommand{\arraystretch}{1.2}
\begin{tabular}{cc|c|c|c}
\toprule
\multicolumn{2}{c|}{}                                                       & CIFAR-10    & MNIST   & Tiny ImageNet         \\ \midrule
\multicolumn{1}{c}{\multirow{2}{*}{DS}}                  & HaS-Net          &  67.2\%    &  68.9\%  &  65.1\%       \\
\multicolumn{1}{c}{}                                     & ~HaS-Net*         &  65.1\%    &  68.3\%  &  66.1\%       \\ \hline
\multicolumn{1}{c}{\multirow{2}{*}{FP}}                  & DEFEAT           &  81.4\%    &  87.6\%  &  80.3\%       \\
\multicolumn{1}{c}{}                                     & ~DEFEAT*          &  83.2\%    &  88.0\%  &  81.9\%       \\ \hline
\multicolumn{1}{c}{\multirow{2}{*}{NAD}}                 & DEFEAT           &  79.2\%    &  81.1\%  &  80.3\%       \\
\multicolumn{1}{c}{}                                     & ~DEFEAT*          &  79.6\%    &  80.3\%  &  80.7\%       \\ \hline
\multicolumn{1}{c}{\multirow{2}{*}{DBD}}                 & IMC              &  54.2\%    &  59.7\%  &  55.0\%       \\
\multicolumn{1}{c}{}                                     & ~IMC*             &  63.5\%    &  64.0\%  &  62.6\%       \\ \hline
\multicolumn{1}{c}{\multirow{2}{*}{GS}}                  & AB               &  64.9\%    &  59.3\%  &  62.1\%       \\
\multicolumn{1}{c}{}                                     & ~AB*              &  65.4\%    &  63.2\%  &  63.5\%       \\ \hline
\multicolumn{1}{c}{\multirow{2}{*}{RAB}}                 & DFST             &  95.3\%    &  94.1\%  &  91.2\%       \\
\multicolumn{1}{c}{}                                     & ~DFST*            &  94.6\%    &  91.2\%  &  91.4\%       \\ \hline
\multicolumn{1}{c}{\multirow{2}{*}{ABL}}                 & AB               &  91.6\%    &  93.6\%  &  92.5\%       \\
\multicolumn{1}{c}{}                                     & ~AB*              &  95.2\%    &  96.7\%  &  93.4\%       \\
\bottomrule
\end{tabular}
\caption{The ASRs of the backdoored models after the mitigation by the backdoor mitigation methods, Note that the high ASR indicates that a backdoor attack is more resilient to the respective defense method. Here, all models are trained using three different datasets, and the backdoored models are infected by different attacks with and without the enhancement of  \ourmethod{}. (The attacks enhanced by \ourmethod{} is denoted by ``*").}
\label{tab: Other defense}
\vspace{-5mm}
\end{table}

\vspace{3pt}\noindent\textbf{Preprocessing-based Defenses}. Preprocessing-based defenses aim to remove the putative poison samples in the training dataset. DeepSweep (DS)~\cite{zeng2020deepsweep} is selected as the representative in this category. We tested the average ASR against the models infected by HaS-Net~\cite{low-conf}, which is a low-confidence backdoor attack. The ASRs of the models poisoned by the \ourmethod{}-enhanced HaS-Net attack (HaS-Net*) after the DeepSweep mitigation is comparable with those on the model poisoned by the HaS-Net attack (Table~\ref{tab: Other defense}), which indicates \ourmethod{} enhancement did not make the Has-Net backdoored models more easily mitigated by the DeepSweep.

\vspace{3pt}\noindent\textbf{Model Reconstruction}. 
Fine-pruning~\cite{fine-purning} and NAD~\cite{NAD} are two typical model reconstruction methods to remove the backdoor.  
We conducted an experiment to compare the performance of Fine-pruning and NAD to defend against DEFEAT~\cite{defeat} with or without \ourmethod{} enhancement (DEFEAT or DEFEAT*). Table~\ref{tab: Other defense} shows the ASRs of the injected trigger after fine-pruning and NAD on the respective backdoored models. Note that the model reconstruction methods typically strike a trade-off between the accuracy of the primary task (ACC) and ASR. In our experiment, to keep the fidelity of the mitigation, we control the mitigation procedure such that the ACC does not  decrease by more than 5\%. We use 10\% of clean training data for mitigation, and for fine-pruning, we set 10\% activation pruning.   

Overall, the ASRs of the Fine-prune and NAD on the DEFEAT* attacked models are comparable with the DEFEAT attacked models, indicating the DEFEAT attack with and without \ourmethod{} enhancement are similarly effective against the Fine-pruning and NAD mitigation methods.

\vspace{3pt}\noindent\textbf{Non-Gradient Based Trigger Synthesis}.
Gangsweep~\cite{zhu2020gangsweep} is a non-gradient-based trigger synthesis defense that used the reconstructed trigger for backdoor mitigation. We performed an experimental study to confirm Adaptive-blend enhanced by \ourmethod{} (Adaptive-Blend*) does not generate the backdoors that can be easily mitigated by Gangsweep (GS). Table~\ref{tab: Other defense} shows the ASRs after the mitigating of Gangsweep on the backdoored models created by using Adaptive-Blend (AB) with or without \ourmethod{} enhancement. The ASR on the model attacked by \ourmethod{}-enhanced AB (AB*) is comparable (for the Tiny ImageNet dataset) or lower (for the CIFAR and MNIST dataset) than the ASRs on the models attacked by AB, indicating \ourmethod{} enhancement does not make the attack more easily to be mitigated by the Gangsweep.

\vspace{3pt}\noindent\textbf{Poison Suppression}.
For poison suppression defense, most methods (e.g., DBD~\cite{DBD}) learn a backbone of a DNN model via self-supervised learning based on training samples without their labels to capture those suspicious training data during the training process. 
We tested the performance of DBD~\cite{DBD} defense against models attacked by IMC\cite{IMC} with and without \ourmethod{} enhancement. As shown in Table \ref{tab: Other defense}, the ASR of the DBD in the IMC* attacked models are higher than the IMC attacked models, indicating \ourmethod{} enhancement does not make the attack more easily to be mitigated by the DBD.

\vspace{3pt}\noindent\textbf{Certified Backdoor Defense}.
As previously discussed in Section~\ref{subsec:Impact of noise}, if the value of $c$ in GRASP is chosen to be too small, specifically leading the trigger effective radius below the model's robustness, \ourmethod{} might be susceptible to nullification by backdoor certification methods. However, the model's robustness is typically much larger than the trigger effective radius, a point we have illustrate and discuss in Section~\ref{subsec:Impact of noise}. Owing to this fact, GRASP can preserve its stealthiness (by significantly reducing $c$) while circumventing nullification by certification methods (by ensuring that the trigger effective radius stays larger than the model's robustness).
RAB~\cite{rab} is a certified defense method that aims to eliminate the backdoor in the target model. We performed an experimental study to confirm \ourmethod{} enhancement does not generate the backdoor that is more easily mitigated by RAB.  Table\ref{tab: Other defense} shows the ASR of the injected trigger after RAB on the same models which attacked by DFST\cite{DFST} and DFST enhanced by \ourmethod{} (DFST*).
The ASR on DFST* is comparable to that on the DFST attacked models (MNIST has the most significant difference, which DFST* is 2.9\% lower than DFST), indicating combing \ourmethod{} backdoored models are not easier to mitigate by the RAB than the model infected by DFST. 

\vspace{3pt}\noindent\textbf{Training procedure Defense}
Apart from the five types of mitigation techniques summarized by \cite{backdoorsurvey}, we observe the emergence of a new mitigation approach that aims to remove backdoors during the training process. Notably, ABL\cite{ABL} seeks to identify backdoored inputs during the early training stages. This is because, in a typical backdoored training scenario, trigger-inserted samples (backdoor task) often learn much faster than clean data (legitimate task), as the backdoor task is generally easier than the legitimate task. Based on this observation, ABL mitigates backdoors by eliminating those training samples with small losses during the early stages of the training process.
However, we argue that \ourmethod{} can potentially evade ABL. This is because training on the \ourmethod{} poisoned dataset increases the loss for both poisoned and enhanced data during the initial phase of training. By injecting both poisoning and enhancement data, \ourmethod{} makes the backdoor task more challenging to learn. Consequently, during the early training phase, \ourmethod{} assists data poisoning in evading mitigation from ABL.

We conducted an experiment to verify that \ourmethod{} enhancement does not produce a backdoor that is more easily mitigated by ABL. Table~\ref{tab: Other defense} displays the ASR of the injected trigger following ABL on the same models attacked by AB~\cite{adaptive-blend} and those enhanced by \ourmethod{} (AB*). The ASR on AB* models is higher than that on the AB-attacked models (with the least significant difference on CIFAR-10, where AB is 1.9\% lower than AB*), indicating that combining \ourmethod{} with backdoored models does not make them more susceptible to ABL-based mitigation than models infected by AB.

In summary, we find that for the attacks that effectively evade backdoor mitigation, the \ourmethod{} enhancement  will not make the mitigation less effective. This means that \ourmethod{} enhancement can effectively fend off the existing backdoor defenses even though it is designed for evading trigger mitigation. For some mitigation methods like DBD, the IMC attack, the \ourmethod{} enhancement in fact increases their effectiveness.

% is less vulnerable to detection by all types of backdoor detection techniques compared with BadNet. It means that \ourmethod{} can fend off the existing backdoor defenses more effectively even given the fact that it is designed specifically for defeating trigger inversion.  
% Also it is important to note that these approaches, even when they appear to be effective on \ourmethod{}, work under different threat models than trigger-inversion based solutions and require more resources, particularly clean samples and poisoned samples, to detect backdoor triggers. For example, Activation clustering, Gangsweep, Anomalous Logits, and RAB require trigger-carrying samples or even the whole training dataset, and Fine-pruning requires a large amount of clean data (at least 10\% of the training data). With the help of these extra resources, they could perform better on \ourmethod{} than trigger inversion. Also for ABS, we can see that the detection accuracy on \ourmethod{} is significantly lower than on BadNet, possibly due to the fact that it includes a trigger inversion component. 

%% file: 7-mitigation-and-limitaion.tex
\section{Mitigation and Limitation}
\label{sec:Mitigation and Limitation}
%Limitaion:***
%Our method is designed to evade trigger inversion algorithms, 
%For 
%
%Discussion:

\begin{figure}[htbp] 
% \captionsetup{font=small}
\centering
  \includegraphics[height=2.0in]{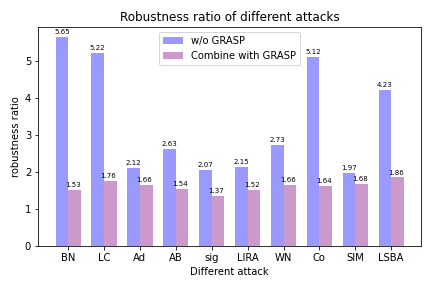}
 
\caption{Blue bars show the ratio between the trigger effective radius and the primary task effective radius on different backdoor attacks. The Red bars show the ratio of different backdoor attacks that are enhanced by \ourmethod{}. }
\label{fig:rob_ratio_hist}
\vspace{-3mm}
\end{figure}

%There is one question may concern about. 
\ourmethod{} can successfully increase the change rate around the trigger-inserted inputs, effectively reducing the trigger effective radius of these inputs. 
Note that the trigger effective radius should not be reduced to be lower than the primary task model robustness (primary task effective radius) since otherwise, the backdoor attack may be defended by a straightforward strategy during inference: one can add the noise at the level above the robust radius of the backdoor task but below the robust radius of the primary task into each input; as such, the trigger-inserted inputs will not be predicted as the target class while the prediction of the benign input will not be changed, indicating the backdoor is removed without affecting the primary task.  
In practice, we found it is easy to reduce the  trigger effective radius while keeping it above the primary task effective radius, as we observed that in the BadNet backdoor attack, the trigger robustness is always much greater than the primary task effective radius (as shown in Fig.~\ref{fig:rob_ratio_hist}).  
Therefore, it is always possible for \ourmethod{} to generate backdoors more effectively to evade the trigger-inversion algorithms while not affecting the performance of the primary task.

%\vspace{3pt}\noindent$\bullet$\textit{~ Limitation: Low confidence trigger}. 
In Section~\ref{sec:GRASP}, we assume the model will always give approximately 100\% confident prediction (as the target class) on all trigger-inserted inputs. In practice, when this assumption does not hold, for example, in~\cite{low-conf}, where a low confidence backdoor is injected into the model by manipulating the logits of the poisoning data, the change rate around a perfect trigger may not be very large. Specifically, %when we have a low confidence trigger, 
the trigger-inserted inputs may be predicted as the target class with the lowest confidence in the backdoored model, which turns out to be a perfect trigger without any constraints on the local Lipschitz constant. For such backdoors, \ourmethod{} cannot further enhance their stealthiness.

%% file: 8-Related-work.tex
% \section{Related work}\label{sec:Related work}
% \Rui{The prior study~\cite{IMC} posits that robust learning with respect to legitimate data (i.e., enhancing the legitimate effective radius) can contribute to the stealthiness of backdoor attacks. However, there are certain inefficiencies associated with this approach compare with \ourmethod{}. Firstly, enhancing the legitimate effective radius to increase the stealthiness of backdoors is often an expensive trade-off. It becomes particularly challenging when further enhancing the legitimate effective radius is difficult, and yet the stealthiness is still insufficient for the backdoor to evade current detection methods, especially those that involve reverse engineering. Secondly, as attackers, having control over the training procedure, which is usually required for this approach, is often impractical in real-world attack scenarios.}

%% file: 9-Discussion.tex
\section{Related Work}\label{sec:Discussion}

The prior study~\cite{IMC} showed that robust learning on the primary task (i.e., to increase its effective radius) can enhance the stealthiness of backdoor attacks. However, comparing with \ourmethod{}, this method is less effective because even when it is difficult to further increase
the effective radius of the primary task, the stealthiness is still insufficient
for the backdoor to evade current detection methods. More importantly, this method requires the adversary to control the model training procedure, which is often impractical in real-world attack scenarios. In comparison, \ourmethod{} can be achieved through data contamination as shown here.

% We also evaluate the effectiveness of trigger inversion using three different optimizers: SGD (first-order), Adam~\cite{adam} (first-order with momentum), and AdamHessian~\cite{adahessian} (second-order with momentum). We follow the same evaluation setup as described in Section~\ref{sec:Against Detection} and use NC to detect trojans in models that may have been attacked by BadNet on the CIFAR-10 dataset. For the two optimizers with momentum, we set the momentum equal to $0.9$. 

% \begin{figure}[H]
% \centering
% \captionsetup{font=small}
% \includegraphics[width=0.25\textwidth]{IEEE/optimizer_compare.png}
% \caption{The fold-change of step size when utilizing three different optimizers to run NC for detection. As a second-order optimizer, AdamHessian does not offer a selection for step size, which is why it has the same AUC scores on the graph.}
% \label{Fig: optimizer compare}
% \end{figure}

%% file: 10-Conclusion.tex
\section{Conclusion}
\label{sec:conclusion}
In this paper, we analyzed the efficacy of trigger inversion algorithms in backdoor defense, finding that current backdoor attacks inject noise-robust triggers, facilitating reconstruction via gradient-based algorithms. Consequently, we introduced a gradient shaping (GRASP) strategy to improve backdoor attacks by diminishing trigger robustness using data poisoning, thus evading defenses using trigger inversion. Through theory and experiments, we showed GRASP's enhancement of top stealthy backdoor attacks' effectiveness against trigger inversion, without impairing their resistance to other defenses, including those based on weight analysis.

%%% Local Variables:
%%% mode: latex
%%% TeX-master: "main"
%%% End:

%% file: 99_appendix.tex
\appendix

Due to space constraints, this Appendix contains only a selection of supplementary materials(Section A to E). For the rest set of appendices (Section F to J), please refer to the supplementary document available at \href{https://www.example.com/full-appendix}{this link}\footnote{\url{https://drive.google.com/file/d/12EVjcrznWnhT-lpdLex8_eaZr3g11t_M/view?usp=sharing}}.

\subsection{Proof of Theorem~\ref{thm:MC_proof}}\label{subsec: proof of MC}

\vspace{2pt}\noindent\textbf{Theorem~\ref{thm:MC_proof}}

Given a piece-wise linear function $\ell(\cdot): [a,b] \rightarrow [0,1] $ with a global optimum sit on a convex hull. Assume such a convex hull satisfies the largest update (step size times the largest gradient) in the convex hull is smaller than the shortest linear piece in the convex hull. After $n$ iterations update, a gradient-based optimizer starting from a random initialization converges to the optimum with the probability:
\begin{equation*}
1- B_1^{-1}(b-a)^{-1}(4-B_1B_2)^n(1-B_1B_2)]
\end{equation*}

\vspace{2pt}\noindent\textbf{Proof:}

As the input space is one-dimension, a gradient-based optimization on a piece-wise linear loss function can then be considered as a Markov chain(MC)~\cite{li2017training}; If we use $\mathcal{A}$ to denote the equivalent MC, each linear piece represents a state (or node) in $\mathcal{A}$. The transition probability between two nodes is the probability that after one update step, the optimization could jump from the first node to the other one. Specifically, consider any two nodes (linear-piece), $i$ and $j$, in $\mathcal{A}$, the transition probability of $i^{th}$ node to the $j^{th}$ means when the optimization is in the $i^{th}$ node, the probability that after one step update, the optimization could move to the $j^{th}$ linear piece. We use $i \text{ connected with } j$ to denote if the transition probability from $i$ to $j$ is not equal to zero. Then the  adjacent matrix $A$ can be written as:

\begin{equation}
    A_{i,j} = \begin{cases}
\frac{l_j}{\alpha \nabla_i}  \hspace{
    5mm}& i \text{ connected with } j \\
0 & o.w \end{cases}
\end{equation}

where $A_{i,j}$ indicate the transition probability of $i^{th}$ node to the $j^{th}$, $\alpha$ is the update step size, $l_j$ is the length of $j^{th}$ linear piece in domain $[a,b]$, and $\nabla_i$ is the gradient of $i^{th}$ linear piece. 

The probabilities that the optimization converges to each linear piece could then be computed by the stationary distribution $\mathcal{S}$ of $\mathcal{A}$:
\begin{equation}
    \mathcal{P} = \lim_{n\rightarrow\infty}\pi_0 A ^n 
\end{equation}
where $\pi_0$ is the initial distribution, which is the initial probability for each linear piece: $\frac{l_i}{b-a}$.

Directly computing $A^n$ is not easy; it requires diagonalization on a conditional Adjencent matrix. We then simplify $\mathcal{A}$ to a two-state MC.

Since the largest update (step size times the largest gradient) in the desired convex hull is smaller than the shortest linear piece in the desired convex hull. This indicates all nodes which represent those linear pieces in the desired convex hull formed a recurrent state; The transition probability that nodes from the desired convex hull to the node outside the desired convex hull are zero. We then can collapse every linear region in the desired convex hull into one state $\mathbb{P}$.

Similarly, we can collapse those linear regions which are not in the desired convex hull into one another state $\mathbb{Q}$; this is because our goal is to compute the stationary probability of $\mathbb{P}$, the details of the stationary probability for each linear regions which are not in the desired convex hull is not necessary. Then we can simplify our $\mathcal{A}$ into two states $\mathbb{P,Q}$. The simplified MC is shown in Fig~\ref{Fig: MC2}:

\begin{figure}[ht]
\centering
% \captionsetup{font=small}
\includegraphics[width=0.12\textwidth]{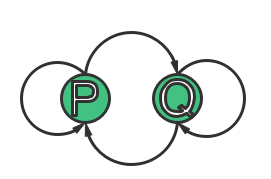} 
\caption{A Two-state Markov Chain. State $\mathcal{P}$ represents the linear regions in the desired convex hull, and state $\mathcal{Q}$ represents the linear region outside the desired convex hull.}
\label{Fig: MC2} 

\end{figure}

The initial distribution can be written as follows:
\begin{equation}
    \pi_0 = [\pi_0^{\mathbb{(P)}},\pi_0^{\mathbb{(Q)}}] = [\frac{\sum\limits_{\mathbb{i\in P}}l_i}{b-a}, \frac{\sum\limits_{\mathbb{i\in Q}}l_i}{b-a} ]
\end{equation}

We now consider the $2 \times 2$ adjacent matrix $A_{\mathbb{P,Q}}$.  Since the desired convex hull satisfies the largest update in the convex hull is smaller than the shortest linear piece in the convex hull. It can be infers that $A_{\mathbb{P \rightarrow P}} = 1$, and $A_{\mathbb{P \rightarrow Q}} = 0$. 

And now let’s consider the entry $A_{\mathbb{Q \rightarrow P}}$, 

\begin{equation}
    \begin{split}
A_{\mathbb{Q \rightarrow P}} &= \frac{\sum\limits_{i \in \mathbb{Q} \text{ \& connect } \mathbb{P}}l_i}{\sum\limits_{i \in \mathbb{Q}}l_i}\cdot \sum\limits_{i \in \mathbb{Q} \text{ \& connect } \mathbb{P}}\frac{l_i}{\sum\limits_{j \in \mathbb{Q} \text{ \& connect } \mathbb{P}}l_j}\cdot\frac{l_i}{\alpha\nabla_i} \\ & = \frac{1}{\sum\limits_{i \in \mathbb{Q}}l_i}\cdot \sum\limits_{i \in \mathbb{Q} \text{ \& connect } \mathbb{P}}l_i\cdot\frac{l_i}{\alpha\nabla_i} \\ &= \frac{1}{\sum\limits_{i \in \mathbb{Q}}l_i}\cdot \sum\limits_{i \in \mathbb{Q} \text{ \& connect } \mathbb{P}}\frac{l_i^2}{\alpha\nabla_i}\end{split}
\end{equation}

The last entry $A_{\mathbb{Q \rightarrow Q}}$ then becomes: 

$A_{\mathbb{Q \rightarrow Q}} = 1- \frac{1}{\sum\limits_{i \in \mathbb{Q}}l_i}\cdot \sum\limits_{i \in \mathbb{Q} \text{ \& connect } \mathbb{P}}\frac{l_i^2}{\alpha\nabla_i}$

Then the adjacent matrix becomes:
\begin{equation*}
A_{\mathbb{P,Q}} =
\end{equation*}
\begin{equation*}
\begin{pmatrix}
1 & 0 \\
\frac{1}{\sum\limits_{i \in \mathbb{Q}}l_i}\cdot \sum\limits_{i \in \mathbb{Q} \text{ \& connect } \mathbb{P}}\frac{l_i^2}{\alpha\nabla_i} & 1-\frac{1}{\sum\limits_{i \in \mathbb{Q}}l_i}\cdot \sum\limits_{i \in \mathbb{Q} \text{ \& connect } \mathbb{P}}\frac{l_i^2}{\alpha\nabla_i} 
\end{pmatrix}
\end{equation*}

As a $2\times2$ matrix, We can apply Hamilton-Cayley theorem to compute $A_{\mathbb{P,Q}}^n$:

\begin{equation}
    \begin{split}A_{\mathbb{P,Q}}^n &= \text{Tr}^n(A_{\mathbb{P,Q}}) \cdot A_{\mathbb{P,Q}} \\ &= (4-\frac{1}{\sum\limits_{i \in \mathbb{Q}}l_i}\cdot \sum\limits_{i \in \mathbb{Q} \text{ \& connect } \mathbb{P}}\frac{l_i^2}{\alpha^2\nabla^2_i})^n\cdot A_{\mathbb{P,Q}} \end{split}
\end{equation}

To make the equation more compact, we denote 

$B_1 = \frac{1}{\sum\limits_{i \in \mathbb{Q}}l_i}$, which indicate the extent of area under the convex hull

$B_2 = \sum\limits_{i \in \mathbb{Q} \text{ \& connect } \mathbb{P}}\frac{l_i^2}{\alpha\nabla_i}$, indicates the extent of the likelihood the linear pieces outside the convex hull can jump into the convex hull.

Then the stationary distribution of $\mathcal{A}$ after $n$ iteration will be:

\begin{equation}
    \begin{split}\mathcal{P} &= \pi_0 A^n_{\mathbb{P,Q}} \\ & = \pi_0\begin{pmatrix} (4-B_1B_2)^n & 0 \\ (4-B_1B_2)^nB_1B_2 & (4-B_1B_2)^n(1-B_1B_2) \end{pmatrix} \\ &= [1-B_1^{-1}(b-a)^{-1},B_1^{-1}(b-a)^{-1}] \cdot  \\ & \begin{pmatrix} (4-B_1B_2)^n & 0 \\ (4-B_1B_2)^nB_1B_2 & (4-B_1B_2)^n(1-B_1B_2) \end{pmatrix} \\ & = [\frac{(4-B_1B_2)^{2n}(B_2-B_1^{-1})}{b-a},\\ & B_1^{-1}(b-a)^{-1}(4-B_1B_2)^n(1-B_1B_2)]\end{split}
\end{equation}

The stationary probability for state $\mathbb{Q}$ is equal to $B_1^{-1}(b-a)^{-1}(4-B_1B_2)^n(1-B_1B_2)$. This shows a negative relationship between the stationary probability for state $\mathbb{Q}$ and $B_1, B_2$. Similarly, stationary probability for state $\mathbb{P}$ is equal to $1- B_1^{-1}(b-a)^{-1}(4-B_1B_2)^n(1-B_1B_2)$. It shows a positive relationship between the stationary probability for state $\mathbb{P}$ and $B_1, B_2$.

After the optimization jump into the desired convex hull, state $\mathcal{P}$ and the step size followed by the assumption in the theorem, as long as $n$ is large enough and the step size followed by the assumption in the theorem, the optimization will converge to the desired optimum.

\begin{figure}[H]
\centering
\captionsetup{font=small}
\includegraphics[width=0.45\textwidth]{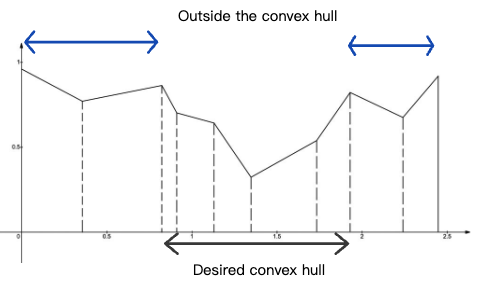} 
\caption{A toy example of piece-wise linear function. The region indicated by the blue arrow represents the area outside of the desired convex hull, the size of which is controlled by $B_2$. The region indicated by the black arrow represents the desired convex hull itself, with its size controlled by $B_1$.}
\label{Fig:piecewiselinear} 
\end{figure}

We further illustrate Theorem~\ref{thm:MC_proof}
using a piecewise linear loss function in one-dimensional input space (Fig~\ref{Fig:piecewiselinear}):

Each linear piece represents one state on the MC.
\begin{figure}[H]
\centering
\captionsetup{font=small}
\includegraphics[width=0.25\textwidth]{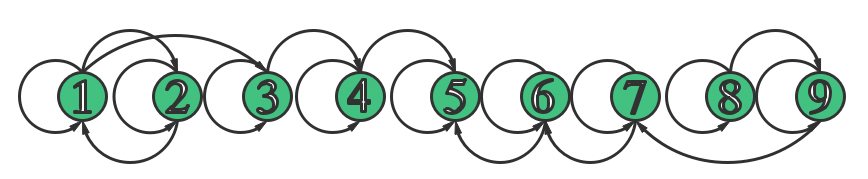} 
\label{Fig: MC9} 
\end{figure}

We can further reduce the MC by collapsing the desired convex hull into one single state $\mathbb{P}$.
\begin{figure}[H]
\centering
\captionsetup{font=small}
\includegraphics[width=0.17\textwidth]{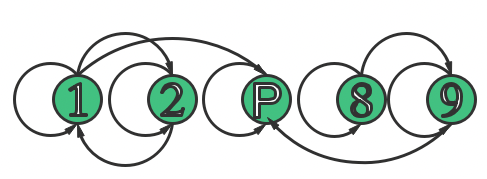} 
\label{Fig: MC5} 
\end{figure}

For the rest of the nodes not in the desired convex hull, we can reduce them into one state $\mathbb{Q}$ (Fig ~\ref{Fig: MC2}). 

End of proof.

\subsection{Proof of Theorem~\ref{thm1a}}\label{subsec: proof of thm2}
\noindent Before we elaborate our theorem, we need to first formally define some concepts and a Lemma from \cite{yang2020adversarial}: 

\begin{definitionbold}[Astuteness]
A classifier $f: \mathcal{X} \rightarrow \mathcal{Y}$ is {\em astute} at an input sample $x$, if the predicted label by $f$ is the same as the true label: $\hat{y}=z(f(x))=y$. 

\end{definitionbold}
\begin{definitionbold}[r-local minimum]
A function $f: \mathcal{X} \rightarrow \mathbb{R}$ has a (unique) $r$-local minimum at $x^{\star}$, if there is no other $x$ on which $f$ gets lower or equal value than what can get on $x^*$, within the ball centered on $x^*$ with radius $r$, 
i.e., $f(x) > f(x^*), \forall x, \|x-x^*\|_2 \leq r$.
%i.e., $||x - x^{\star}|| \leq r$.

\end{definitionbold}
\begin{definitionbold}[Increasing rate and relaxation function]
Given a function $f: \mathcal{X} \rightarrow \mathbb{R}$ with a $r$-local minimum at $x^{\star}$, we define that $f$ has an increasing rate of $\kappa$ at $x^{\star}$, if there exists some $\kappa \geq 0$ and $c_{\kappa} \geq 0$, such that
$f(x) - f\left(x^{\star}\right) \geq \sup_{c_{\kappa},\kappa} c_{\kappa} \cdot\left\|x-x^{\star}\right\|_{2}^{\kappa}$, when $\|x -x^{\star}\| \leq r$. Accordingly, we refer the function $\bar{g}(x) = c_{\kappa} \cdot\left\|x-x^{\star}\right\|_{2}^{\kappa}$ as the relaxation function of $f$ at $x$. 
\end{definitionbold}

\begin{definitionbold}[Local Lipschitz constant]
For a function $f: \mathcal{X} \rightarrow \mathcal{Y}$, a given input $x$ and a pre-defined radius $r$, if $L(f,\mathcal{X}_{x,r})$ exists and is finite, where $\mathcal{X}_{x,r} = \{x':\|x'-x\|_2 < r\}$ and
\begin{equation}
\begin{array}{r@{\quad}l}
 L(f,\mathcal{X}_{x,r}) = \sup_{x_1,x_2 \in \mathcal{X}_{x,r}} \frac{||f(x_2) - f(x_1)||_2}{||x_2-x_1||_2},
\end{array}
\end{equation}
we define $L(f,\mathcal{X}_{x,r})$ as the local Lipschitz constant of $x$ with radius $r$ for function $f$.
\end{definitionbold}

\begin{lemmabold}
Consider the data distribution $X$, and assume the minimum $l_2$ norm between any two different class data is $r$. If a function is astuteness in $X$, then $f$ has a local Lipschitz constant of $r'$ around any $x \in X$ such that $r'\geq r$ 
\end{lemmabold}

\vspace{2pt}\noindent\textbf{Theorem~\ref{thm1a}}

If the noise $\epsilon \sim \mathcal{N}(0,1)$ (i.e., the white noise), and $c < \|x'-x\|_2 \cdot \frac{\Gamma\left(\frac{|m^*|}{2}\right)}{\sqrt{2} \Gamma\left(\frac{|m^*|+1}{2}\right)}$, where $|m^*|$ is the $l_1$ norm (i.e., the size) of the trigger, $\Gamma$ is the Euler’s gamma function. a model backdoor attacked by a backdoor attack and enhanced by \ourmethod{} using the training data points $(x,y),(x',y_t)$ and $(x^*,y)$) has a greater local Lipschitz constant around $x$ than the model backdoored by the same attack without the \ourmethod{} enhancement using the training data points $(x,y),(x',y_t)$.

Similarly, if $\epsilon \sim uniform(-1,1)$, and $c < \|x'-x\|_2$,
the \ourmethod{}-enhanced model has greater local Lipschitz constant around $x$ than the model without the enhancement.

\vspace{2pt}\noindent\textbf{Proof:}

Consider in the BadNet data contamination, recall that trigger $l_1$ norm is $m^*$, the subspace $V \in \mathbb{R}^{m^*}$ is set of those dimensions which the mask matrix $\boldsymbol{M}$ has non-zero entry.   
$$
E(||A(\boldsymbol{x}, \boldsymbol{M},\boldsymbol{\Delta}) - x||_2) = 2r_{BadNet}
$$
where $r_{BadNet}$ is the expectation of robust radius, which is defined in model contaminated by BadNet and trigger-inserted data

 In GRASP, we can choose a random noise $\epsilon$. First let's consider $\epsilon = c\mathcal{N}(0,I)$ is added to trojan input only on the subspace $V$. Then, the expectation of magnitude of this noise is:

$$
     E(\|\epsilon\|_2)=\frac{\sqrt{2} \Gamma\left(\frac{|m^*|+1}{2}\right)}{\Gamma\left(\frac{|m^*|}{2}\right)} cI = 2r_{GRASP}
$$
which implies:

$$
r_{GRASP} = c\cdot \frac{\sqrt{2} \Gamma\left(\frac{|m^*|+1}{2}\right)}{2\Gamma\left(\frac{|m^*|}{2}\right)}
$$
Similarly, when $\epsilon = c \cdot unif(-1,1)$:
$$
     E(\|\epsilon\|_2)= \frac{c}{2} = 2r_{GRASP} \\
$$
And
$$
     r_{GRASP} = \frac{c}{4}
$$

where $\Gamma$ is the Euler’s gamma function, $r_{GRASP}$ is the expectation of robust radius, which is defined in model contaminated by \ourmethod{} and trigger-inserted data. And $c$ is the noise scalar parameter.  

When $\epsilon = c\mathcal{N}(0,I)$ and let $c$ :

$$
c< ||x'-x||_2  \cdot \frac{\Gamma\left(\frac{|m^*|}{2}\right)}{\sqrt{2} \Gamma\left(\frac{|m^*|+1}{2}\right)}
$$

when $\epsilon = c \cdot unif(-1,1)$, we let:

$$
c< \frac{||x'-x||_2}{4}
$$

 Then we have,

\[
\begin{aligned}
2r_{GRASP} & = E(\|\epsilon\|_2) \\
& < E(||A(\boldsymbol{x}, \boldsymbol{M},\boldsymbol{\Delta}) - x||_2) = 2r_{BadNet}
\end{aligned}
\]

According to the Lemma 1, we have the local lipchitz constant around $A(\boldsymbol{x}, \boldsymbol{M},\boldsymbol{\Delta})$  for GRASP is $\frac{1}{r_{GRASP}}$, and for the BadNet contaminating backdoor attack is $\frac{1}{r_{BadNet}}$. And since $r_{GRASP} < r_{BadNet}$, so $\frac{1}{r_{GRASP}} > \frac{1}{r_{BadNet}}$.  So GRASP can achieve a larger local lipschitz constant around trojan data than badNet.\\

\noindent End of proof.

\subsection{PL Condition and Convergence Rate in High-Dimensional Non-Convex Optimization} \label{sec: PL condition}
Now we consider the target function as high-dimensional non-convex but satisfies the proximal-PL condition~\cite{PL_condition}, which is often considered in the theoretical analysis of neural networks. Formally, the proximal-PL condition is defined below.

\begin{definitionbold} [Proximal-PL condition]
We consider the optimization problem in the form:
\begin{equation}\label{formula:problem_proximal-PL}
    \underset{x \in \mathbb{R}^d}{\operatorname{argmin}} F(x)=f(x)+g(x),
\end{equation}
\noindent where $f$ is a differentiable function with an $L$-Lipschitz continuous gradient and $g$ is a simple but potentially non-smooth convex function \footnote{Typical examples of the simple function $g$ include a scaled $\ell_1$-norm of the parameter vectors (the size of the trigger), $g(x)=\lambda\|x\|_1$, and indicator functions that are zero if $x$ lies in a simple convex set, and are infinity otherwise.}. To analyze the proximal-gradient algorithms (i.e., a more general form of the Projected Gradient Descent (PGD)), a natural %(though not particularly intuitive) 
generalization of the PL inequality is that there exists $\mu>0$ satisfying:

\begin{equation}\label{formula:proximal-PL}
    \frac{1}{2} \mathcal{D}_g(x, L) \geq \mu\left(F(x)-F^*\right)
\end{equation}
where
\begin{equation*}
\scriptstyle
\mathcal{D}_g(x, \alpha) \equiv -2 \alpha \min _y\left[\langle\nabla f(x), y-x\rangle+\frac{\alpha}{2}\|y-x\|^2+g(y)-g(x)\right] .    
\end{equation*}
\end{definitionbold}

%We call this the proximal-PL condition. Below is the 
We now present Theorem \ref{thm:PL_condition} from \cite{PL_condition}, which demonstrates that the proximal-PL condition is sufficient for the proximal-gradient method to achieve a global linear convergence rate.

\begin{theorembold}\label{thm:PL_condition}
Consider the optimization problem in Eq.~\ref{formula:problem_proximal-PL}, where $f$ has an L-Lipschitz continuous gradient (Eq.~\ref{formula:proximal-PL}), $F$ has a non-empty solution set $\mathcal{X}^*, g$ is convex, and $F$ satisfies the proximal-PL inequality. Then the proximal gradient method with a step size of $1 / L$ converges linearly to the optimal value $F^*$:
\begin{equation}\label{eq: PL convergence rate}
    F\left(x_k\right)-F^* \leq\left(1-\frac{\mu}{L}\right)^k\left[F\left(x_0\right)-F^*\right\rceil
\end{equation}

\end{theorembold}
Theorem \ref{thm:PL_condition} indicates a negative relationship between the Lipschitz constant ($L$) of the target function and the convergence rate to the local optimum (trigger), i.e., the difference in target function values between the optimal point and the actual trigger-inserted point. Previous research~\cite{cartis2018worst} has shown that the second-order optimizer has the same lower bound of convergence rate as the first-order optimizer under the PL condition. Many existing studies \cite{yang2020adversarial,terjek2019adversarial,krishnan2020lipschitz} have demonstrated that in neural networks, a lower Lipschitz constant implies higher robustness of the model. Therefore, combining these findings with Theorem~\ref{thm:PL_condition}, we conclude that gradient-based trigger inversion methods perform well on triggers with large effective radius.

% \subsection{Proof of Theorem~\ref{thm:weight analysis}}

\subsection{Impact on $\beta$ with More Complexity Trigger Type}

\begin{figure}[H]
\centering
% \captionsetup{font=small}
\includegraphics[width=0.4\textwidth]{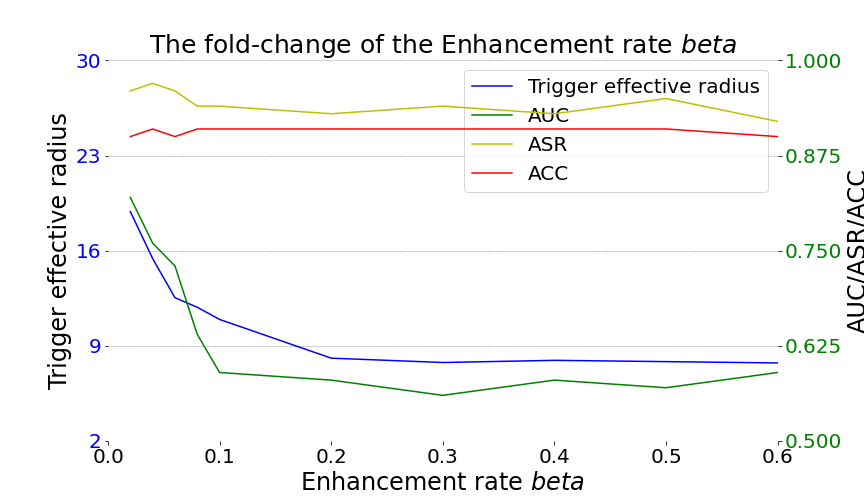} 
\caption{Impact of $\beta$ with the composite~\cite{composite} trigger pattern. }
\label{Fig: beta2} 
\end{figure}

\begin{figure}[H]
\centering
% \captionsetup{font=small}
\includegraphics[width=0.4\textwidth]{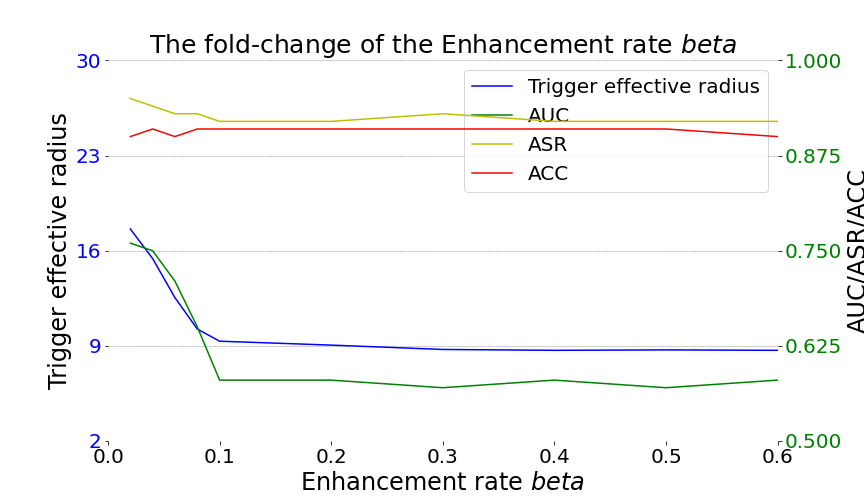} 
\caption{Impact of $\beta$ with the Reflection~\cite{sig} trigger pattern }
\label{Fig: beta3} 
\end{figure}

In this section, we primarily continue the discussion about the choice of $\beta$ from Section~\ref{subsec:Impact of enhancement rate}, and further test two more complex trigger patterns (composite~\cite{composite}, and reflection~\cite{sig}). As depicted in Fig~\ref{Fig: beta2},~\ref{Fig: beta3}, the complexity of the trigger pattern does not significantly impact the choice of $\beta$. A $\beta$ value of 10\% remains suitable even with these complex triggers.

% \subsection{Impact on Learning Optimizer}\label{subsec:optimizer}
% In this section, we evaluate the effectiveness of trigger inversion using three different optimizers: SGD (first-order), Adam~\cite{adam} (first-order with momentum), and AdamHessian~\cite{adahessian} (second-order with momentum). We follow the same evaluation setup as described in Section~\ref{sec:Against Detection} and use NC to detect trojans in models that may have been attacked by BadNet on the CIFAR-10 dataset. For the two optimizers with momentum, we set the momentum equal to $0.9$. 

% \begin{figure}[H]
% \centering
% \captionsetup{font=small}
% \includegraphics[width=0.35\textwidth]{IEEE/optimizer_compare.png}
% \caption{The fold-change of step size when utilizing three different optimizers to run NC for detection. As a second-order optimizer, AdamHessian does not offer a selection for step size, which is why it has the same AUC scores on the graph.}
% \label{Fig: optimizer compare}
% \end{figure}

\subsection{Impact on Different Type of Corruption }\label{subsec:Appendix corruption}

\begin{figure}[H]
    \centering
    \includegraphics[width=0.4\textwidth]{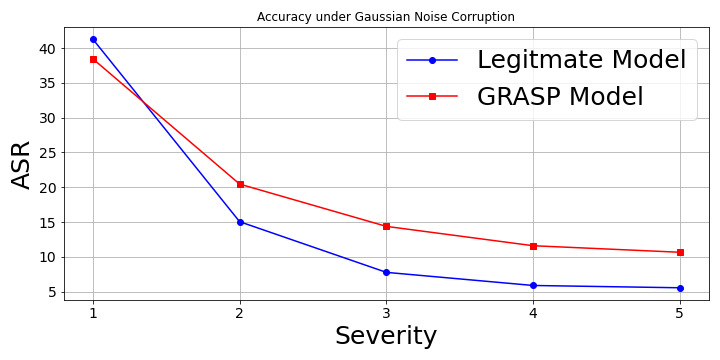}
    \caption{Gaussian Noise}
    \label{fig:gaussian_noise}
\end{figure}

\begin{figure}[H]
    \centering
    \includegraphics[width=0.4\textwidth]{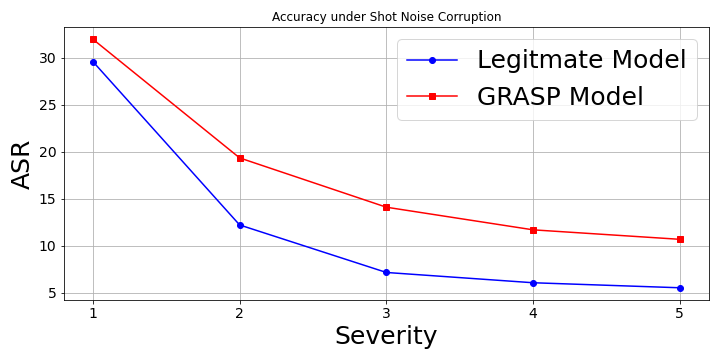}
    \caption{Shot Noise}
    \label{fig:shot_noise}
\end{figure}

\begin{figure}[H]
    \centering
    \includegraphics[width=0.4\textwidth]{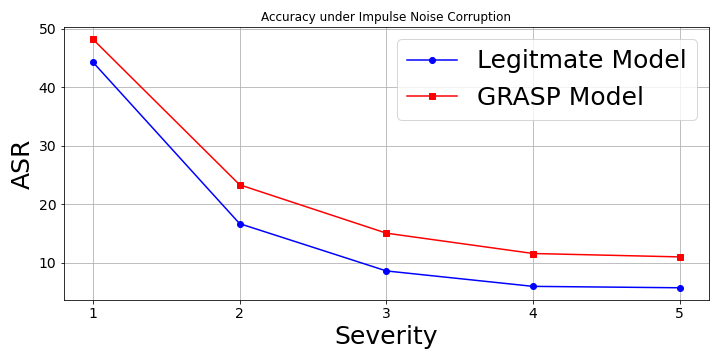}
    \caption{Impulse Noise}
    \label{fig:impulse_noise}
\end{figure}

\begin{figure}[H]
    \centering
    \includegraphics[width=0.4\textwidth]{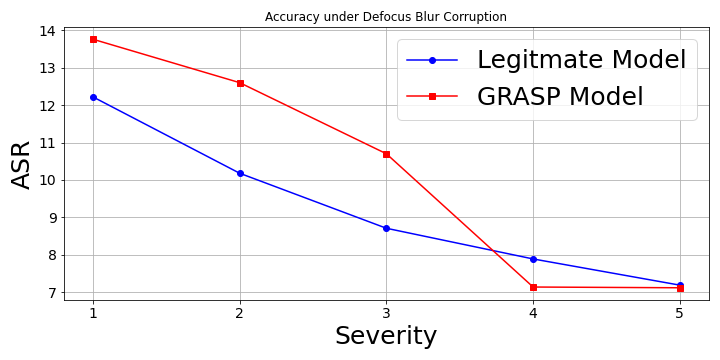}
    \caption{Defocus Blur}
    \label{fig:defocus_blur}
\end{figure}

\begin{figure}[H]
    \centering
    \includegraphics[width=0.4\textwidth]{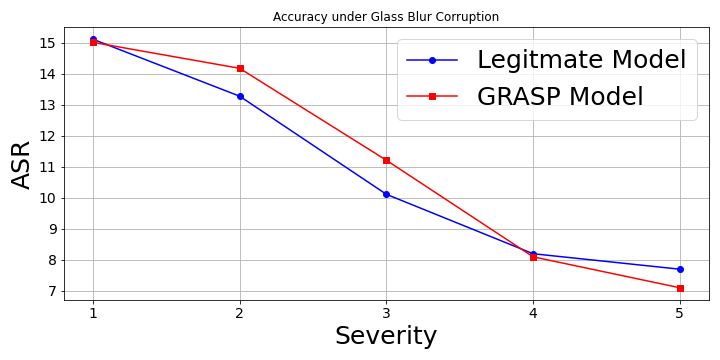}
    \caption{Glass Blur}
    \label{fig:glass_blur}
\end{figure}

\begin{figure}[H]
    \centering
    \includegraphics[width=0.4\textwidth]{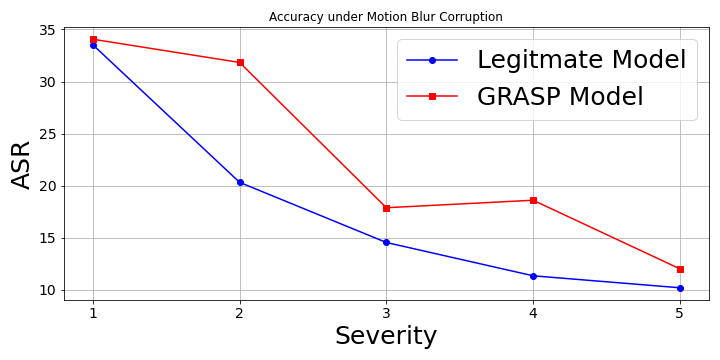}
    \caption{Motion Blur}
    \label{fig:motion_blur}
\end{figure}

\begin{figure}[H]
    \centering
    \includegraphics[width=0.4\textwidth]{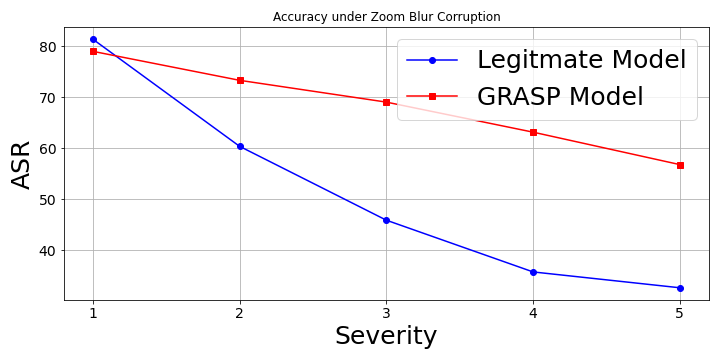}
    \caption{Zoom Blur}
    \label{fig:zoom_blur}
\end{figure}

\begin{figure}[H]
    \centering
    \includegraphics[width=0.4\textwidth]{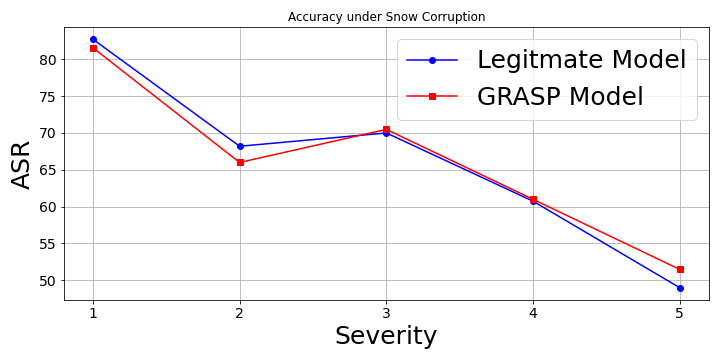}
    \caption{Snow}
    \label{fig:snow}
\end{figure}

\begin{figure}[H]
    \centering
    \includegraphics[width=0.4\textwidth]{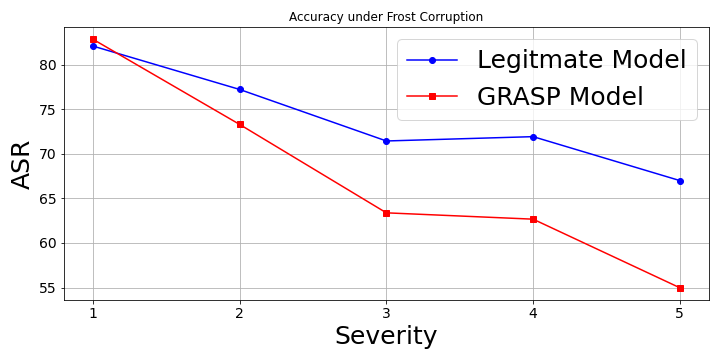}
    \caption{Forst}
    \label{fig:frost}
\end{figure}

\begin{figure}[H]
    \centering
    \includegraphics[width=0.4\textwidth]{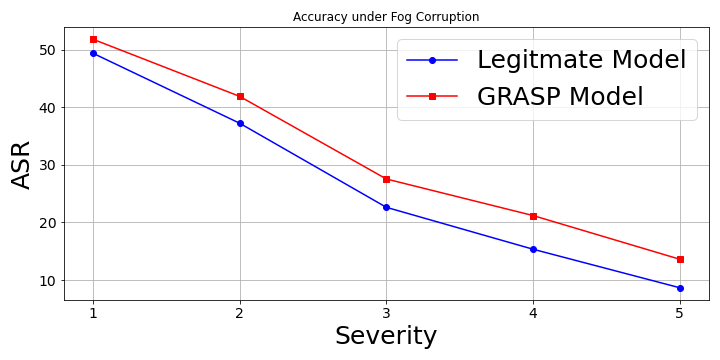}
    \caption{Fog}
    \label{fig:fog}
\end{figure}

\begin{figure}[H]
    \centering
    \includegraphics[width=0.4\textwidth]{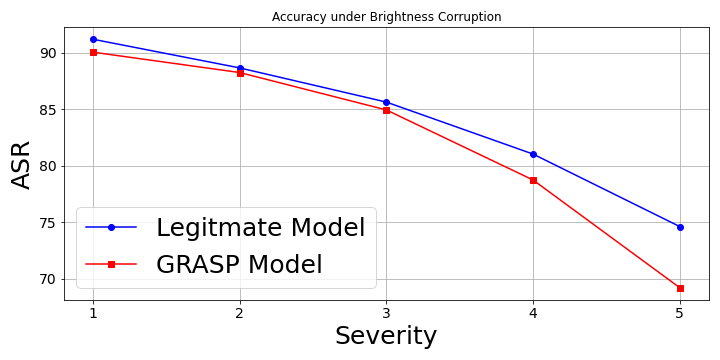}
    \caption{Brightness}
    \label{fig:brightness}
\end{figure}

\begin{figure}[H]
    \centering
    \includegraphics[width=0.4\textwidth]{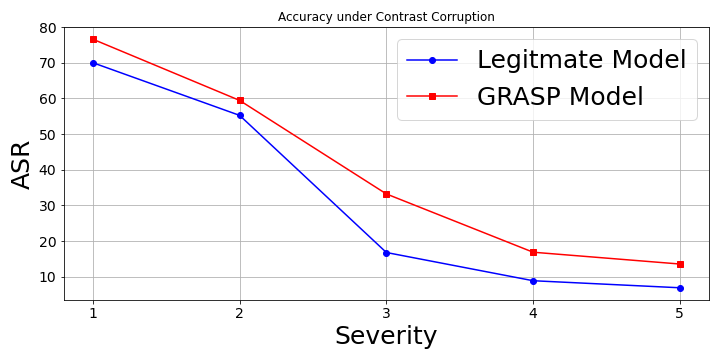}
    \caption{Contrast}
    \label{fig:contrast}
\end{figure}

\begin{figure}[H]
    \centering
    \includegraphics[width=0.4\textwidth]{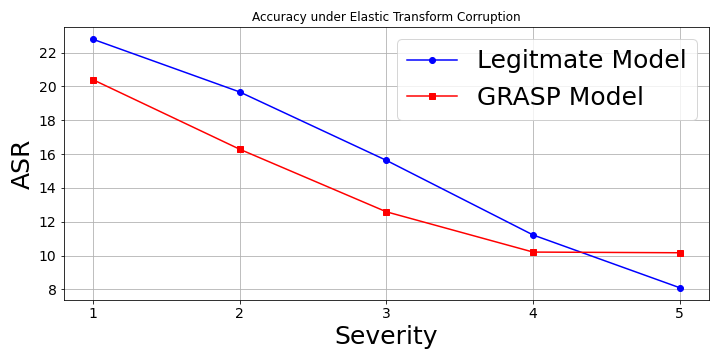}
    \caption{Elastic Transform}
    \label{fig:elastic_transform}
\end{figure}

\begin{figure}[H]
    \centering
    \includegraphics[width=0.4\textwidth]{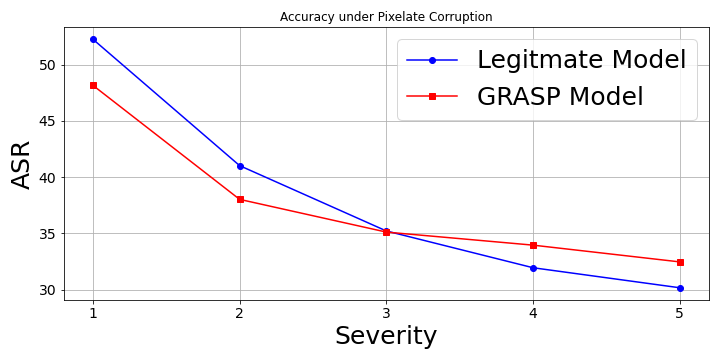}
    \caption{Pixelate}
    \label{fig:pixelate}
\end{figure}

\begin{figure}[H]
    \centering
    \includegraphics[width=0.4\textwidth]{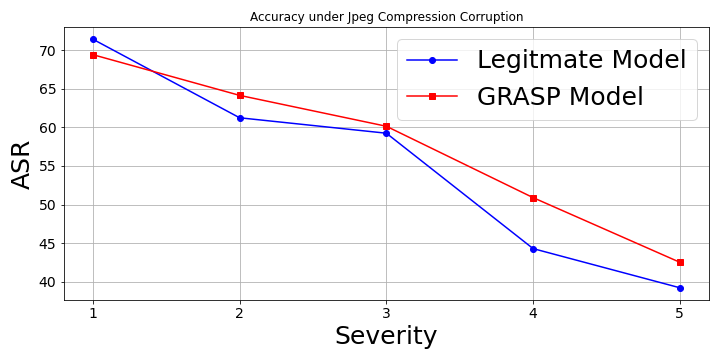}
    \caption{jpeg Compression}
    \label{fig:jpeg_compression}
\end{figure}

Figures~\ref{fig:gaussian_noise} to ~\ref{fig:jpeg_compression} document the comparative impact of 15 different types of corruptions from the imagecorruptions benchmark on a GRASP-enhanced BadNet versus BadNet. More specifically, we employed the ResNet-101 neural network architecture to train two distinct models on the CIFAR-10 dataset: one contaminated with a backdoor using the BadNet methodology, and another utilizing a backdoor implemented via the GRASP-enhanced BadNet approach.  we present the ASR for these corrupted testing data, subjected to 15 corruptions in five varying degrees of severity on trigger-inserted testing data.

\newpage

% \title{Online document of \textit{Gradient Shaping: Enhancing Backdoor Attack Against Reverse Engineering}}
\twocolumn[
\begin{@twocolumnfalse}  % 开始一个伪单栏环境
\begin{center}
\vspace*{\stretch{1}}  % 在标题前添加垂直空间
\Huge Online document of \textit{Gradient Shaping: Enhancing Backdoor Attack Against Reverse Engineering}
\vspace*{\stretch{2}}  % 在标题后添加垂直空间
\end{center}
\end{@twocolumnfalse}
]

\subsection{Additional Comparison Results }
\label{subsec:additional eps4}

\para{Additional network architecture}.
Due to space constraints in the main text, we were unable to include all experimental results. In this subsection, we provide a detailed record of the experimental outcomes for ResNet-101, ShuffleNet, and ResNet18, maintaining the same experimental setup as in Fig~\ref{tab:trigger_AUC} from the main text, albeit with different network architectures. Specifically, Fig~\ref{tab:trigger_AUC_Res101}, Fig~\ref{tab:trigger_AUC_ShuffleNet}, and Fig~\ref{tab:trigger_AUC_ResNet18} respectively present the results for ResNet-101, ShuffleNet, and ResNet18 under the $\epsilon_4$ scenario. 

\para{Effectiveness on unlearning}.
Furthermore, we have incorporated an additional set of comparative results with the blind backdoor attack~\cite{bagdasaryan2021blind} in Table~\ref{tab:auc_blind_backdoor}, assessing the performance differences between ~\ourmethod{} and the blind backdoor when evading more advanced reverse engineering techniques like Pixel~\cite{pixelbackdoor}. This comparison maintaining the experimental setup as outlined in Table~\ref{tab:trigger_AUC}.  Observations from the results across three distinct datasets reveal that the GRASP-enhanced BadNet consistently achieves lower AUC scores, signifying a more effective evasion of the Pixel technique than blind backdoor.

It is important to highlight that in this comparison, we focus on the ~\ourmethod{}-enhanced BadNet, a black-box backdoor attack, in contrast to the blind backdoor, which necessitates control over the training process. This distinction underscores the broader applicability and generalizability of our method in real-world scenarios, as it does not require the stringent conditions that the Blind Backdoor depends on for successful implementation.

\begin{table}[h]
    \centering
    \caption{ASR Comparison after NC Unlearning on Different Datasets.}
    \label{tab:unlearning_asr_comparison}
    \begin{tabular}{@{}cccc@{}}
        \toprule
        Attack/Database & CIFAR-10 & MNIST & Tiny ImageNet \\ 
        \midrule
        BadNet  & 11.1\% & 8.9\%  & 15.2\% \\ 
        ~BadNet* & 87.2\% & 90.1\% & 77.4\% \\ 
        \bottomrule
    \end{tabular}
\end{table}

\begin{table}[h]
    \centering
    \small
    \caption{AUC scores of blind backdoor and GRASP-enhanced BadNet under Pixel~\cite{pixelbackdoor} detect on various datasets.}
    \label{tab:auc_blind_backdoor}
    \begin{tabular}{ccc}
        \toprule
        Dataset       & Blind Backdoor & GRASP-enhanced BadNet \\
        \midrule
        CIFAR-10      & 0.89           & 0.80 \\
        MNIST         & 0.93           & 0.83 \\
        Tiny-ImageNet & 0.87           & 0.79 \\
        \bottomrule
    \end{tabular}
\end{table}

\begin{table*}[ht]
\centering
%\footnotesize
\small
\renewcommand{\arraystretch}{1.2}
\begin{tabular}{c|ccccc|cccc|cccc}
\toprule
\multicolumn{2}{c|}{}       & \multicolumn{4}{c|}{CIFAR-10} & \multicolumn{4}{c|}{MNIST} & \multicolumn{4}{c}{Tiny ImageNet}\\
\midrule
\multicolumn{2}{c|}{}       & NC  & Tabor  & K-arm  & Pixel & NC & Tabor & K-arm & Pixel & NC & Tabor & K-arm & Pixel \\ 

\midrule
\multicolumn{13}{c}{$\epsilon_4$: AUC }\\
\midrule
%                                   CIFAR-10                                        MNIST                                          Tiny ImageNet      
%                                   & NC        & Tabor     & ABS       & Pixel     & NC       & Tabor     & ABS       & Pixel     & NC        & Tabor     & ABS       & Pixel \\ 
%Patch
\multicolumn{2}{c|}{BadNet}         & 79.8\% & 83.2\% & 84.7\% & 92.1\% & 77.1\% & 81.3\% & 83.9\% & 89.6\% & 74.3\% & 78.7\% & 80.4\% & 85.2\% \\
\multicolumn{2}{c|}{~BadNet*}        & 54.0\% & 55.9\% & 58.5\% & 79.9\% & 53.7\% & 54.6\% & 60.1\% & 82.9\% & 56.3\% & 57.9\% & 57.1\% & 78.9\% \\
\multicolumn{2}{c|}{LSBA}           & 67.9\% & 67.6\% & 71.9\% & 81.2\% & 68.0\% & 69.2\% & 70.5\% & 78.5\% & 63.5\% & 69.1\% & 70.9\% & 85.9\% \\
\multicolumn{2}{c|}{~LSBA*}          & 54.9\% & 56.5\% & 59.7\% & 63.0\% & 52.2\% & 57.5\% & 55.8\% & 63.2\% & 55.7\% & 51.9\% & 57.0\% & 63.6\% \\
\multicolumn{2}{c|}{Composite}      & 67.4\% & 65.9\% & 69.5\% & 84.2\% & 65.2\% & 64.5\% & 68.5\% & 83.7\% & 64.8\% & 64.1\% & 66.1\% & 81.7\% \\
\multicolumn{2}{c|}{~Composite*}     & 53.7\% & 58.9\% & 61.1\% & 72.6\% & 53.0\% & 52.3\% & 59.9\% & 70.8\% & 54.5\% & 53.3\% & 58.3\% & 71.4\% \\

%Clean label
\multicolumn{2}{c|}{Latent}         & 79.0\% & 76.7\% & 79.2\% & 87.1\% & 79.8\% & 79.9\% & 81.4\% & 88.9\% & 72.9\% & 79.3\% & 75.4\% & 83.9\% \\
\multicolumn{2}{c|}{~Latent*}        & 53.5\% & 55.8\% & 58.7\% & 75.6\% & 53.3\% & 55.0\% & 59.2\% & 74.4\% & 54.0\% & 55.5\% & 56.7\% & 71.9\% \\
%Imperceptible
\multicolumn{2}{c|}{DEFEAT}         & 64.7\% & 64.2\% & 77.7\% & 69.7\% & 66.9\% & 68.5\% & 80.4\% & 71.8\% & 62.9\% & 66.8\% & 76.1\% & 67.5\% \\
\multicolumn{2}{c|}{~DEFEAT*}        & 59.2\% & 59.9\% & 72.0\% & 61.9\% & 59.1\% & 58.0\% & 70.9\% & 59.8\% & 58.3\% & 59.0\% & 72.1\% & 63.2\% \\
\multicolumn{2}{c|}{IMC}            & 68.1\% & 65.3\% & 76.0\% & 79.2\% & 67.9\% & 68.7\% & 77.9\% & 80.0\% & 68.6\% & 72.9\% & 77.5\% & 78.0\% \\
\multicolumn{2}{c|}{~IMC*}           & 55.2\% & 54.5\% & 72.8\% & 71.9\% & 54.0\% & 53.2\% & 74.1\% & 73.9\% & 63.9\% & 65.3\% & 71.0\% & 74.8\% \\
%Latent space similarity
\multicolumn{2}{c|}{Adaptive-Blend} & 66.1\% & 67.5\% & 67.9\% & 77.0\% & 59.1\% & 62.9\% & 65.3\% & 81.0\% & 61.9\% & 64.9\% & 65.8\% & 77.5\% \\
\multicolumn{2}{c|}{~Adaptive-Blend*}& 54.9\% & 53.8\% & 54.2\% & 68.5\% & 52.5\% & 58.5\% & 61.2\% & 75.6\% & 55.7\% & 53.3\% & 60.2\% & 68.9\% \\
\bottomrule

\end{tabular}
\caption{For ResNet-101 Architecture, the AUCs of backdoor detection by trigger inversion methods on the backdoored models poisoned by different backdoor attacks with and without the enhancement of \ourmethod{}. }
\label{tab:trigger_AUC_Res101}
\vspace{2mm}
\end{table*}

\begin{table*}[ht]
\centering
%\footnotesize
\small
\renewcommand{\arraystretch}{1.2}
\begin{tabular}{c|ccccc|cccc|cccc}
\toprule
\multicolumn{2}{c|}{}       & \multicolumn{4}{c|}{CIFAR-10} & \multicolumn{4}{c|}{MNIST} & \multicolumn{4}{c}{Tiny ImageNet}\\
\midrule
\multicolumn{2}{c|}{}       & NC  & Tabor  & K-arm  & Pixel & NC & Tabor & K-arm & Pixel & NC & Tabor & K-arm & Pixel \\ 

\midrule
\multicolumn{13}{c}{$\epsilon_4$: AUC }\\
\midrule
%                                   CIFAR-10                                        MNIST                                          Tiny ImageNet      
%                                   & NC        & Tabor     & ABS       & Pixel     & NC       & Tabor     & ABS       & Pixel     & NC        & Tabor     & ABS       & Pixel \\ 
%Patch
\multicolumn{2}{c|}{BadNet}         & 79.8\%    & 83.0\%    & 85.4\%    & 92.1\%    & 77.2\%   & 80.9\%    & 83.9\%    & 89.9\%    & 74.1\%    & 77.8\%    & 80.3\%    & 85.9\%   \\
\multicolumn{2}{c|}{~BadNet*}        & 53.5\%    & 56.0\%    & 58.5\%    & 79.8\%    & 54.1\%   & 54.7\%    & 59.2\%    & 83.1\%    & 55.9\%    & 57.8\%    & 58.0\%    & 78.5\%   \\
\multicolumn{2}{c|}{LSBA}           & 66.5\%    & 68.0\%    & 71.8\%    & 81.9\%    & 68.0\%   & 69.2\%    & 70.5\%    & 78.3\%    & 62.9\%    & 70.1\%    & 71.2\%    & 86.3\%   \\
\multicolumn{2}{c|}{~LSBA*}          & 54.9\%    & 55.5\%    & 59.9\%    & 63.1\%    & 53.0\%   & 57.1\%    & 56.8\%    & 64.2\%    & 55.1\%    & 52.2\%    & 57.5\%    & 64.1\%   \\
\multicolumn{2}{c|}{Composite}      & 67.9\%    & 65.8\%    & 68.5\%    & 84.7\%    & 66.2\%   & 63.5\%    & 69.5\%    & 83.6\%    & 64.5\%    & 65.2\%    & 66.6\%    & 81.5\%   \\
\multicolumn{2}{c|}{~Composite*}     & 53.9\%    & 59.6\%    & 62.1\%    & 72.1\%    & 52.5\%   & 52.9\%    & 60.1\%    & 71.3\%    & 53.5\%    & 54.0\%    & 57.2\%    & 71.8\%   \\
\multicolumn{2}{c|}{Latent}         & 78.2\%    & 76.9\%    & 80.2\%    & 87.2\%    & 80.5\%   & 79.1\%    & 81.0\%    & 88.5\%    & 72.8\%    & 80.2\%    & 74.8\%    & 84.4\%   \\
\multicolumn{2}{c|}{~Latent*}        & 52.5\%    & 55.6\%    & 59.5\%    & 76.0\%    & 53.0\%   & 55.8\%    & 58.2\%    & 73.9\%    & 53.9\%    & 54.6\%    & 56.9\%    & 71.9\%   \\
\multicolumn{2}{c|}{DEFEAT}         & 65.2\%    & 63.2\%    & 77.8\%    & 69.8\%    & 67.9\%   & 68.7\%    & 79.5\%    & 71.2\%    & 62.4\%    & 67.1\%    & 76.1\%    & 67.9\%   \\
\multicolumn{2}{c|}{~DEFEAT*}        & 60.2\%    & 59.9\%    & 71.6\%    & 61.5\%    & 58.2\%   & 57.6\%    & 71.9\%    & 59.9\%    & 57.5\%    & 59.7\%    & 72.8\%    & 63.0\%   \\
\multicolumn{2}{c|}{IMC}            & 67.5\%    & 65.6\%    & 75.8\%    & 79.2\%    & 68.0\%   & 69.3\%    & 76.7\%    & 80.1\%    & 67.9\%    & 73.5\%    & 77.3\%    & 76.8\%   \\
\multicolumn{2}{c|}{~IMC*}           & 56.1\%    & 54.2\%    & 72.7\%    & 71.8\%    & 54.7\%   & 53.1\%    & 74.0\%    & 74.6\%    & 63.6\%    & 64.8\%    & 71.0\%    & 75.2\%   \\
\multicolumn{2}{c|}{Adaptive-Blend} & 67.1\%    & 67.0\%    & 68.0\%    & 77.9\%    & 59.8\%   & 63.6\%    & 65.6\%    & 81.5\%    & 62.9\%    & 64.1\%    & 66.5\%    & 77.5\%   \\
\multicolumn{2}{c|}{~Adaptive-Blend*}& 55.9\%    & 53.5\%    & 54.9\%    & 69.4\%    & 51.8\%   & 58.5\%    & 61.2\%    & 75.6\%    & 55.7\%    & 53.3\%    & 60.2\%    & 68.9\% \\
\bottomrule

\end{tabular}
\caption{For ShuffleNet Architecture, the AUCs of backdoor detection by trigger inversion methods on the backdoored models poisoned by different backdoor attacks with and without the enhancement of \ourmethod{}. }
\label{tab:trigger_AUC_ShuffleNet}
\vspace{2mm}
\end{table*}

\begin{table*}[ht]
\centering
%\footnotesize
\small
\renewcommand{\arraystretch}{1.2}
\begin{tabular}{c|ccccc|cccc|cccc}
\toprule
\multicolumn{2}{c|}{}       & \multicolumn{4}{c|}{CIFAR-10} & \multicolumn{4}{c|}{MNIST} & \multicolumn{4}{c}{Tiny ImageNet}\\
\midrule
\multicolumn{2}{c|}{}       & NC  & Tabor  & K-arm  & Pixel & NC & Tabor & K-arm & Pixel & NC & Tabor & K-arm & Pixel \\ 

\midrule
\multicolumn{13}{c}{$\epsilon_4$: AUC }\\
\midrule
%                                   CIFAR-10                                        MNIST                                          Tiny ImageNet      
%                                   & NC        & Tabor     & ABS       & Pixel     & NC       & Tabor     & ABS       & Pixel     & NC        & Tabor     & ABS       & Pixel \\ 
%Patch
\multicolumn{2}{c|}{BadNet}         & 79.8\%    & 83.0\%    & 84.8\%    & 92.1\%    & 77.7\%   & 81.7\%    & 83.0\%    & 89.5\%    & 75.0\%    & 78.6\%    & 80.9\%    & 85.5\%   \\
\multicolumn{2}{c|}{~BadNet*}        & 54.1\%    & 55.1\%    & 59.0\%    & 79.5\%    & 53.5\%   & 54.4\%    & 59.7\%    & 83.5\%    & 56.2\%    & 57.6\%    & 57.5\%    & 78.1\%   \\
\multicolumn{2}{c|}{LSBA}           & 67.1\%    & 67.5\%    & 71.3\%    & 81.7\%    & 68.1\%   & 68.8\%    & 71.1\%    & 78.9\%    & 63.3\%    & 69.4\%    & 70.9\%    & 86.0\%   \\
\multicolumn{2}{c|}{~LSBA*}          & 54.5\%    & 56.1\%    & 59.0\%    & 63.4\%    & 52.7\%   & 57.6\%    & 56.0\%    & 63.6\%    & 55.2\%    & 51.6\%    & 57.1\%    & 64.0\%   \\
\multicolumn{2}{c|}{Composite}      & 67.0\%    & 65.4\%    & 69.1\%    & 84.6\%    & 65.6\%   & 64.1\%    & 69.1\%    & 83.2\%    & 64.1\%    & 64.4\%    & 66.2\%    & 82.0\%   \\
\multicolumn{2}{c|}{~Composite*}     & 53.3\%    & 59.2\%    & 61.3\%    & 72.5\%    & 53.0\%   & 52.1\%    & 59.4\%    & 71.0\%    & 54.1\%    & 53.2\%    & 57.8\%    & 71.2\%   \\
%Clean label
\multicolumn{2}{c|}{Latent}         & 78.6\%    & 76.2\%    & 79.6\%    & 87.3\%    & 80.0\%   & 79.4\%    & 81.5\%    & 89.2\%    & 73.0\%    & 79.7\%    & 75.1\%    & 84.0\%   \\
\multicolumn{2}{c|}{~Latent*}        & 53.2\%    & 55.3\%    & 59.1\%    & 75.4\%    & 53.8\%   & 55.0\%    & 58.9\%    & 74.0\%    & 54.3\%    & 55.0\%    & 56.3\%    & 71.5\%   \\
%Imperceptible
\multicolumn{2}{c|}{DEFEAT}         & 64.6\%    & 63.8\%    & 77.6\%    & 69.2\%    & 67.2\%   & 69.1\%    & 80.1\%    & 71.6\%    & 63.0\%    & 66.6\%    & 76.5\%    & 67.2\%   \\
\multicolumn{2}{c|}{~DEFEAT*}        & 59.4\%    & 59.5\%    & 72.5\%    & 62.0\%    & 58.8\%   & 58.3\%    & 71.0\%    & 59.5\%    & 58.0\%    & 59.1\%    & 72.2\%    & 63.5\%   \\
\multicolumn{2}{c|}{IMC}            & 68.2\%    & 65.1\%    & 76.0\%    & 79.4\%    & 67.4\%   & 69.1\%    & 77.3\%    & 79.6\%    & 68.1\%    & 73.1\%    & 77.2\%    & 77.4\%   \\
\multicolumn{2}{c|}{~IMC*}           & 55.4\%    & 54.8\%    & 72.3\%    & 71.6\%    & 54.5\%   & 53.7\%    & 74.2\%    & 74.0\%    & 64.1\%    & 65.1\%    & 71.4\%    & 75.2\%   \\
%Latent space similarity
\multicolumn{2}{c|}{Adaptive-Blend} & 66.4\%    & 67.2\%    & 67.5\%    & 77.3\%    & 59.4\%   & 63.0\%    & 65.2\%    & 81.1\%    & 62.3\%    & 64.4\%    & 66.1\%    & 77.0\%   \\
\multicolumn{2}{c|}{~Adaptive-Blend*}& 55.4\%    & 53.0\%    & 54.1\%    & 68.1\%    & 52.9\%   & 62.3\%    & 64.3\%    & 72.2\%    & 52.1\%    & 51.1\%    & 58.8\%    & 69.7\%   \\
\bottomrule

\end{tabular}
\caption{For ResNet18 Architecture, the AUCs of backdoor detection by trigger inversion methods on the backdoored models poisoned by different backdoor attacks with and without the enhancement of \ourmethod{}. }
\label{tab:trigger_AUC_ResNet18}
\vspace{2mm}
\end{table*}

\subsection{Feasibility of Black-box Attacks and Feature-space Attacks}

In this section, we discuss the feasibility and potential mitigations when ~\ourmethod{} is employed as an enhancement to Black-box Attacks, such as GRASP-enhanced BadNet. Additionally, we discuss the feasibility associated with using ~\ourmethod{} as an enhancement for sample-specific backdoor attacks, as illustrated by GRASP-enhanced IMC.

\para{Feasibility of Black-box Attacks}.
Firstly, when considering ~\ourmethod{} as an enhancement to Black-box Attacks, an intuitive idea might be to screen the training data prior to model training, anticipating potential anomalies. Specifically, while the trigger-inserted data and its corresponding augmented data may only differ in the trigger region, maintaining high similarity, their labels are distinct. Such a trivial screening seems capable of filtering out GRASP-related poisoning data. However, this defense strategy incurs a computational cost of \(O(m^2n^2)\), where \(n\) is the size of the training dataset, $m$ is the input dimension, substantially elevating the defense expenses against GRASP. Additionally, GRASP might introduce subtle noise not only in the trigger areas but also in non-trigger regions, rendering this simplistic pre-processing approach less effective. Furthermore, in Section 6, we comprehensively assessed GRASP's robustness against this type of pre-processing defense, affirming its efficacy.

\para{Feasibility of Feature-space Backdoor}.
As we mentioned in the Section 4, the enhancement of GRASP can be applied to any backdoor where a trigger amending function($A$) can be defined, including for feature-space triggers like sample-specific triggers. For these triggers, Algorithm 1 can still be used for enhancement. 

Typically, feature-space backdoor attacks such as those discussed in ~\cite{adaptive, adaptive-blend, latent, defeat}, often operate under a threat model that requires the attacker to have white-box access to the victim model or the ability to control the entire training dataset. In the case of GRASP, a technique for backdoor enhancement, while it can further obscure the backdoor attack, making it more stealthy as elaborated in Section~\ref{sec:Against Detection}, the enhanced attack inherits a similarly specific and less general threat model akin to those of feature-space backdoors. This parallels the compromise in generality for the sake of enhanced stealthiness and attack efficacy.

\subsection{Additional results from Section~\ref{sec:GRASP}}\label{subsec:additional result of sec3}

In this section, we aim to supplement the illustrative experiments presented in Section~\ref{sec:GRASP}. Specifically, while all the experiments in Section~\ref{sec:GRASP} were conducted using Neural Cleanse (NC)~\cite{NC} as the reverse engineering tool to demonstrate and validate the impact of various factors, in this section, we replicate these experiments using Tabor as outlined in Table~\cite{tabor}. Our findings indicate that the observed trends and results are largely consistent, exhibiting a remarkable similarity between utilizing Tabor and NC as the reverse engineering tools.

\begin{figure}[H]
    \centering
    \includegraphics[width=0.3\textwidth]{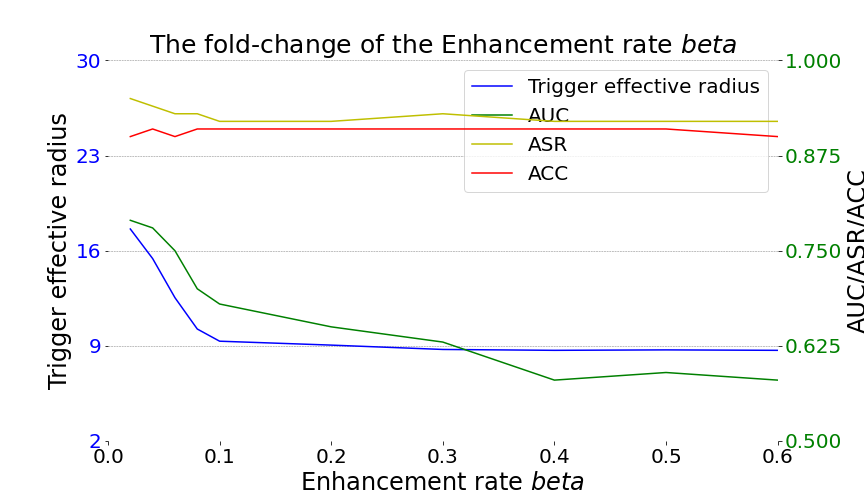}
    \caption{Impact on $\beta$ while using Tabor as detector}
    \label{fig:Enhancement_Level_tabor}
\end{figure}

\para{Impact of Enhancement Rate in GRASP}.
Figure~\ref{fig:Enhancement_Level_tabor} serves a similar purpose to Figure~\ref{Fig: impact of Enhancement rate} in the Section~\ref{sec:GRASP}, with the key distinction being that Tabor is employed here as the reverse engineering detector. This enables us to provide a more comprehensive validation and to ensure the robustness of our findings across different reverse engineering tools.

\begin{figure*}
  \centering
  \begin{subfigure}[b]{0.30\textwidth}
    \centering
    \includegraphics[width=\linewidth]{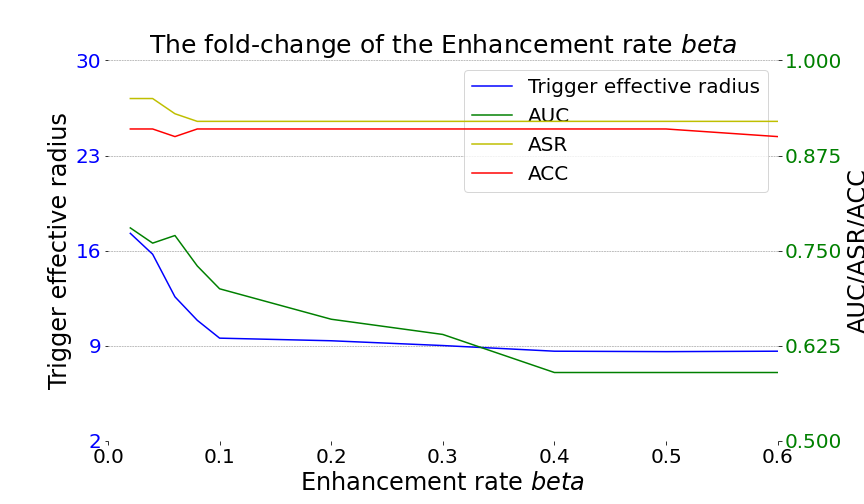}
    \caption{Tabor}
  \end{subfigure}
  \hfill
  \begin{subfigure}[b]{0.30\textwidth}
    \centering
    \includegraphics[width=\linewidth]{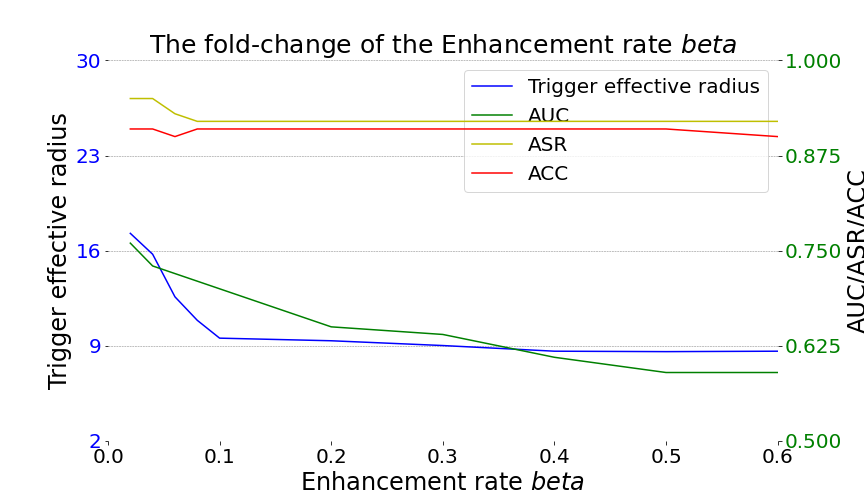}
    \caption{NC}
  \end{subfigure}
  \hfill
  \begin{subfigure}[b]{0.30\textwidth}
    \centering
    \includegraphics[width=\linewidth]{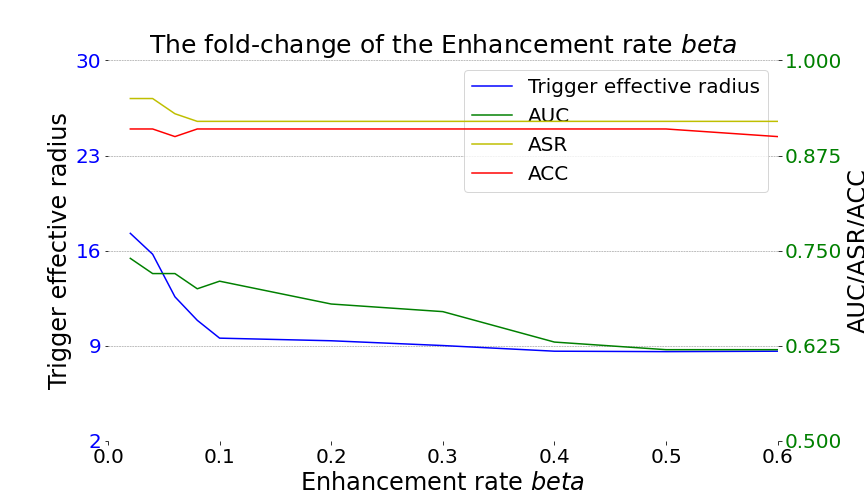}
    \caption{ABS}
  \end{subfigure}
  \caption{Impact on enhancement rate while using NC, Tabor, and ABS as detector}
  \label{fig:additional_Enhancement_Level_tabor}
\end{figure*}

In addition, we extend our evaluation to the ResNet-16 network architecture, incorporating NC, Tabor, and ABS as detectors to further validate the effectiveness of our approach. The results of this comprehensive analysis are presented in Fig~\ref{fig:additional_Enhancement_Level_tabor}, providing insights into the performance and robustness of our method across different settings.

\begin{figure}[H]
    \centering
    \includegraphics[width=0.3\textwidth]{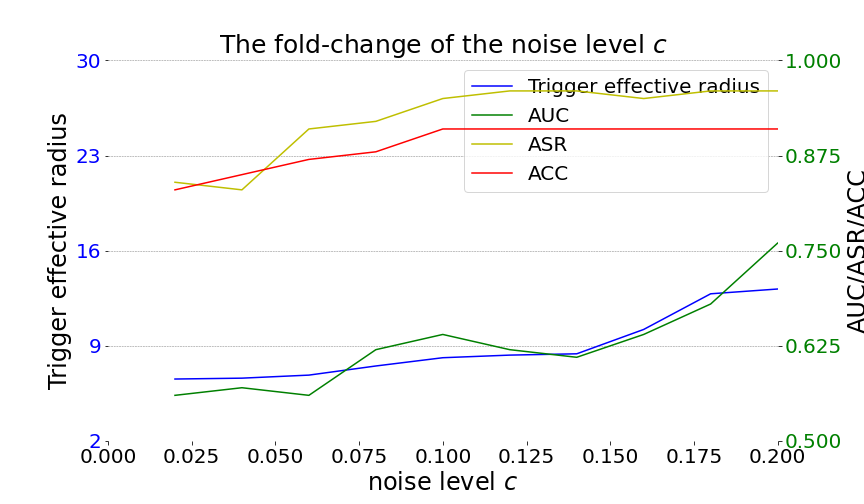}
    \caption{Impact on $c$ while using Tabor as detector}
    \label{fig:Noise_Level_tabor}
\end{figure}

\para{Impact of Noise Level in GRASP}.
Figure~\ref{fig:Noise_Level_tabor} serves a similar purpose to Figure~\ref{Fig: c local lip} in the Section~\ref{sec:GRASP}, with the key distinction being that Tabor is employed here as the reverse engineering detector. This enables us to provide a more comprehensive validation and to ensure the robustness of our findings across different reverse engineering tools.

\begin{figure*}
  \centering
  \begin{subfigure}[b]{0.30\textwidth}
    \centering
    \includegraphics[width=\linewidth]{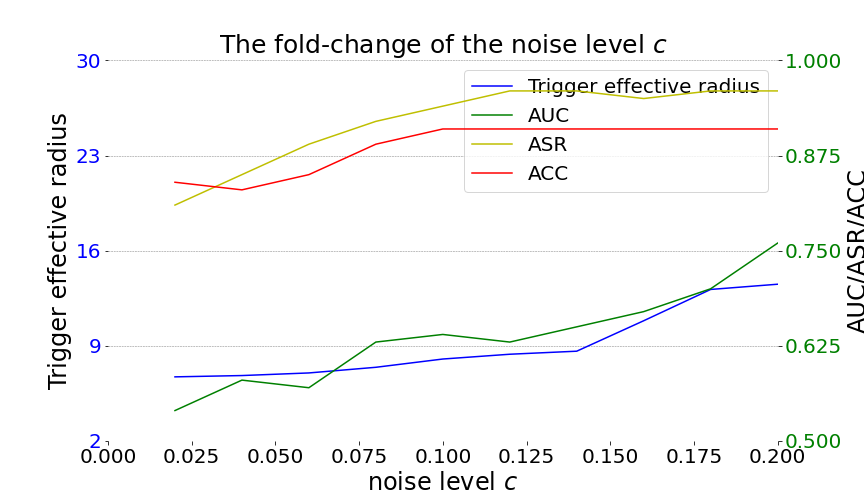}
    \caption{Tabor}
  \end{subfigure}
  \hfill
  \begin{subfigure}[b]{0.30\textwidth}
    \centering
    \includegraphics[width=\linewidth]{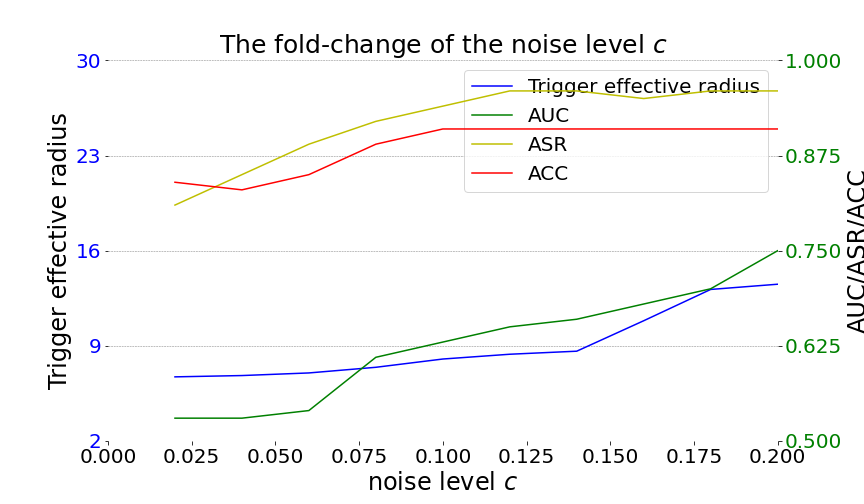}
    \caption{NC}
  \end{subfigure}
  \hfill
  \begin{subfigure}[b]{0.30\textwidth}
    \centering
    \includegraphics[width=\linewidth]{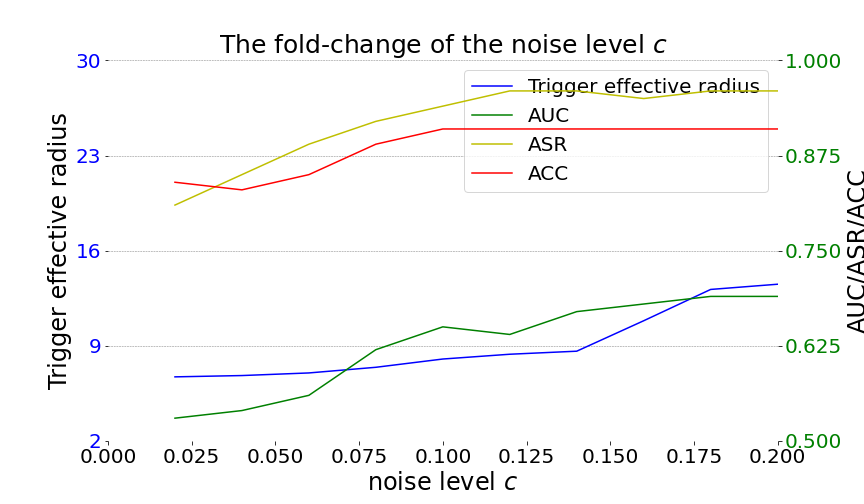}
    \caption{ABS}
  \end{subfigure}
  \caption{Impact on noise level while using NC, Tabor, and ABS as detector}
  \label{fig:additional_Noise_Level_tabor}
\end{figure*}

In addition, we extend our evaluation to the ResNet-16 network architecture, incorporating NC, Tabor, and ABS as detectors to further validate the effectiveness of our approach. The results of this comprehensive analysis are presented in Fig~\ref{fig:additional_Noise_Level_tabor}, providing insights into the performance and robustness of our method across different settings.

\begin{figure}[H]
    \centering
    \includegraphics[width=0.3\textwidth]{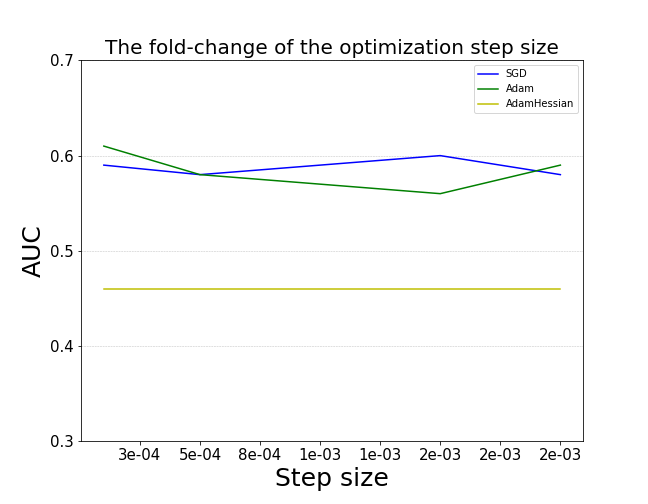}
    \caption{Impact on optimizer while using Tabor as detector}
    \label{fig:optimizer_tabor}
\end{figure}

\para{Impact of Optimizer}.
Figure~\ref{fig:optimizer_tabor} serves a similar purpose to Figure~\ref{Fig: optimizer compare} in the Section~\ref{sec:GRASP}, with the key distinction being that Tabor is employed here as the reverse engineering detector. This enables us to provide a more comprehensive validation and to ensure the robustness of our findings across different reverse engineering tools.

\begin{figure*}
  \centering
  \begin{subfigure}[b]{0.30\textwidth}
    \centering
    \includegraphics[width=\linewidth]{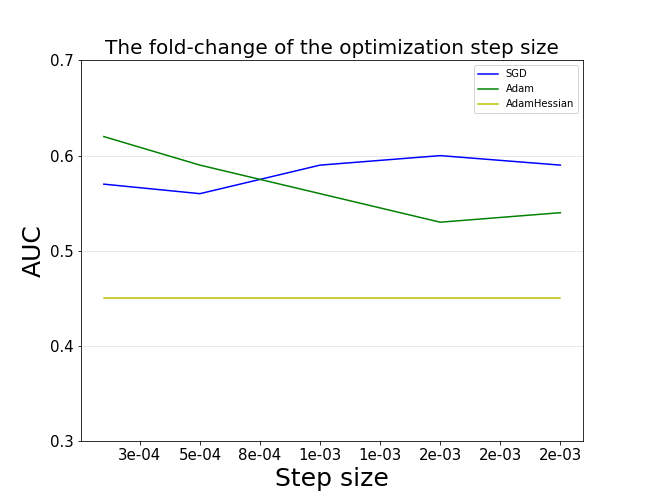}
    \caption{Tabor}
  \end{subfigure}
  \hfill
  \begin{subfigure}[b]{0.30\textwidth}
    \centering
    \includegraphics[width=\linewidth]{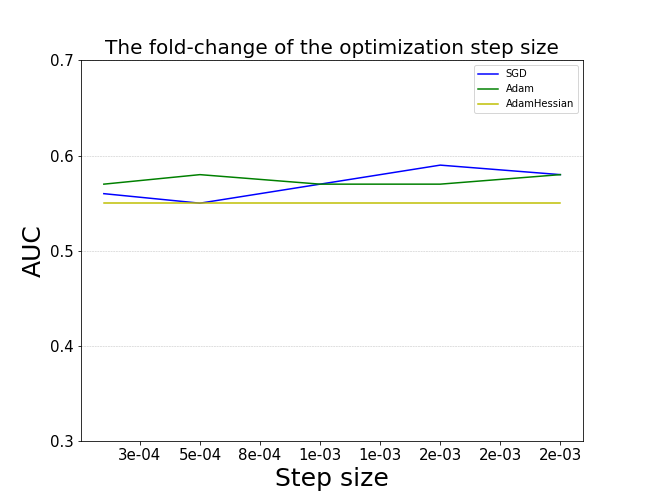}
    \caption{NC}
  \end{subfigure}
  \hfill
  \begin{subfigure}[b]{0.30\textwidth}
    \centering
    \includegraphics[width=\linewidth]{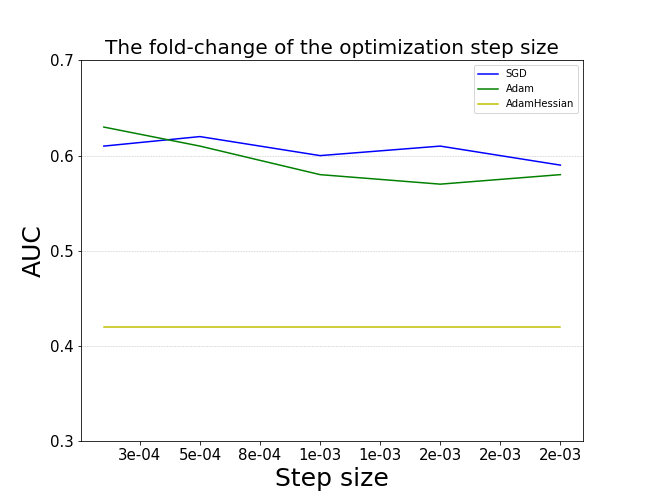}
    \caption{ABS}
  \end{subfigure}
  \caption{Impact on different optimizers and learning rate while using NC, Tabor, and ABS as detector}
  \label{fig:additional_optimizer_tabor}
\end{figure*}

In addition, we extend our evaluation to the ResNet-16 network architecture, incorporating NC, Tabor, and ABS as detectors to further validate the effectiveness of our approach. The results of this comprehensive analysis are presented in Fig~\ref{fig:additional_optimizer_tabor}, providing insights into the performance and robustness of our method across different settings.



\subsection{Additional Figures}\label{subsec:trigger pattern}

\begin{figure}[ht]
\centering
\captionsetup{font=tiny}
\subfloat[Square]{\includegraphics[height=0.4in]{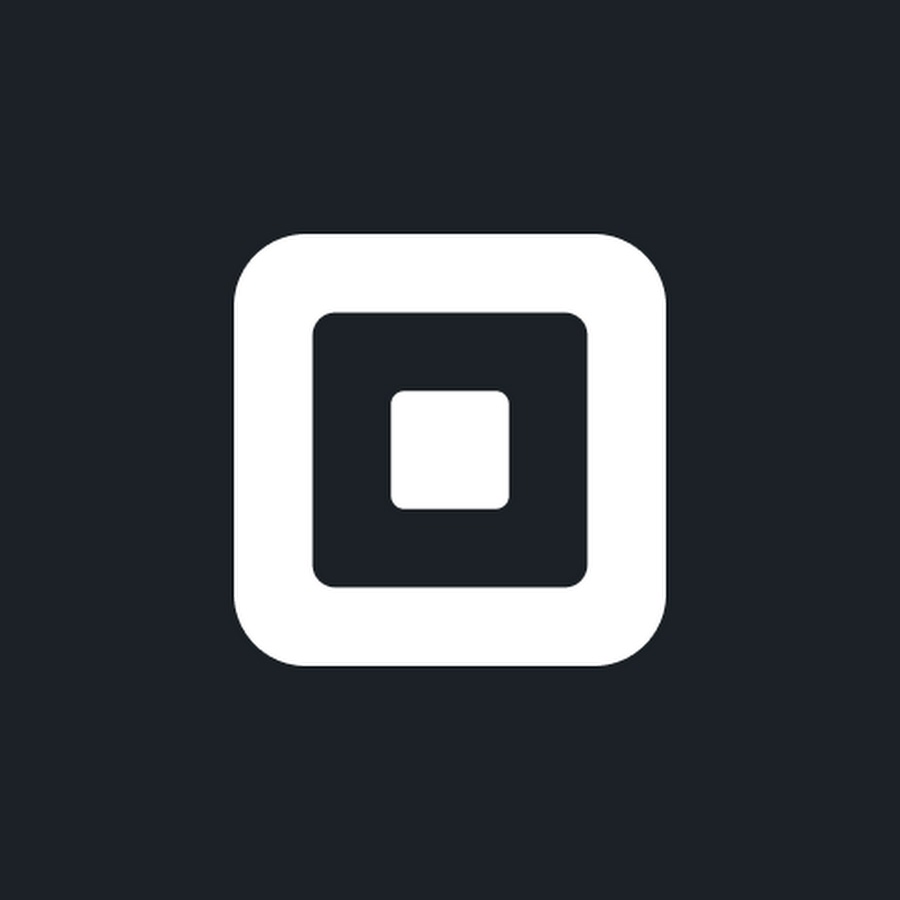}}
\enspace\enspace\enspace\enspace\enspace
\subfloat[Watermark]{\includegraphics[height=0.4in]{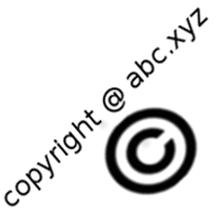}}
\vspace{0cm}
\captionsetup{font=small}
\caption{Two trigger patterns used in the experiments.}
\label{fig:trigger type}
\end{figure}

\begin{figure}[htbp]
\captionsetup{font=small}
\centering
\includegraphics[height=0.8\linewidth]{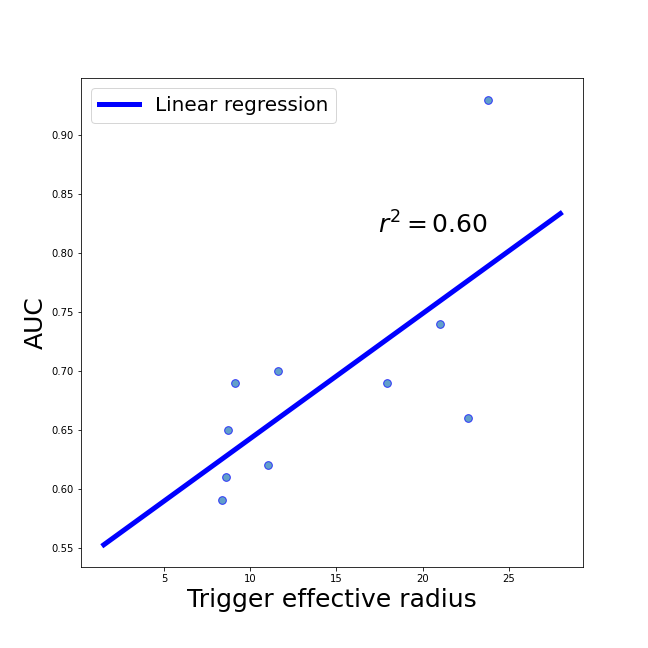}
\caption{ResNet-18 results on the relationship between the trigger
effective radius and the effectiveness of ten attacks to evade the NC
backdoor detection}
\label{fig:rob_scatter_ResNet}

\end{figure}

\begin{figure}[H]
\centering
\captionsetup{font=small}
\includegraphics[width=0.35\textwidth]{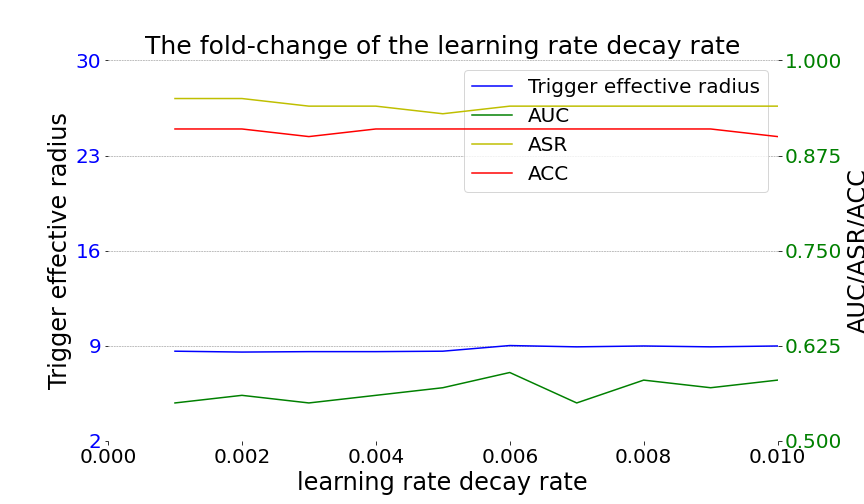} 
\caption{Impact of learning rate decay rate }
\label{Fig: decay} 
\end{figure}

\subsection{Putative Trigger Effectiveness}
The entire evaluation results of Section~\ref{subsec: Trigger accuracy} shows in Fig.~\ref{tab:trigger_accuracy_1}.
\begin{table*}[ht]
\centering
\footnotesize
\begin{tabular}{c|ccccc|cccc|cccc}
\toprule
\multicolumn{2}{c|}{}       & \multicolumn{4}{c|}{CIFAR-10} & \multicolumn{4}{c|}{MNIST} & \multicolumn{4}{c}{Tiny ImageNet}  \\ \midrule
\multicolumn{2}{c|}{}       & NC  & Tabor  & k-arm  & Pixel & NC & Tabor & k-arm & Pixel & NC & Tabor & k-arm & Pixel \\ 

\midrule
\multicolumn{13}{c}{$\epsilon_1$: effectiveness of unlearning}\\
\midrule
                                        % & NC      & Tabor      & k-arm     & Pixel       & NC       & Tabor       & k-arm      & Pixel       & NC       & Tabor       & k-arm       & Pixel
\multicolumn{2}{c|}{BadNet}             & 80.1\%    & 97.5\%     & 94.6\%    & 96.2\%      & 82.3\%   & 92.1\%      & 92.5\%     & 95.7\%      & 80.3\%   & 86.3\%      & 81.3\%      & 85.1\%    \\
\multicolumn{2}{c|}{~BadNet*}            & 2.20\%    & 1.50\%     & 22.5\%    & 58.6\%      & 1.10\%   & 3.10\%      & 23.0\%     & 56.2\%      & 2.90\%   & 7.31\%      & 21.5\%      & 53.6\%    \\
\multicolumn{2}{c|}{IMC}                & 42.1\%    & 62.6\%     & 67.5\%    & 69.5\%      & 32.2\%   & 68.4\%      & 62.0\%     & 58.5\%      & 41.6\%   & 60.9\%      & 53.5\%      & 55.2\%    \\
\multicolumn{2}{c|}{~IMC*}               & 2.90\%    & 2.70\%     & 23.1\%    & 46.7\%      & 2.50\%   & 5.10\%      & 21.6\%     & 45.3\%      & 4.10\%   & 7.40\%      & 21.6\%      & 24.6\%    \\

\multicolumn{2}{c|}{Composite}          & 45.1\%    & 67.2\%     & 71.4\%    & 71.2\%      & 41.6\%   & 71.3\%      & 65.2\%     & 59.1\%      & 43.2\%   & 60.1\%      & 55.2\%      & 54.0\%    \\
\multicolumn{2}{c|}{~Composite*}         & 3.20\%    & 3.10\%     & 32.2\%    & 52.1\%      & 5.20\%   & 7.20\%      & 27.3\%     & 40.0\%      & 6.15\%   & 7.40\%      & 26.1\%      & 38.1\%    \\

\multicolumn{2}{c|}{Latent}             & 79.7\%    & 66.5\%     & 58.2\%    & 83.3\%      & 82.6\%   & 85.2\%      & 81.2\%     & 88.3\%      & 70.1\%   & 74.1\%      & 69.2\%      & 83.0\%    \\
\multicolumn{2}{c|}{~Latent*}            & 8.90\%    & 4.72\%     & 23.5\%    & 55.6\%      & 3.70\%   & 5.70\%      & 26.2\%     & 56.2\%      & 4.80\%   & 12.1\%      & 12.1\%      & 63.2\%    \\
\multicolumn{2}{c|}{Adaptive-Blend}     & 6.20\%    & 9.50\%     & 31.5\%    & 37.0\%      & 10.1\%   & 13.2\%      & 13.3\%     & 11.1\%      & 6.30\%   & 6.22\%      & 9.30\%      & 19.4\%    \\
\multicolumn{2}{c|}{~Adaptive-Blend*}    & 2.10\%    & 1.04\%     & 2.40\%    & 16.7\%      & 2.40\%   & 2.10\%      & 4.00\%     & 15.7\%      & 2.42\%   & 4.45\%      & 5.30\%      & 16.1\%    \\

\multicolumn{2}{c|}{DEFEAT}             & 6.14\%    & 8.21\%     & 35.2\%    & 31.2\%      & 15.2\%   & 16.1\%      & 15.2\%     & 13.4\%      & 5.17\%   & 7.10\%      & 10.2\%      & 16.4\%    \\
\multicolumn{2}{c|}{~DEFEAT*}            & 3.12\%    & 3.54\%     & 2.50\%    & 22.3\%      & 2.45\%   & 5.15\%      & 8.20\%     & 22.3\%      & 4.12\%   & 6.24\%      & 9.33\%      & 11.2\%    \\

\multicolumn{2}{c|}{LSBA}               & 5.20\%    & 7.90\%     & 1.50\%    & 18.6\%      & 1.70\%   & 8.30\%      & 3.20\%     & 9.50\%      & 5.21\%   & 4.30\%      & 3.20\%      & 19.1\%    \\
\multicolumn{2}{c|}{~LSBA*}              & 1.90\%    & 2.40\%     & 1.30\%    & 0.60\%      & 0.70\%   & 1.20\%      & 2.10\%     & 1.20\%      & 1.99\%   & 0.70\%      & 1.73\%      & 2.10\%    \\

\midrule
\multicolumn{13}{c}{$\epsilon_2$: distance between trigger masks}\\
\midrule

                                        % & NC    & Tabor     & k-arm    & Pixel     & NC     & Tabor     & k-arm    & Pixel     & NC     & Tabor     & k-arm     & Pixel
\multicolumn{2}{c|}{BadNet}             & 0.54    & 0.34      & 0.61     & 0.24      & 0.47   & 0.32      & 0.33     & 0.04      & 0.65   & 0.51      & 0.65      & 0.64     \\
\multicolumn{2}{c|}{~BadNet*}            & 0.00    & 0.00      & 0.00     & 0.05      & 0.07   & 0.00      & 0.10     & 0.03      & 0.02   & 0.04      & 0.02      & 0.06   \\

\multicolumn{2}{c|}{IMC}                & 0.31    & 0.21      & 0.59     & 0.24      & 0.28   & 0.41      & 0.51     & 0.25      & 0.63   & 0.69      & 0.65      & 0.23     \\
\multicolumn{2}{c|}{~IMC*}               & 0.00    & 0.00      & 0.00     & 0.06      & 0.03   & 0.01      & 0.01     & 0.02      & 0.02   & 0.04      & 0.02      & 0.03\\

\multicolumn{2}{c|}{Composite}          & 0.32    & 0.36      & 0.53     & 0.26      & 0.24   & 0.47      & 0.55     & 0.31      & 0.55   & 0.67      & 0.67      & 0.26     \\
\multicolumn{2}{c|}{~Composite*}         & 0.00    & 0.00      & 0.00     & 0.03      & 0.02   & 0.01      & 0.04     & 0.05      & 0.02   & 0.05      & 0.04      & 0.03\\

\multicolumn{2}{c|}{Latent}             & 0.37    & 0.55      & 0.41     & 0.27      & 0.67   & 0.28      & 0.46     & 0.22      & 0.16   & 0.16      & 0.30      & 0.21     \\
\multicolumn{2}{c|}{~Latent*}            & 0.07    & 0.01      & 0.02     & 0.00      & 0.07   & 0.13      & 0.18     & 0.02      & 0.02   & 0.01      & 0.02      & 0.02     \\

\multicolumn{2}{c|}{Adaptive-Blend}     & 0.12    & 0.11      & 0.12     & 0.18      & 0.11   & 0.13      & 0.13     & 0.16      & 0.11   & 0.12      & 0.06      & 0.20\\
\multicolumn{2}{c|}{~Adaptive-Blend*}    & 0.05    & 0.04      & 0.06     & 0.12      & 0.11   & 0.12      & 0.06     & 0.05      & 0.04   & 0.09      & 0.04      & 0.04\\

\multicolumn{2}{c|}{DEFEAT}             & 0.14    & 0.10      & 0.11     & 0.18      & 0.10   & 0.12      & 0.12     & 0.13      & 0.10   & 0.08      & 0.05      & 0.22\\
\multicolumn{2}{c|}{~DEFEAT*}            & 0.06    & 0.05      & 0.07     & 0.13      & 0.11   & 0.12      & 0.07     & 0.07      & 0.09   & 0.10      & 0.03      & 0.08\\

\multicolumn{2}{c|}{LSBA}               & 0.07    & 0.08      & 0.22     & 0.18      & 0.02   & 0.07      & 0.33     & 0.16      & 0.05   & 0.03      & 0.50      & 0.19\\
\multicolumn{2}{c|}{~LSBA*}              & 0.01    & 0.02      & 0.02     & 0.08      & 0.01   & 0.01      & 0.08     & 0.13      & 0.02   & 0.02      & 0.03      & 0.05\\

\midrule
\multicolumn{13}{c}{$\epsilon_3$:ASR of the reconstructed triggers on a clean model}\\
\midrule

                                        % & NC      & Tabor     & k-arm    & Pixel       & NC     & Tabor      & k-arm     & Pixel     & NC     & Tabor     & k-arm     & Pixel
\multicolumn{2}{c|}{BadNet}               & 0.05\%  & 0.04\%    & 0.03\%   & 0.07\%      & 0.03\% & 0.04\%     & 0.05\%    & 0.05\%    & 0.02\% & 0.02\%    & 0.02\%    & 0.02\%    \\
\multicolumn{2}{c|}{~BadNet*}              & 88.8\%  & 88.2\%    & 49.3\%   & 12.2\%      & 92.5\% & 91.7\%     & 42.2\%    & 17.2\%    & 51.2\% & 61.5\%    & 54.2\%    & 23.6\%    \\

\multicolumn{2}{c|}{IMC}                  & 11.5\%  & 15.1\%    & 4.71\%   & 0.29\%      & 15.2\% & 21.5\%     & 0.22\%    & 0.32\%    & 11.5\% & 11.5\%    & 0.23\%    & 0.25\%    \\
\multicolumn{2}{c|}{~IMC*}                 & 89.2\%  & 83.2\%    & 47.5\%   & 41.3\%      & 82.7\% & 95.3\%     & 46.1\%    & 12.3\%    & 41.2\% &  60.2\%    & 56.6\%    & 21.0\%    \\

\multicolumn{2}{c|}{Composite}            & 12.3\%  & 13.4\%    & 5.41\%   & 2.11\%      & 11.4\% & 23.1\%     & 1.34\%    & 1.30\%    & 12.3\% & 11.0\%    & 2.32\%    & 2.23\%    \\
\multicolumn{2}{c|}{~Composite*}           & 88.4\%  & 82.4\%    & 49.1\%   & 42.5\%      & 86.7\% & 92.1\%     & 44.4\%    & 15.3\%    & 47.2\% & 64.1\%    & 54.7\%    & 27.9\%    \\

\multicolumn{2}{c|}{Latent}               & 1.20\%  & 1.30\%    & 1.2\%    & 1.10\%      & 1.10\% & 1.10\%     & 1.90\%    & 1.90\%    & 2.20\% &1.50\%     & 2.10\%    & 5.2\%      \\
\multicolumn{2}{c|}{~Latent*}              & 89.2\%  & 84.3\%    & 45.1\%   & 41.0\%      & 91.1\% & 91.0\%     & 41.0\%    & 10.9\%    & 62.1\% & 59.4\%    & 34.1\%    & 22.3\% \\

\multicolumn{2}{c|}{Adaptive-Blend}       & 22.3\%  & 11.4\%    & 22.4\%   & 28.3\%      & 22.0\% & 13.2\%     & 13.7\%    & 13.4\%    & 10.0\% & 21.0\%    & 26.1\%    & 19.2\%\\
\multicolumn{2}{c|}{~Adaptive-Blend*}      & 58.1\%  & 55.2\%    & 51.5\%   & 32.1\%      & 52.3\% & 51.2\%     & 53.2\%    & 39.1\%    & 51.5\% & 55.4\%    & 57.1\%    & 39.2\% \\

\multicolumn{2}{c|}{DEFEAT}               & 21.4\%  & 15.5\%    & 26.4\%   & 24.2\%      & 26.0\% & 16.6\%     & 19.1\%    & 13.1\%    & 18.1\% & 26.1\%    & 24.6\%    & 21.9\%\\
\multicolumn{2}{c|}{~DEFEAT*}              & 59.2\%  & 59.3\%    & 56.1\%   & 35.2\%      & 53.3\% & 55.1\%     & 56.1\%    & 30.7\%    & 54.3\% & 51.1\%    & 51.1\%    & 31.1\% \\

\multicolumn{2}{c|}{LSBA}                 & 21.2\%  & 25.2\%    & 32.5\%   & 21.3\%      & 37.2\% & 31.3\%     & 35.2\%    & 15.6\%    & 30.2\% & 34.9\%    & 32.6\%    & 29.2\%\\
\multicolumn{2}{c|}{~LSBA*}                & 88.8\%  & 88.2\%    & 89.3\%   & 42.1\%      & 92.5\% & 91.7\%     & 82.2\%    & 40.1\%    & 52.0\% & 61.4\%    & 61.6\%    & 41.5\% \\
\bottomrule

\end{tabular}
\caption{The evaluation of the effectiveness of unlearning ($\epsilon_1$), the distance between trigger masks ($\epsilon_2$), and ASR of the reconstructed triggers on a clean model ($\epsilon_3$) on the trigger inversion methods against the backdoored models poisoned by different backdoor attacks with and without the \ourmethod{} enhancement. The notation of the attacks follows Table 3.}
\label{tab:trigger_accuracy_1}
\end{table*}

\begin{table*}[ht]
\centering
\footnotesize
\begin{tabular}{c|ccccc|cccc|cccc}
\toprule
\multicolumn{2}{c|}{}       & \multicolumn{4}{c|}{CIFAR-10} & \multicolumn{4}{c|}{MNIST} & \multicolumn{4}{c}{Tiny ImageNet}  \\ \midrule
\multicolumn{2}{c|}{}       & ABS  & MNTD  & TS  & AC & ABS  & MNTD  & TS  & AC & ABS  & MNTD  & TS  & AC \\ 

\midrule
\multicolumn{13}{c}{AUC}\\
\midrule
%                                   CIFAR-10                                        MNIST                                          Tiny ImageNet      
%                                   & ABS       & MNTD      & TS        & AC        & ABS       & MNTD     & TS        & AC        & ABS       & MNTD      & TS        & AC \\ 
%Patch
\multicolumn{2}{c|}{BadNet}         & 84.4\%    & 76.5\%    & 76.3\%    & 71.2\%    & 81.4\%   & 80.6\%    & 77.3\%    & 74.1\%    & 79.6\%    & 75.1\%    & 71.0\%    & 72.4\%   \\
\multicolumn{2}{c|}{~BadNet*}        & 65.1\%    & 71.4\%    & 72.5\%    & 63.7\%    & 64.0\%   & 72.8\%    & 61.4\%    & 60.4\%    & 66.3\%    & 69.5\%    & 68.4\%    & 59.0\%   \\
\multicolumn{2}{c|}{LSBA}           & 69.3\%    & 66.5\%    & 59.1\%    & 61.3\%    & 67.1\%   & 65.0\%    & 55.2\%    & 61.5\%    & 67.4\%    & 59.9\%    & 54.1\%    & 57.0\%   \\
\multicolumn{2}{c|}{~LSBA*}          & 66.3\%    & 65.4\%    & 58.4\%    & 59.4\%    & 62.5\%   & 58.2\%    & 51.3\%    & 57.5\%    & 56.5\%    & 59.4\%    & 54.6\%    & 58.2\%   \\
\multicolumn{2}{c|}{Composite}      & 71.1\%    & 71.4\%    & 70.2\%    & 71.6\%    & 69.9\%   & 69.3\%    & 65.3\%    & 66.6\%    & 71.3\%    & 72.4\%    & 64.7\%    & 70.1\%   \\
\multicolumn{2}{c|}{~Composite*}     & 66.3\%    & 70.3\%    & 69.4\%    & 68.2\%    & 67.4\%   & 65.3\%    & 61.0\%    & 60.5\%    & 65.2\%    & 65.0\%    & 62.5\%    & 64.4\%   \\
%Clean label
\multicolumn{2}{c|}{Latent}         & 73.6\%    & 64.3\%    & 65.9\%    & 66.4\%    & 70.1\%   & 70.4\%    & 61.4\%    & 62.2\%    & 73.0\%    & 70.5\%    & 66.2\%    & 65.1\%   \\
\multicolumn{2}{c|}{~Latent*}        & 56.5\%    & 53.3\%    & 55.1\%    & 65.2\%    & 56.0\%   & 60.2\%    & 58.8\%    & 60.1\%    & 58.4\%    & 61.5\%    & 63.8\%    & 61.1\%   \\
%Imperceptible
\multicolumn{2}{c|}{DEFEAT}         & 69.6\%    & 69.2\%    & 68.9\%    & 65.2\%    & 68.2\%   & 73.1\%    & 67.3\%    & 61.9\%    & 70.3\%    & 71.3\%    & 67.0\%    & 62.5\%   \\
\multicolumn{2}{c|}{~DEFEAT*}        & 63.4\%    & 66.0\%    & 60.5\%    & 63.2\%    & 62.4\%   & 72.9\%    & 68.0\%    & 58.3\%    & 65.8\%    & 71.7\%    & 70.2\%    & 60.2\%   \\
\multicolumn{2}{c|}{DFST}           & 67.4\%    & 64.0\%    & 65.2\%    & 61.3\%    & 65.0\%   & 62.4\%    & 63.4\%    & 59.2\%    & 68.4\%    & 69.4\%    & 58.2\%    & 59.0\%   \\
\multicolumn{2}{c|}{~DFST*}          & 63.1\%    & 65.2\%    & 62.3\%    & 60.3\%    & 62.7\%   & 60.2\%    & 61.4\%    & 60.2\%    & 62.0\%    & 66.8\%    & 57.0\%    & 61.2\%   \\
%Latent space similarity
\multicolumn{2}{c|}{Adaptive-Blend} & 76.8\%    & 70.2\%    & 78.2\%    & 68.4\%    & 77.3\%   & 69.1\%    & 75.2\%    & 69.1\%    & 74.3\%    & 68.2\%    & 69.0\%    & 67.1\%   \\
\multicolumn{2}{c|}{~Adaptive-Blend*}& 63.2\%    & 69.3\%    & 70.1\%    & 57.2\%    & 62.3\%   & 63.4\%    & 69.5\%    & 59.0\%    & 67.3\%    & 62.5\%    & 66.3\%    & 60.3\%   \\
\bottomrule

\end{tabular}
\caption{The AUCs of backdoor detection by weight analysis-based methods on the backdoored models poisoned by different backdoor attacks with and without \ourmethod{} enhancement. The attack with the \ourmethod{} enhancement is denoted by the symbol ``*" appended to the name of the respective attack.}
\label{tab:weight analysis AUC entire}
\end{table*}

\subsection{Theoratical Analysis on GRASP Against Weight Analysis Detection}\label{subsec:theory_weight_ana}

Formally, given a training dataset $D$, and a backdoor attack, we train $t$ benign ML models $\{f_{\theta_{(i)}}| i \in \{1,2,...,t\}\}$ on $D$, and $t$ backdoored ML models $\{f_{\hat{\theta}_{(i)}}| i \in \{1,2,...,t\}\}$ on $D$ with the given backdoor attack.  $z(f_\theta(\cdot)):\mathcal{X}^m \rightarrow \mathcal{Y}$, where $\theta \in \mathbb{R}^K$ represent the set ($K$) parameters in the model $f_{\theta}$. A weight analysis methods then build a classifier $g(\cdot): \mathbb{R}^K \rightarrow [0,1]$, which is trained on the dataset $D_{\theta} = \{(\theta_{(1)},0), (\theta_{(2)},0),...,(\theta_{(t)},0) \} \cup \{(\hat{\theta}_{(1)},1), (\hat{\theta}_{(2)},1),...,(\hat{\theta}_{(t)},1) \}$ where the label $``0"$ indicates the parameters of benign models, and $``1”$ indicates the parameters of backdoored models.

Next, we show why \ourmethod{} does not reduce the attack effectiveness to evade the weight analysis. Consider a neural network with any initialization $f_{\theta_0}$ is trained on the dataset $D_{benign}$, the backdoor dataset $D_{backdoor}$, and the \ourmethod{}-enhanced backdoor dataset $D_{GRASP}$, respectively. Specifically, we denote $D_{benign} = \{D_{ori}, D^*_{troj}, D^*_{Aug} \}$, $D_{backdoor} = \{D_{ori}, \hat{D}_{troj}, \hat{D}_{Aug} \}$, and $D_{GRASP} = \{D_{ori}, \hat{D}_{troj}, D^*_{Aug} \}$, where $D_{ori}$ is the legitimate training dataset used for training all three models, $\hat{D}_{troj}$ and $D^*_{troj}$ represent the set of trigger inserted samples labeled by the target and the source (legitimate) class, respectively, and $\hat{D}_{Aug}$ and $D^*_{Aug}$ represent the set of augmented samples (i.e., the trigger-inserted samples with added noise) labeled by the target and the source (legitimate) classes, respectively.

Theorem~\ref{thm:weight analysis} below indicates the models trained on $D_{GRASP}$ (enhanced by \ourmethod{}) are not easier to be distinguished by the weight analysis from the benign models (trained on $D_{benign}$) than the models trained on $D_{backdoor}$ (by the backdoor attack without \ourmethod{} enhancement).

\begin{theorembold}
\label{thm:weight analysis}
Consider a $L$ - layer neural network $f_{\theta}(\cdot): \mathcal{X}^m \rightarrow \bar{\mathcal{Y}}$. Given an input $x\in \mathcal{X}^m$, in the $l^{th}$ layer with $K_{(l)}$ neurons, where $1<l<L$, let $\phi(x)^{(l)}_{k}$ denote the output of the $k^{th}$ neuron before activation, and $\sigma (x)^{(l)}_{k}$ denote denote the output after activation. Let $\theta ^{(l)}_{(p,q)}$ denote the weight connecting $q^{th}$ neuron in the $(l-1)^{th}$ layer and $p^{th}$ neuron in the $l^{th}$ layer.

We assume:

\begin{equation}
 \sum\limits^{D_{ori}}_i \sigma(x_i)^{(l)}_k \cdot \sum\limits^{D_{troj}^*}_i \sigma(x_i)^{(l)}_k\cdot \sum\limits^{D_{Aug}^*}_k \sigma(x_i)^{(l)}_i \neq 0
\end{equation}

\noindent and the square loss function $C(\theta) = \frac{1}{2m} \sum\limits^n_i (f(x_i;\theta)-y_i)^2$
is used for training.
Then for any set of parameters $\theta$, the gradient of the loss function w.r.t any parameter $\theta^{(l)}_{(p,q)}$ in the model $f_{\theta}$ on the three datasets satisfy:

\begin{equation}\label{eq:weight analysis}
\scriptsize\mathop{\nabla}\limits_{D_{benign}}\theta^{(l)}_{(p,q)} - \mathop{\nabla}\limits_{D_{backdoor}}\theta^{(l)}_{(p,q)}  >   \mathop{\nabla}\limits_{D_{benign}}\theta^{(l)}_{(p,q)} - \mathop{\nabla}\limits_{D_{GRASP}}\theta^{(l)}_{(p,q)}
\end{equation}

\end{theorembold}

The proof of Theorem~\ref{thm:weight analysis} is given below. This theorem shows the difference of gradient on the parameters between the backdoored models poisoned by a \ourmethod{}-enhanced attack and the benign models is always smaller than the difference between the backdoored models poisoned by the same attack without \ourmethod{} enhancement, which implies it is not easier to distinguish the \ourmethod{} poisoned models from benign models than to distinguish the backdoored models without \ourmethod{} enhancement from the benign models.

% \vspace{2pt}\noindent\textbf{Theorem~\ref{thm:weight analysis}}
% Consider a $L$ - layer neural network $f_{\theta}(\cdot): \mathcal{X}^m \rightarrow \bar{\mathcal{Y}}$. Given an input $x\in \mathcal{X}^m$, in the $l^{th}$ layer with $K_{(l)}$ neurons, where $1<l<L$, let $\phi(x)^{(l)}_{k}$ denote the output of the $k^{th}$ neuron before activation, and $\sigma (x)^{(l)}_{k}$ denote denote the output after activation. Let $\theta ^{(l)}_{(p,q)}$ denote the weight connecting $q^{th}$ neuron in the $(l-1)^{th}$ layer and $p^{th}$ neuron in the $l^{th}$ layer.

% We assume:

% \begin{equation}
%  \sum\limits^{D_{ori}}_i \sigma(x_i)^{(l)}_k \cdot \sum\limits^{D_{troj}^*}_i \sigma(x_i)^{(l)}_k\cdot \sum\limits^{D_{Aug}^*}_k \sigma(x_i)^{(l)}_i \neq 0
% \end{equation}

% \noindent and the square loss function $C(\theta) = \frac{1}{2m} \sum\limits^n_i (f(x_i;\theta)-y_i)^2$
% is used for training.
% Then for any set of parameters $\theta$, the gradient of the loss function w.r.t any parameter $\theta^{(l)}_{(p,q)}$ in the model $f_{\theta}$ on the three datasets satisfy:

% \begin{equation}\label{eq:weight analysis}
% \scriptsize\mathop{\nabla}\limits_{D_{benign}}\theta^{(l)}_{(p,q)} - \mathop{\nabla}\limits_{D_{backdoor}}\theta^{(l)}_{(p,q)}  >   \mathop{\nabla}\limits_{D_{benign}}\theta^{(l)}_{(p,q)} - \mathop{\nabla}\limits_{D_{GRASP}}\theta^{(l)}_{(p,q)}
% \end{equation}

\vspace{3pt}\noindent\textbf{Proof outline:}

We first we proof the eq.~\ref{eq:weight analysis} hold when weights are from the linear layer which is weights between the penultimate layer and last layer. Then we will proof the same thing but between hidden layers. Then we finish the proof.

\vspace{3pt}\noindent\textbf{Proof:}

\begin{figure}[H]
\centering
\captionsetup{font=small}
\includegraphics[width=0.4\textwidth]{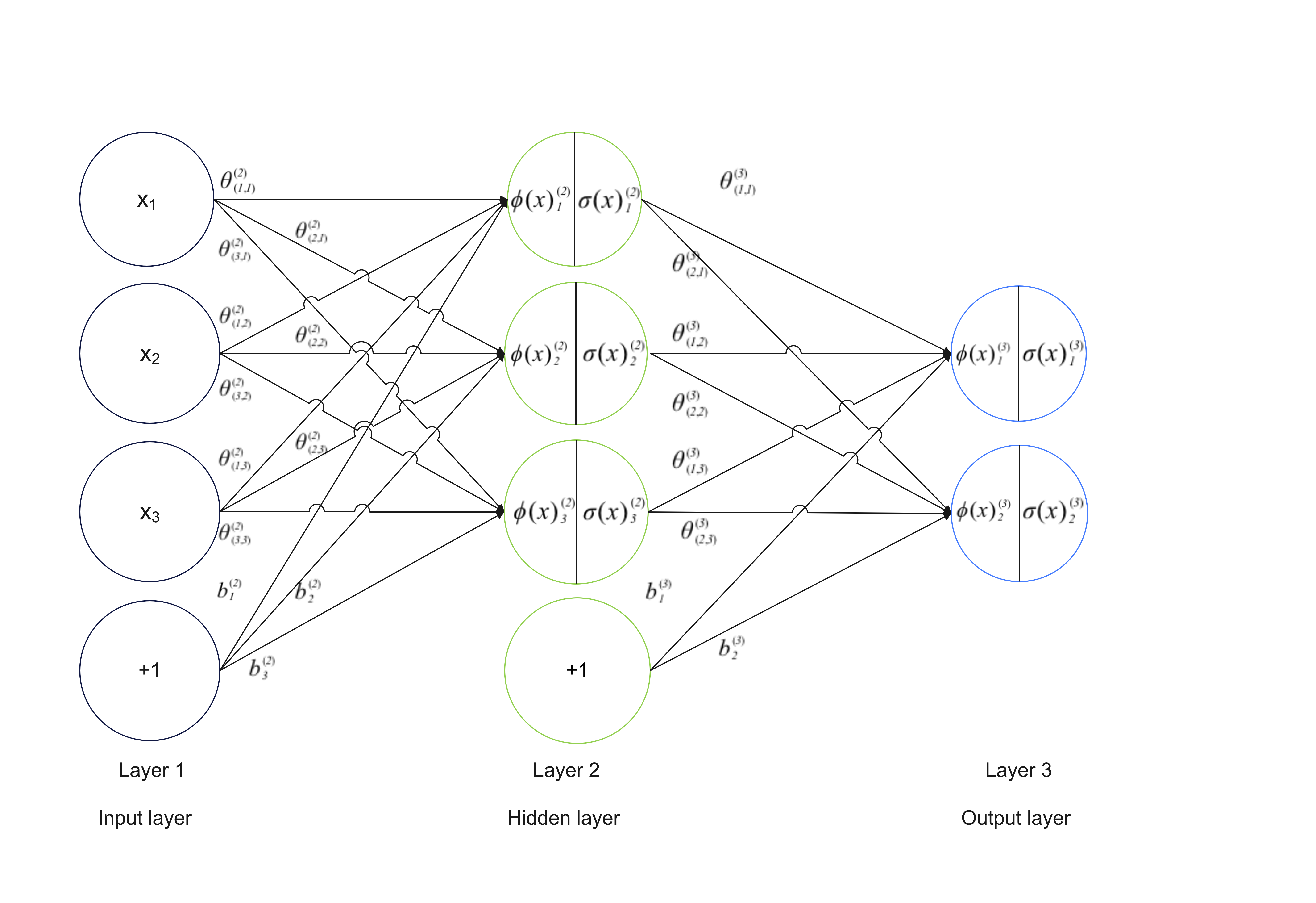} 
\caption{The notations used in theorem~\ref{thm:weight analysis}}
\label{Fig: weight ana notation} 

\end{figure}

\vspace{3pt}\noindent\textbf{Weights from linear layer}.
Without loss of generality, the loss function has gradient w.r.t $\theta^{(L-1)}_{(p,q)}$ by training on three different dataset;
\begin{equation}
\begin{split}\mathop{\nabla}\limits_{D_{benign}} \theta^{(L-1)}_{(p,q)} =& \sum\limits^{D_{ori}}_i(y_i-f_{\theta}(x_i))\sigma(x)^{(L-1)}_{(p,q)} +  \\
    & \sum\limits^{D^*_{troj}}_i(y_i-f_{\theta}(x_i)) \sigma(x)^{(L-1)}_{(p,q)} +  \\
    & \sum\limits^{D^*_{Aug}}_i(y_i-f_{\theta}x_i)) \sigma(x)^{(L-1)}_{(p,q)} \end{split}
\end{equation}
 
\begin{equation}
\begin{split}\mathop{\nabla}\limits_{D_{backdoor}} \theta^{(L-1)}_{(p,q)} =& \sum\limits^{D_{ori}}_i(y_i-f_{\theta}(x_i)) \sigma(x)^{(L-1)}_{(p,q)} +  \\ &
\sum\limits^{\hat{D}_{troj}}_i(y_i-f_{\theta}(x_i)) \sigma(x)^{(L-1)}_{(p,q)} +  \\ 
& \sum\limits^{\hat{D}_{Aug}}_i(y_i-f_{\theta}x_i)) \sigma(x)^{(L-1)}_{(p,q)}  \end{split}
\end{equation}

\begin{equation}
\begin{split}\mathop{\nabla}\limits_{D_{GRASP}} \theta^{(L-1)}_{(p,q)} =& \sum\limits^{D_{ori}}_i(y_i-f_{\theta}(x_i)) \sigma(x)^{(L-1)}_{(p,q)} + \\
& \sum\limits^{\hat{D}_{troj}}_i(y_i-f_{\theta}(x_i)) \sigma(x)^{(L-1)}_{(p,q)} +  \\
& \sum\limits^{D^*_{Aug}}_i(y_i-f_{\theta}x_i)) \sigma(x)^{(L-1)}_{(p,q)}  \end{split}
\end{equation}

Consider:
\begin{equation} \label{eq: weight ratio}
\frac{\mathop{\nabla}\limits_{D_{benign}} \theta^{(L-1)}_{(p,q)} - \mathop{\nabla}\limits_{D_{backdoor}} \theta^{(L-1)}_{(p,q)}}{ \mathop{\nabla}\limits_{D_{benign}} \theta^{(L-1)}_{(p,q)} - \mathop{\nabla}\limits_{D_{GRASP}} \theta^{(L-1)}_{(p,q)}}   
\end{equation}

If eq.~\ref{eq: weight ratio} is always greater than 1. Then it is equivalent to proof theorem~\ref{thm:weight analysis}.

Expand eq.~\ref{eq: weight ratio}, and replace those poisoned label by target label $y_t$, eq.~\ref{eq: weight ratio} is equal to:
\begin{equation*}
\footnotesize
\begin{split}
=& \frac{\sum\limits^{D^*_{troj}}_i\sigma(x_i)^{(L-1)}_{(p,q)}(y_i-f_{\theta}(x_i) -y_t +f_{\theta}(x_i) + y_i-f_{\theta}(x_i) -y_t +f_{\theta}(x_i))  } {\sum\limits^{D^*_{troj}}_i\sigma(x_i)^{(L-1)}_{(p,q)}(y_i-f_{\theta}(x_i) -y_t +f_{\theta}(x_i))}
\\
=& \frac{\sum\limits^{D^*_{troj}}_i\sigma(x_i)^{(L-1)}_{(p,q)} (2(y_i-y_t))} {\sum\limits^{D^*_{troj}}_i\sigma(x_i)^{(L-1)}_{(p,q)} (y_i-y_t)}
\end{split}
\end{equation*}
Since $  \sum\limits^{D_{ori}}_i \sigma(x_i)^{(l)}_k \cdot \sum\limits^{D_{troj}^*}_i \sigma(x_i)^{(l)}_k\cdot \sum\limits^{D_{Aug}^*}_k \sigma(x_i)^{(l)}_i \neq 0$ for any $1<l<L$, we can infer that:
\begin{equation*}
\footnotesize
\frac{\sum\limits^{D^*_{troj}}_i\sigma(x_i)^{(L-1)}_{(p,q)} (2(y_i-y_t))} {\sum\limits^{D^*_{troj}}_i\sigma(x_i)^{(L-1)}_{(p,q)} (y_i-y_t)} = 2 > 1 
\end{equation*}
Which indicates the weights in the last layer (linear map) always satisfy eq.\ref{eq:weight analysis}

\vspace{3pt}\noindent\textbf{Weights between hidden layers}.
Now we consider the gradient of loss function w.r.t weights in hidden layers. 
During the back-propagation, we use $\delta^{L}_p$ denote the error term from the $L^{th}$ layer to the $p^{th}$ neuron in the $(L-1)^{th}$ layer:
\begin{equation}
\begin{split} \delta^{L}_p =& \frac{\partial C(\theta)}{\partial  \phi(x)^{(L)}_p} \\ =& \frac{\partial C(\theta)}{\partial \sigma(x)^{(L)}_p}\cdot \frac{\partial \sigma(x)^{(L)}_p}{\partial \phi(x)^{(L)}_p}   \end{split}
\end{equation}
As we can see, the error term can be separated into two parts: loss gradient w.r.t the neuron output, the first, and a derivative term, the second. For the first term, as we have proven from the previous analysis, if we use three different datasets, this loss gradient always satisfies eq.\ref{eq:weight analysis}. For the second term, three different datasets has the same value. This is because $D_{benign},D_{backdoor}$, and $D_{GRASP}$ has the same inputs. As a result, the error term $\delta^{L}_p$ always satisfy $\underset{backdoor}{\delta_{p}^{L}} < \underset{GRASP}{\delta_{p}^{L}}$.
Without loss of generality, we consider the weight between $q^{th}$ neuron in $l^{th}$ layer and  $p^{th}$ neuron in $(l+1)^{th}$ layer, $\theta^{(l)}_{(p,q)}$, we have loss function gradient:
\begin{equation}
    \frac{\partial C(\theta)}{\partial \theta^{(l)}_{(p,q)}} = \delta^{(l)}_{(p)} \cdot \sigma(x)^{(l)}_{q}
\end{equation}
 The second term above is the partial derivative of neuron output w.r.t the neuron input for the $q^{th}$ neuron in the $l^{th}$ layer. Note that the second term is only related to the input (independent of the label). For the proof of eq.\ref{eq:weight analysis}, we can ignore this term since they are equal while using three different datasets. The first term $\delta^{(l)}_p$ is the error term back-propagate from $(l+1)^{th}$ layer to the $p^{th}$ neuron in the $l^{th}$ layer:
\begin{equation}
    \begin{split}\delta^{(l)}_p =& \frac{\partial C(\theta)}{\partial \phi(x)^{(l)}_p} \\ =&  \sum\limits^{K_{l+1}}_i\frac{\partial C(\theta)}{\partial \phi(x)^{(l+1)}_i} \cdot \frac{\partial \phi(x)^{(l+1)}_i}{\partial \sigma(x)^{(l)}_q} \cdot \frac{\partial \sigma(x)^{(l)}_q}{\partial \phi(x)^{(l)}_q}\\=& \sum\limits^{K_{l+1}}_i  \ \ \  \underbrace{\delta^{(l+1)}_i}_{\text{Error term}}  \ \ \ \cdot  \ \ \ \underbrace{\theta^{(l+1)}_{(p,q)}}_{\text{Weight}} \    \cdot    \underbrace{\sigma'(x)^{(l)}_q}_{\text{Neuron output derivative}} \end{split}
\end{equation}
Among the three terms above, Error term, weight, and neuron output derivative, only the error term is related to the label. Others are independent of the label. So for $D_{benign},D_{backdoor}$, and $D_{GRASP}$, their loss function gradient w.r.t the weights in hidden layers, $\mathop{\nabla}\limits_{D_{benign}} \theta^{(l)}_{(p,q)}$, $\mathop{\nabla}\limits_{D_{backdoor}} \theta^{(l)}_{(p,q)}$ and $\mathop{\nabla}\limits_{D_{GRASP}} \theta^{(l)}_{(p,q)}$ always satisfy eq.\ref{eq:weight analysis}.
\\
\noindent End of proof

%% file: main.bbl
% Generated by IEEEtranS.bst, version: 1.12 (2007/01/11)
\begin{thebibliography}{10}
\providecommand{\url}[1]{#1}
\csname url@samestyle\endcsname
\providecommand{\newblock}{\relax}
\providecommand{\bibinfo}[2]{#2}
\providecommand{\BIBentrySTDinterwordspacing}{\spaceskip=0pt\relax}
\providecommand{\BIBentryALTinterwordstretchfactor}{4}
\providecommand{\BIBentryALTinterwordspacing}{\spaceskip=\fontdimen2\font plus
\BIBentryALTinterwordstretchfactor\fontdimen3\font minus \fontdimen4\font\relax}
\providecommand{\BIBforeignlanguage}[2]{{%
\expandafter\ifx\csname l@#1\endcsname\relax
\typeout{** WARNING: IEEEtranS.bst: No hyphenation pattern has been}%
\typeout{** loaded for the language `#1'. Using the pattern for}%
\typeout{** the default language instead.}%
\else
\language=\csname l@#1\endcsname
\fi
#2}}
\providecommand{\BIBdecl}{\relax}
\BIBdecl

\bibitem{NIPS_competation}
``Tdc 2022,'' \url{https://trojandetection.ai/}, accessed: 2022-09-30.

\bibitem{low-conf}
H.~Ali, S.~Nepal, S.~S. Kanhere, and S.~Jha, ``Has-nets: A heal and select mechanism to defend dnns against backdoor attacks for data collection scenarios,'' \emph{arXiv preprint arXiv:2012.07474}, 2020.

\bibitem{bagdasaryan2021blind}
E.~Bagdasaryan and V.~Shmatikov, ``Blind backdoors in deep learning models,'' in \emph{30th USENIX Security Symposium (USENIX Security 21)}, 2021, pp. 1505--1521.

\bibitem{cartis2018worst}
C.~Cartis, N.~I. Gould, and P.~L. Toint, ``Worst-case evaluation complexity and optimality of second-order methods for nonconvex smooth optimization,'' in \emph{Proceedings of the International Congress of Mathematicians: Rio de Janeiro 2018}.\hskip 1em plus 0.5em minus 0.4em\relax World Scientific, 2018, pp. 3711--3750.

\bibitem{AC}
B.~Chen, W.~Carvalho, N.~Baracaldo, H.~Ludwig, B.~Edwards, T.~Lee, I.~Molloy, and B.~Srivastava, ``Detecting backdoor attacks on deep neural networks by activation clustering,'' \emph{arXiv preprint arXiv:1811.03728}, 2018.

\bibitem{DFST}
S.~Cheng, Y.~Liu, S.~Ma, and X.~Zhang, ``Deep feature space trojan attack of neural networks by controlled detoxification,'' in \emph{Proceedings of the AAAI Conference on Artificial Intelligence}, vol.~35, no.~2, 2021, pp. 1148--1156.

\bibitem{lira}
K.~Doan, Y.~Lao, W.~Zhao, and P.~Li, ``Lira: Learnable, imperceptible and robust backdoor attacks,'' in \emph{Proceedings of the IEEE/CVF International Conference on Computer Vision}, 2021, pp. 11\,966--11\,976.

\bibitem{du2019robust}
M.~Du, R.~Jia, and D.~Song, ``Robust anomaly detection and backdoor attack detection via differential privacy,'' \emph{arXiv preprint arXiv:1911.07116}, 2019.

\bibitem{trojan_signiture}
G.~Fields, M.~Samragh, M.~Javaheripi, F.~Koushanfar, and T.~Javidi, ``Trojan signatures in dnn weights,'' in \emph{Proceedings of the IEEE/CVF International Conference on Computer Vision}, 2021, pp. 12--20.

\bibitem{badnet}
T.~Gu, B.~Dolan-Gavitt, and S.~Garg, ``Badnets: Identifying vulnerabilities in the machine learning model supply chain,'' \emph{arXiv preprint arXiv:1708.06733}, 2017.

\bibitem{tabor}
W.~Guo, L.~Wang, X.~Xing, M.~Du, and D.~Song, ``Tabor: A highly accurate approach to inspecting and restoring trojan backdoors in ai systems,'' \emph{arXiv preprint arXiv:1908.01763}, 2019.

\bibitem{DBD}
K.~Huang, Y.~Li, B.~Wu, Z.~Qin, and K.~Ren, ``Backdoor defense via decoupling the training process,'' \emph{arXiv preprint arXiv:2202.03423}, 2022.

\bibitem{PL_condition}
H.~Karimi, J.~Nutini, and M.~Schmidt, ``Linear convergence of gradient and proximal-gradient methods under the polyak-{\l}ojasiewicz condition,'' in \emph{Joint European conference on machine learning and knowledge discovery in databases}.\hskip 1em plus 0.5em minus 0.4em\relax Springer, 2016, pp. 795--811.

\bibitem{trojai}
K.~Karra, C.~Ashcraft, and N.~Fendley, ``The trojai software framework: An opensource tool for embedding trojans into deep learning models,'' \emph{arXiv preprint arXiv:2003.07233}, 2020.

\bibitem{adam}
D.~P. Kingma and J.~Ba, ``Adam: A method for stochastic optimization,'' \emph{arXiv preprint arXiv:1412.6980}, 2014.

\bibitem{krishnan2020lipschitz}
V.~Krishnan, A.~Makdah, A.~AlRahman, and F.~Pasqualetti, ``Lipschitz bounds and provably robust training by laplacian smoothing,'' \emph{Advances in Neural Information Processing Systems}, vol.~33, pp. 10\,924--10\,935, 2020.

\bibitem{krizhevsky2009learning}
A.~Krizhevsky, G.~Hinton \emph{et~al.}, ``Learning multiple layers of features from tiny images,'' 2009.

\bibitem{TinyImage}
Y.~Le and X.~S. Yang, ``Tiny imagenet visual recognition challenge,'' 2015.

\bibitem{lecun1998gradient}
Y.~LeCun, L.~Bottou, Y.~Bengio, and P.~Haffner, ``Gradient-based learning applied to document recognition,'' \emph{Proceedings of the IEEE}, vol.~86, no.~11, pp. 2278--2324, 1998.

\bibitem{li2017training}
H.~Li, S.~De, Z.~Xu, C.~Studer, H.~Samet, and T.~Goldstein, ``Training quantized nets: A deeper understanding,'' \emph{Advances in Neural Information Processing Systems}, vol.~30, 2017.

\bibitem{ABL}
\BIBentryALTinterwordspacing
Y.~Li, X.~Lyu, N.~Koren, L.~Lyu, B.~Li, and X.~Ma, ``Anti-backdoor learning: Training clean models on poisoned data,'' in \emph{Advances in Neural Information Processing Systems 34: Annual Conference on Neural Information Processing Systems 2021, NeurIPS 2021, December 6-14, 2021, virtual}, 2021, pp. 14\,900--14\,912. [Online]. Available: \url{https://proceedings.neurips.cc/paper/2021}
\BIBentrySTDinterwordspacing

\bibitem{NAD}
\BIBentryALTinterwordspacing
------, ``Neural attention distillation: Erasing backdoor triggers from deep neural networks,'' in \emph{9th International Conference on Learning Representations, {ICLR} 2021, Virtual Event, Austria, May 3-7, 2021}, 2021. [Online]. Available: \url{https://openreview.net/forum?id=9l0K4OM-oXE}
\BIBentrySTDinterwordspacing

\bibitem{backdoorsurvey}
Y.~Li, B.~Wu, Y.~Jiang, Z.~Li, and S.-T. Xia, ``Backdoor learning: A survey,'' \emph{arXiv preprint arXiv:2007.08745}, 2020.

\bibitem{benchmark}
Y.~Li, Y.~Li, B.~Wu, L.~Li, R.~He, and S.~Lyu, ``Backdoorbench: a comprehensive benchmark of backdoor attack and defense methods,'' in \emph{https://github.com/SCLBD/BackdoorBench}, 2021.

\bibitem{composite}
J.~Lin, L.~Xu, Y.~Liu, and X.~Zhang, ``Composite backdoor attack for deep neural network by mixing existing benign features,'' in \emph{Proceedings of the 2020 ACM SIGSAC Conference on Computer and Communications Security}, 2020, pp. 113--131.

\bibitem{liu1989limited}
D.~C. Liu and J.~Nocedal, ``On the limited memory bfgs method for large scale optimization,'' \emph{Mathematical programming}, vol.~45, no. 1-3, pp. 503--528, 1989.

\bibitem{fine-purning}
K.~Liu, B.~Dolan-Gavitt, and S.~Garg, ``Fine-pruning: Defending against backdooring attacks on deep neural networks,'' in \emph{International Symposium on Research in Attacks, Intrusions, and Defenses}.\hskip 1em plus 0.5em minus 0.4em\relax Springer, 2018, pp. 273--294.

\bibitem{abs}
Y.~Liu, W.-C. Lee, G.~Tao, S.~Ma, Y.~Aafer, and X.~Zhang, ``Abs: Scanning neural networks for back-doors by artificial brain stimulation,'' in \emph{Proceedings of the 2019 ACM SIGSAC Conference on Computer and Communications Security}, 2019, pp. 1265--1282.

\bibitem{sig}
Y.~Liu, X.~Ma, J.~Bailey, and F.~Lu, ``Reflection backdoor: A natural backdoor attack on deep neural networks,'' in \emph{European Conference on Computer Vision}.\hskip 1em plus 0.5em minus 0.4em\relax Springer, 2020, pp. 182--199.

\bibitem{ma2022beatrix}
W.~Ma, D.~Wang, R.~Sun, M.~Xue, S.~Wen, and Y.~Xiang, ``The" beatrix''resurrections: Robust backdoor detection via gram matrices,'' \emph{arXiv preprint arXiv:2209.11715}, 2022.

\bibitem{michaelis2019dragon}
C.~Michaelis, B.~Mitzkus, R.~Geirhos, E.~Rusak, O.~Bringmann, A.~S. Ecker, M.~Bethge, and W.~Brendel, ``Benchmarking robustness in object detection: Autonomous driving when winter is coming,'' \emph{arXiv preprint arXiv:1907.07484}, 2019.

\bibitem{wanet}
A.~Nguyen and A.~Tran, ``Wanet--imperceptible warping-based backdoor attack,'' \emph{arXiv preprint arXiv:2102.10369}, 2021.

\bibitem{IMC}
R.~Pang, H.~Shen, X.~Zhang, S.~Ji, Y.~Vorobeychik, X.~Luo, A.~Liu, and T.~Wang, ``A tale of evil twins: Adversarial inputs versus poisoned models,'' in \emph{Proceedings of the 2020 ACM SIGSAC Conference on Computer and Communications Security}, 2020, pp. 85--99.

\bibitem{lsba}
M.~Peng, Z.~Xiong, M.~Sun, and P.~Li, ``Label-smoothed backdoor attack,'' \emph{arXiv preprint arXiv:2202.11203}, 2022.

\bibitem{adaptive-blend}
X.~Qi, T.~Xie, S.~Mahloujifar, and P.~Mittal, ``Circumventing backdoor defenses that are based on latent separability,'' \emph{arXiv preprint arXiv:2205.13613}, 2022.

\bibitem{adaptive}
A.~Rajabi, B.~Ramasubramanian, and R.~Poovendran, ``Trojan horse training for breaking defenses against backdoor attacks in deep learning,'' \emph{arXiv preprint arXiv:2203.15506}, 2022.

\bibitem{simtrojan}
Y.~Ren, L.~Li, and J.~Zhou, ``Simtrojan: Stealthy backdoor attack,'' in \emph{2021 IEEE International Conference on Image Processing (ICIP)}.\hskip 1em plus 0.5em minus 0.4em\relax IEEE, 2021, pp. 819--823.

\bibitem{latent}
A.~Shafahi, W.~R. Huang, M.~Najibi, O.~Suciu, C.~Studer, T.~Dumitras, and T.~Goldstein, ``Poison frogs! targeted clean-label poisoning attacks on neural networks,'' \emph{Advances in neural information processing systems}, vol.~31, 2018.

\bibitem{k-arm}
G.~Shen, Y.~Liu, G.~Tao, S.~An, Q.~Xu, S.~Cheng, S.~Ma, and X.~Zhang, ``Backdoor scanning for deep neural networks through k-arm optimization,'' in \emph{International Conference on Machine Learning}.\hskip 1em plus 0.5em minus 0.4em\relax PMLR, 2021, pp. 9525--9536.

\bibitem{augmentationtraining}
C.~Shorten and T.~M. Khoshgoftaar, ``A survey on image data augmentation for deep learning,'' \emph{Journal of big data}, vol.~6, no.~1, pp. 1--48, 2019.

\bibitem{shumailov2021manipulating}
I.~Shumailov, Z.~Shumaylov, D.~Kazhdan, Y.~Zhao, N.~Papernot, M.~A. Erdogdu, and R.~J. Anderson, ``Manipulating sgd with data ordering attacks,'' \emph{Advances in Neural Information Processing Systems}, vol.~34, pp. 18\,021--18\,032, 2021.

\bibitem{szegedy2013intriguing}
C.~Szegedy, W.~Zaremba, I.~Sutskever, J.~Bruna, D.~Erhan, I.~Goodfellow, and R.~Fergus, ``Intriguing properties of neural networks,'' \emph{arXiv preprint arXiv:1312.6199}, 2013.

\bibitem{pixelbackdoor}
G.~Tao, G.~Shen, Y.~Liu, S.~An, Q.~Xu, S.~Ma, P.~Li, and X.~Zhang, ``Better trigger inversion optimization in backdoor scanning.''

\bibitem{terjek2019adversarial}
D.~Terj{\'e}k, ``Adversarial lipschitz regularization,'' \emph{arXiv preprint arXiv:1907.05681}, 2019.

\bibitem{NC}
B.~Wang, Y.~Yao, S.~Shan, H.~Li, B.~Viswanath, H.~Zheng, and B.~Y. Zhao, ``Neural cleanse: Identifying and mitigating backdoor attacks in neural networks,'' in \emph{2019 IEEE Symposium on Security and Privacy (SP)}.\hskip 1em plus 0.5em minus 0.4em\relax IEEE, 2019, pp. 707--723.

\bibitem{rab}
M.~Weber, X.~Xu, B.~Karla{\v{s}}, C.~Zhang, and B.~Li, ``Rab: Provable robustness against backdoor attacks,'' \emph{arXiv preprint arXiv:2003.08904}, 2020.

\bibitem{MNTD}
X.~Xu, Q.~Wang, H.~Li, N.~Borisov, C.~A. Gunter, and B.~Li, ``Detecting ai trojans using meta neural analysis,'' in \emph{2021 IEEE Symposium on Security and Privacy (SP)}.\hskip 1em plus 0.5em minus 0.4em\relax IEEE, 2021, pp. 103--120.

\bibitem{yang2019federated}
Q.~Yang, Y.~Liu, Y.~Cheng, Y.~Kang, T.~Chen, and H.~Yu, ``Federated learning,'' \emph{Synthesis Lectures on Artificial Intelligence and Machine Learning}, vol.~13, no.~3, pp. 1--207, 2019.

\bibitem{yang2020adversarial}
Y.-Y. Yang, C.~Rashtchian, H.~Zhang, R.~Salakhutdinov, and K.~Chaudhuri, ``Adversarial robustness through local lipschitzness,'' 2020.

\bibitem{adahessian}
Z.~Yao, A.~Gholami, S.~Shen, M.~Mustafa, K.~Keutzer, and M.~Mahoney, ``Adahessian: An adaptive second order optimizer for machine learning,'' in \emph{Proceedings of the AAAI Conference on Artificial Intelligence}, vol.~35, no.~12, 2021, pp. 10\,665--10\,673.

\bibitem{zeng2020deepsweep}
Y.~Zeng, H.~Qiu, S.~Guo, T.~Zhang, M.~Qiu, and B.~Thuraisingham, ``Deepsweep: An evaluation framework for mitigating dnn backdoor attacks using data augmentation,'' \emph{arXiv e-prints}, pp. arXiv--2012, 2020.

\bibitem{defeat}
Z.~Zhao, X.~Chen, Y.~Xuan, Y.~Dong, D.~Wang, and K.~Liang, ``Defeat: Deep hidden feature backdoor attacks by imperceptible perturbation and latent representation constraints,'' in \emph{Proceedings of the IEEE/CVF Conference on Computer Vision and Pattern Recognition}, 2022, pp. 15\,213--15\,222.

\bibitem{zhu2020gangsweep}
L.~Zhu, R.~Ning, C.~Wang, C.~Xin, and H.~Wu, ``Gangsweep: Sweep out neural backdoors by gan,'' in \emph{Proceedings of the 28th ACM International Conference on Multimedia}, 2020, pp. 3173--3181.

\end{thebibliography}
